\documentclass[aps,pra,floatfix,onecolumn,tightenlines,superscriptaddress,notitlepage,nofootinbib]{revtex4-1}
\usepackage{amsmath}  % was for \begin{align}, but caused error and align was not actually used
\usepackage{amssymb}
\usepackage{color,soul}
\usepackage{graphicx}
\usepackage[utf8]{inputenc} % unicode characters
\usepackage{subcaption} 
\usepackage{enumitem}
\usepackage{geometry}
\usepackage{verbatim}
\usepackage[title]{appendix}
 
\usepackage[TS1,T1]{fontenc}
\usepackage{textcomp}
\usepackage[dvipsnames]{xcolor}
\usepackage{url}
\usepackage{hyperref}  % for links in the table of contents.
\hypersetup{
    colorlinks=true, %set true if you want colored links
    linktoc=all,     %set to all if you want both sections and subsections linked
    linkcolor=Blue,  %choose some color if you want links to stand out
}

\usepackage{etoolbox}
\makeatletter
\def\@mkboth#1#2{}
\newlength\appendixwidth
\preto\appendix{\addtocontents{toc}{\protect\patchl@section}}
\newcommand{\patchl@section}{%
  \settowidth{\appendixwidth}{\textbf{Appendix }}%
  \addtolength{\appendixwidth}{1.5em}%
  \patchcmd{\l@section}{1.5em}{\appendixwidth}{}{\ddt}%
}
\begin{document}
\setcounter{footnote}{0} 
\setcounter{section}{0} 
\title{The Deep Space Quantum Link: Prospective Fundamental Physics Experiments using Long-Baseline Quantum Optics}

\author{Makan Mohageg}
\email[Corresponding author: ]{E-mail: makan.mohageg@jpl.nasa.gov}
\affiliation{Jet Propulsion Laboratory, California Institute of Technology, Pasadena, California, U.S.A.}

\author{Luca Mazzarella}
\affiliation{Jet Propulsion Laboratory, California Institute of Technology, Pasadena, California, U.S.A.}

\author{Dmitry V. Strekalov}
\affiliation{Jet Propulsion Laboratory, California Institute of Technology, Pasadena, California, U.S.A.}

\author{Nan Yu}
\affiliation{Jet Propulsion Laboratory, California Institute of Technology, Pasadena, California, U.S.A.}

\author{Aileen Zhai}
\affiliation{Jet Propulsion Laboratory, California Institute of Technology, Pasadena, California, U.S.A.}

\author{Spencer Johnson}
\affiliation{Dept. of Physics, University of Illinois at Urbana-Champaign; Illinois Quantum Information Science \& Technology Center, Urbana, Illinois, U.S.A.}

\author{Charis Anastopoulos}
\affiliation{Department of Physics, University of Patras, Patras, Greece}

\author{Jason Gallicchio}
\affiliation{Department of Physics, Harvey Mudd College, Claremont, California, U.S.A.}

\author{ Bei Lok Hu}
\affiliation{Maryland Center for Fundamental Physics and Joint Quantum Institute, University of
Maryland, College Park, Maryland,  U.S.A.}

\author{ Thomas Jennewein}
\affiliation{Institute for Quantum Computing and Dep. of Physics and Astronomy, University of Waterloo, Waterloo, Canada}

\author{ Shih-Yuin Lin}
\affiliation{Department of Physics, National Changhua University of Education, Changhua, Taiwan}

\author{Alexander Ling}
\affiliation{Centre for Quantum Technologies and Department of Physics, National University of Singapore}

\author{Christoph Marquardt}
\affiliation{Max Planck Institute for the Science of Light, Erlangen, Germany}

\author{Matthias Meister}
\affiliation{Institute of Quantum Technologies, German Aerospace Center (DLR), Ulm, Germany}

\author{Albert Roura}
\affiliation{Institute of Quantum Technologies, German Aerospace Center (DLR), Ulm, Germany}

\author{Lisa Wörner}
\affiliation{Institute of Quantum Technologies, German Aerospace Center (DLR), Ulm, Germany}

\author{Wolfgang P. Schleich}
\affiliation{Institute of Quantum Technologies, German Aerospace Center (DLR), Ulm, Germany}
\affiliation{Institut für Quantenphysik and Center for Integrated Quantum Science and Technology (IQ$^{\rm ST}$), Universität Ulm, Ulm, Germany}
\affiliation{Hagler Institute for Advanced Study at Texas A$\&$M University, Texas A$\&$M AgriLife Research, Institute for Quantum Science and Engineering (IQSE), and Department of Physics and Astronomy, Texas A$\&$M University, College Station, Texas, U.S.A}

\author{Raymond Newell}
\affiliation{Los Alamos National Laboratory, Los Alamos, New Mexico, U.S.A.}

\author{Christian Schubert}
\affiliation{Institute for Satellite Geodesy and Inertial Sensing, German Aerospace Center (DLR), Hanover, Germany}
\affiliation{Hanover, Germany Leibniz University Hannover, Institute for Quantum Optics, Hanover,
Germany}

\author{Giuseppe Vallone}
\affiliation{Dipartimento di Ingegneria dell'Informazione, Università degli Studi di Padova, Padova, Italy}
\affiliation{Padua Quantum Technologies Research Center, Università degli Studi di Padova, Padova, Italy}
\affiliation{Dipartimento di Fisica e Astronomia, Università di Padova, Padova, Italy}

\author{Paolo Villoresi}
\affiliation{Dipartimento di Ingegneria dell'Informazione, Università degli Studi di Padova, Padova, Italy}
\affiliation{Padua Quantum Technologies Research Center, Università degli Studi di Padova, Padova, Italy}

\author{Paul Kwiat}
\email{kwiat@illinois.edu}
\affiliation{Dept. of Physics, University of Illinois at Urbana-Champaign; Illinois Quantum Information Science \& Technology Center, Urbana, Illinois, U.S.A.}

%\date{\today{}}

\begin{abstract}
The National Aeronautics and Space Administration's Deep Space Quantum Link mission concept enables a unique set of science experiments by establishing robust quantum optical links across extremely long baselines.  Potential mission configurations include establishing a quantum link between the Lunar Gateway moon-orbiting space station and nodes on or near the Earth. In this publication, we summarize the principal experimental goals of the Deep Space Quantum Link mission.  These include long-range teleportation, tests of gravitational coupling to quantum states, and advanced tests of quantum nonlocality.

\end{abstract}

\maketitle

\tableofcontents

%\newpage

\section{Introduction: The Case for Deep Space Quantum Optics}
Space-based quantum optical links support future networking applications for quantum sensing, quantum communications, and quantum information science \cite{Bedington:17, Sidhu2021,qtechspace,belenchia2021quantum}.  In addition, such links enable new scientific experiments impossible to reach in terrestrial experiments \cite{rideout_2012b}.  The Deep Space Quantum Link (DSQL) is a spacecraft mission concept that aims to use extremely long-baseline quantum optical links to test fundamental quantum physics in novel special and general relativistic regimes 
\cite{Mohageg:18,DSQLspie,DSQLOsa}.  DSQL is currently in the pre-project developmental phase, with expected mission integration planned to begin in the late 2020's.  One or more nodes of a DSQL network could deploy on deep space platforms, such as the Lunar Gateway (LG) \cite{Gateway} or Orion modules.  

A key challenge of contemporary physics is the reconciliation of gravity and quantum mechanics.  Quantum Field Theory in Curved Spacetime (QFTCST), established in the 1970's, is the  most reliable theory combining two well-established theories: quantum field theory for matter and general relativity for spacetime dynamics.  It predicts effects like Hawking radiation from black holes \cite{Hawking74}, and, with extension to semi-classical gravity, provides the theoretical framework for inflationary cosmology which foretells a spatially-flat universe \cite{Guth81}. However, most tests of QFTCST, either in the laboratory setting of analog gravity, or set in strong-field astrophysical processes, are indirect tests. DSQL aims to conduct a series of direct tests of QFTCST in the weak-field regime, accessible through deployment of long-baseline quantum optical links in the Earth-Moon system. 

The essence of the planned experiments with DSQL is to transmit photons between inertial reference frames and across gravitational gradients to implement long-baseline teleportation, test Bell’s inequality, conduct photonic tests of the Weak Equivalence Principle, and validate relativistic predictions through single- and two-photon interference.  In this context, the fundamental question is how to describe the evolution of the photonic state between creation and measurement.  The simple definition of a trajectory which makes sense in classical physics is not a suitable notion for the description of the motion of quantum particles. In quantum physics, the fundamental concepts are time-evolving wave functions, or more generally, consistent histories.  Quantum systems can be prepared in highly delocalized states such as entangled states. When discussing particle propagation in curved spacetime, it is necessary to express particles in terms of quantum fields.  This is because QFTCST is currently the only theory that provides a consistent and general way of coupling quantum matter to background gravitational fields.  Furthermore, the quantum field description is essential for describing higher-order coherences of the electromagnetic field, and for the formulation of a photodetection theory.

The weak equivalence principle states that all objects ``fall'' with the same acceleration regardless of their mass and composition. As embodied in general relativity, it says that all neutral objects, massive and massless, follow geodesic trajectories of the metric, dependent only on initial conditions of position and velocity, and not on mass or composition. The extension of the equivalence principle established in classical mechanics to quantum physics is an important challenge for both theory and experiment \cite{AnHu18a,Roura2020}. Trajectories emerge for quantum histories as an approximation that is valid only for specific quantum states (quasi-localized), under specific conditions (sufficiently decohered) and in specific experimental setups with suitable measurement protocols. Therefore, it is necessary to phrase the equivalence principle in terms of quantum mechanical notions, namely, preparations of the quantum state and statistics of measurement outcomes. For photons, the equivalence principle must be expressed in terms of photodetection probabilities.

Any Colella-Overhauser-Werner (COW) \cite{COW} type test of the equivalence principle that touches upon quantum properties of photons or massive particles will have the significance of being the first direct test of QFTCST.  This benchmark experiment is proposed to be conducted with photons using the DSQL. 
The first  photonic COW test planned for  DSQL will use weak coherent pulses.  Single-photon superposition state tests will follow.  Subsequently, increasingly complex tests using entangled and hyper-entangled states will be conducted.  This sequence of experiments conclude with measuring the gravitationally induced phase shift of frequency-entangled photons using intrinsically nonclassical interference in a Hong-Ou-Mandel (HOM) interferometer.   

Furthermore, the DSQL could test for new physics of gravitational
origin manifested as non-unitary channels in the time
evolution   of photons and/or matter waves. Such channels have
been proposed in different contexts, including 
Ghirardi-Rimini-Weber-type of collapse models \cite{GRWreview},
Diosi-Penrose gravity-induced collapse model \cite{Diosi0,Penrose}, and the
gravity-induced decoherence mechanisms \cite{AHGravDec, Miles}. These processes typically result in a loss of visibility in interference experiments. 
The predicted effects are too weak in Earth-based or space-station experiments, but deep-space experiments with long unbalanced interferometers could provide noticeable constraints that could be used to test the viability of alternative quantum theories.

Long-baseline links enabled by deep-space platforms create a high-latency quantum communication network, sufficient to incorporate human decision-making outside of the light cone of synthesis events.  DSQL could use this to perform a long-range Bell test with and without human involvement.  Such an experiment would test for local hidden variables across long distances, and address the ``free will'' Bell test loophole.

The DSQL could perform tests of quantum optical teleportation and entanglement swapping between inertial frames. The inertial frames are defined by the relative velocity of the nodes in the communication network, and their positions with respect to the central mass. Quantum states depend on the choices of reference frame and gauge condition, and these are not physical entities; only the measurement outcomes are physical. Consider the scenario where an entangled-photon source is centered between two receivers, roughly equidistant on either side, arranged to travel either towards each other or away from each other. In the rest-frame of the source, the subsequent two detection events are simultaneous and therefore space-like separated. These scenarios are interesting because it is possible to construct models of physics where the detection event at one measurement apparatus seemingly instantaneously prepares the measurement outcome of the other sub-system, as initially discussed by Aharonov and Albert in 1981 \cite{AharanovAlbert} and by Moffat in 1997 \cite{Moffat97a}. However, models of this type would be falsified if quantum correlations satisfy the standard Bell tests.

Beyond their intrinsic scientific interest, demonstrations of long-baseline Bell tests and photonic teleportation are key technological achievements with direct relevance to the future deployment of global-scale quantum networks and sensor systems. The large relative velocities and network latency due to long baselines could degrade network performance if left uncorrected.  Effects of gravity would also directly impact timing synchronization for high-rate quantum communications.  The proposed DSQL tests of long-baseline quantum teleportation between different inertial frames, long-baseline Bell tests, and tests of gravitational phase shifts will thus inform future network architectures.

Accomplishing these experiments requires access to different mission configurations.  Some of the experiments can be accomplished with a single spacecraft and ground station, while others require multiple spacecraft, or a pair of ground stations.  The orbital configurations optimized against the experimental goals include LEO, MEO, Lunar orbits, and exotic orbits about the Earth-Moon Lagrange points.  The flight- and ground-system technology required to achieve the experimental goals are also of direct relevance to terrestrial quantum networking: high-rate, high-fidelity, entangled photon pair sources; high count-rate, low-jitter, and low dark count-rate single-photon counters; narrow-band optical filters; temporal synchronization and time transfer for the quantum channels; and, in some cases, quantum memory development.  

In the following sections we highlight potential science experiments that could be conducted by the DSQL mission, and the relevant technology development required.  We also present a link model to estimate the efficacy of deep space optical links under different experiment configurations.  The article concludes with a survey of graduated space- and ground-based opportunities to advance the science objectives of the DSQL mission.

\section{Deep Space Quantum Link Science Experiments}

This section describes the experiments proposed for the DSQL mission, which among other outcomes, would enable testing  theories predicting novel coupling mechanisms between a photonic quantum state and gravity, with the goal of bounding these predictions. The experiments are categorized by the similarity of their scientific objectives.  These are: GR effects on light and tests of the equivalence principle, long-baseline Bell tests, long-baseline quantum teleportation, and applications of squeezed light.  Individual experiments are achievable using a diverse set of different mission architectures.  These architectures could involve multiple ground and/or space network nodes, all with different instrumentation packages.  Subsequent sections summarize these implementation options.

\subsection{GR Effects on Light and Tests of the Equivalence Principle\label{sec:equivalence}}
\setcounter{footnote}{0} 
Light propagating across a weak gravity field is subject to phase delays caused by the characteristics of the local spacetime \cite{Scully1982}.  The DSQL mission will characterize these delays in the quantum optical regime, in order to test the underlying predictions from general relativity.  Another interpretation is that the proposed experiments will test quantum optical interference phenomena in the regime where gravitational effects are resolvable.  These tests embody the notion of testing predictions of the weak-field regime of QFTCST.  Essential background on Einstein's equivalence principle (EEP), and calculations of the magnitude of the predicted general relativistic effects, are summarized prior to description of the specific DSQL experiments.

Einstein's equivalence principle states that the outcome of any local experiment in a freely falling frame will be the same as in Minkowski spacetime, i.e., in the absence of gravitational fields, provided that all relevant length scales are much smaller than the characteristic curvature radius of spacetime.  This guarantees that gravitational tidal effects will not affect the outcome of the local experiment.  
The equivalence principle, which applies to situations where self-gravitation effects are negligible, comprises three different (but intertwined) aspects \cite{Will2014}:
\begin{itemize}
\item universality of free fall (UFF),
\item local Lorentz invariance (LLI),
\item local position invariance, also referred to as universality of gravitational redshift (UGR).
\end{itemize}
These have implications on the propagation of electromagnetic wave packets in curved spacetime, as well as on the comparison of clocks and frequency references at different locations.
In particular, since the relevant modes of the electromagnetic field will have a transverse size far smaller than the spacetime curvature radius, their propagation is well described by the eikonal approximation, with light rays corresponding to null geodesics.

While this approach has been thoroughly tested for classical electromagnetic waves, DSQL would offer a unique opportunity to test it for various quantum states of light, including true single-photon states, quantum superposition states, and entangled and hyperentangled states. 

For optical frequencies, quantum mechanics and general relativity predict the same effect on single photons as on classical light, i.e., the effect is on the \emph{mode} of the field, independent of the precise quantum mechanical excitation of that mode. The proposed DSQL experiments test different aspects of these predictions, using long-range quantum optical channels between inertial frames.

The superposition of quantum states of a single photon was  experimentally tested along space channels by observing the single-photon interference at a ground station due to the coherent superposition of two temporal modes \cite{Vallone2016b}. The measurements were carried out using an unbalanced interferometer to create the two temporal modes, and a matching one at the receiver to recombine them, thereby recovering the fringe visibility. The space channel was realized by exploiting retroreflectors mounted on several spacecraft, as originally exploited for the first single-photon exchange in space \cite{Villoresi2008a,Dequal2016} and recently extended to 20,000 km for MEO orbits \cite{Calderaro2018a}.   In contrast to the experiments conducted to date, which have classical analogs, the tests proposed for the DSQL mission have the advantage of controlled and verifiable quantum statistics and entanglement, leading to measurements that are intrinsically nonclassical.

In addition to verifying that frequency, superposition, and entanglement do not affect the photon propagation, DSQL could measure independent relativistic effects in controlled experiments. Experiments could be designed to independently examine the following effects:
\begin{itemize}
    \item ``Moving clocks run slow'' from special relativity
    \item ``Clocks located down a gravitational well run slow'' from general relativity
    \item Doppler effect from differences in path length between successive clock ticks
    \item Experimental offsets due to drifting equipment
\end{itemize}

These effects are important to the operation of global navigation satellite systems and were calculated in the design-phase of GPS \cite{alleyGPS}. The effects are large enough to matter, but small enough to be treated as first- and second-order corrections. For example, a clock  near the surface of Earth will measure an extra 10 ns/day for every kilometer of altitude \cite{Burns2017}. More generally, when we compare the rate of a clock near the rotating surface of Earth to one on an orbiting satellite, the fractional change is given by simple dimensionless correction factors involving the fraction of the speed of light $v/c$ and the Schwarzschild radius of the gravitational potential well (in this case for the Earth), given by $R_\textrm{Schwarzschild} = 2 G_N M_\textrm{Earth}/c^2 \approx 1\,\textrm{cm}$. The ratio of the  clock-tick interval on the satellite to one moving with the surface of Earth is \cite{taylor2000exploring}

\begin{equation}
 \frac{dt_\textrm{satellite}}{dt_\textrm{Earth}} \approx 1
    + \underbrace{ 
        \frac{1}{2} \left( \frac{R_\textrm{Schwarzschild}}{r_\textrm{Earth}} +  \frac{v^2_\textrm{Earth}}{c^2} \right) 
    }_{\epsilon_\textrm{observatory}} - \underbrace{
        \frac{1}{2} \left( \frac{R_\textrm{Schwarzschild}}{r_\textrm{satellite}} + \frac{v^2_\textrm{satellite}}{c^2} \right)
    }_{\epsilon_\textrm{satellite}}.
    \label{eqnGR_clock_slowing}
\end{equation}

The first correction is constant at $7\times10^{-10}$, with $10^{-13}$ variation for different elevations.

(This is for an observatory at the equator, but for observatories elsewhere, the reduction in velocity is compensated exactly by accounting for the Earth geoid.) The second correction depends on the satellite orbit. For a satellite in a circular orbit, a good approximation to this order is to balance centripetal force against Newton's gravitational force, in which case

\begin{equation}
    \frac{m \, v^2_\textrm{satellite}}{r_\textrm{satellite}}
   = \frac{G_N M_\textrm{Earth} \, m}{r^2_\textrm{satellite}} 
    \qquad \textrm{(circular orbit)} .
\end{equation}
Writing this in terms of $R_\textrm{Schwarzschild}$, we recover the somewhat-well-known result for GPS that the time dilation from special relativity is half of that from general relativity's gravitational effect. The overall satellite correction is then simply
\begin{equation}
    \epsilon_\textrm{satellite} = \frac{3}{4} \frac{R_\textrm{Schwarzschild}}{r_\textrm{satellite}}
    \qquad \textrm{(circular orbit)} .
\end{equation}
and is a small effect given that $R_\textrm{Schwarzschild} \approx 1\,\mathrm{cm}$. This term is $10^{-9}$ for a LEO orbit like ISS and falls to $1.6\times10^{-10}$ for a GEO orbit. The observatory and satellite terms completely cancel at $r_\textrm{satellite} = 10^{7}\,\textrm{m}$, which is in the inner Van Allen Belt.

\begin{figure*}
\centering
    \includegraphics[width=0.9\linewidth]{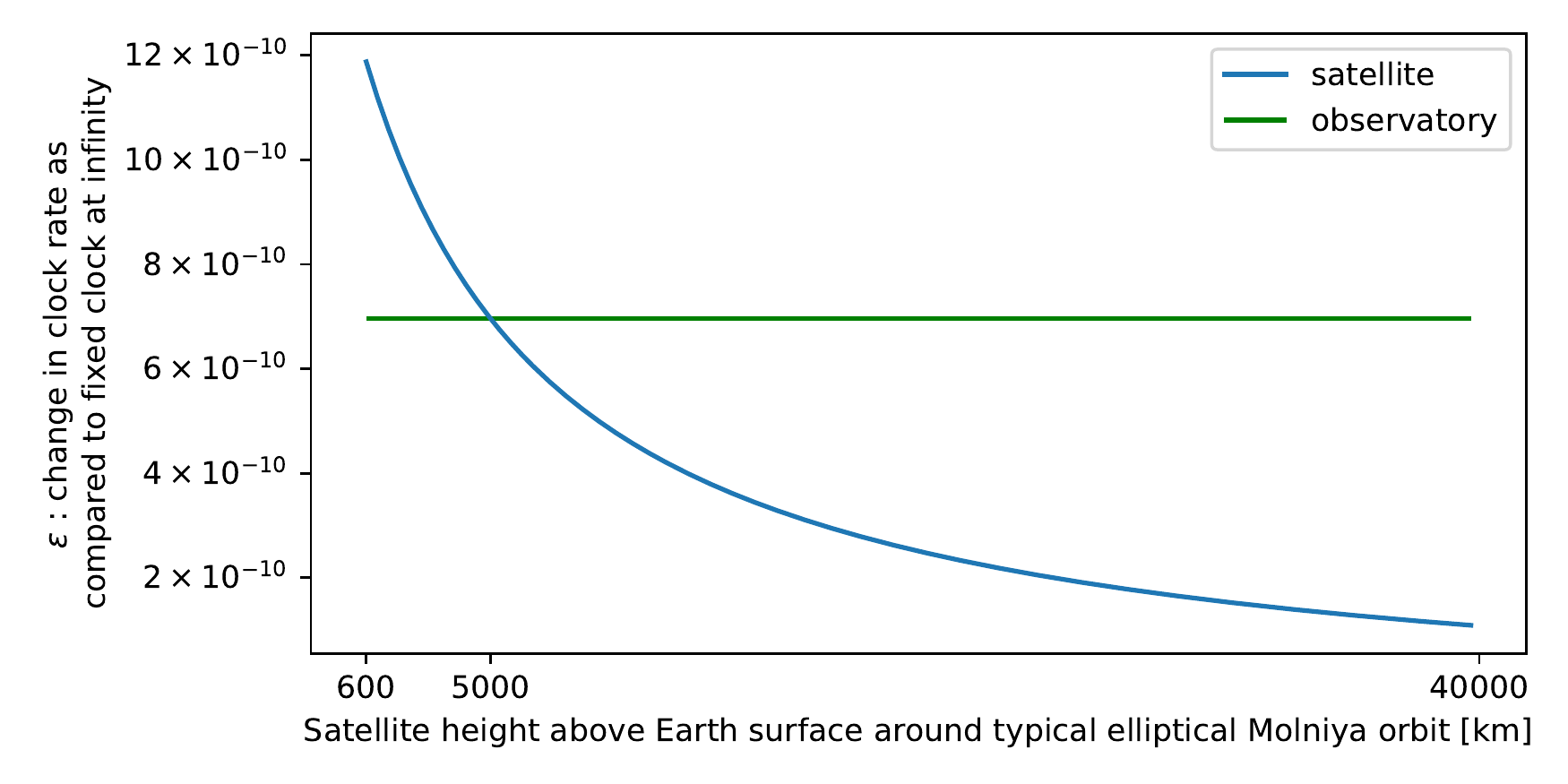}
    \caption{Total relativistic time dilation (from both gravitational and velocity effects) along a highly elliptical satellite orbit (blue) and at a fixed observatory on the surface of Earth (green). These effects subtract (Equation \ref{eqnGR_clock_slowing}) to get the relative rate of clocks in the two locations. Note that the sign of the relative rate can be positive or negative, depending on the satellite altitude.}
    \label{fig:GR_epsilon_clock_Molniya_orbit}
\end{figure*}

For a spacecraft in an elliptic orbit with semi-major axis $a$, the relation between velocity $v$ and distance $r$ is
\begin{equation}
    v^2 = GM\left( \frac{2}{r} - \frac{1}{a} \right)  
    \qquad \textrm{(elliptic orbit)} ,
\end{equation}
or in terms of dimensionless ratios and $R_\textrm{Schwarzschild}$,
\begin{equation}
    \frac{v^2_\textrm{satellite}}{c^2} = R_\textrm{Schwarzschild} \left( \frac{1}{r_\textrm{satellite}} - \frac{1}{2a} \right) \qquad \textrm{(elliptic orbit)} .
\end{equation}
The overall satellite correction for a clock in an elliptic orbit (with semi-major axis $a$), including both special and general relativistic effects is then
\begin{equation}
    \epsilon_\textrm{satellite} = R_\textrm{Schwarzschild} \left( \frac{1}{r_\textrm{satellite}} - \frac{1}{4a} \right)
    \qquad  ,
\end{equation}
which reduces to the circular-orbit case when $a=r_\textrm{satellite}$, but varies throughout an elliptic orbit as the satellite approaches and recedes from Earth. Thus, a highly elliptical orbit breaks the degeneracy of constant offsets due to experimental drift. Even the sign of the total relativistic correction may change throughout the orbit. An example for the ``typical Molniya orbit'' from Wikipedia\footnote{\url{https://en.wikipedia.org/wiki/Molniya_orbit}} is shown in Figure \ref{fig:GR_epsilon_clock_Molniya_orbit}. If the atmosphere is deemed to be a problem, the effect can be enhanced by comparing two Earth-orbiting satellites, one near apogee and one near perigee. Note that for the Lunar Gateway's near-rectilinear halo orbit\footnote{\url{https://www.esa.int/ESA_Multimedia/Images/2019/07/Orbiting_lunar_Gateway}}, the values of $\epsilon_\textrm{satellite}$ are more than 50 times smaller due to the 80 times smaller mass of the moon, while the proposed low lunar orbit of the Lunar Gateway results in $\epsilon_\textrm{satellite}$ about twice as small.  

This order-of-magnitude analysis shows that an Earth-Moon link would {\itshape not} be an ideal starting point for measuring gravitationally induced phase shifts in quantum light. The two-body gravitational field of the Earth-Moon system reaches an inflection point near the first Lagrange point.  Future experiments may leverage Earth-Moon links  to test for modulation, or even full cancellation of the phase shift of a photon propagating across this region \cite{Mohageg:18}.  In contrast, other quantum optical experiments, such as the tests of Bells inequality and tests of quantum teleportation, described in detail below, are enhanced through use of an Earth-Moon baseline.

These calculations capture the difference in rates that the clocks run on the satellite as compared to on Earth. As we will show, these ``clocks'' can be the time bins of pulses or the optical-frequency oscillations of the light itself. The gravitational blueshift (redshift) experienced by the photons as they fall toward (climb away from) Earth is already included in these expressions and should not be double counted. The same applies to length contraction in any fiber delay line or interferometer---Einstein's light-clock thought experiments shows that these effects are already included in the slow-down of the clock ticks that an external observer sees.

In addition to these relativistic effects, there is a large Doppler shift caused by successive pulses or successive crests of waves being sent and received from different places and therefore taking different amounts of time to arrive at the observer. In practice, this will be the dominant effect, but can be separated from the intrinsic time dilation by its different orbit dependence, i.e., the sign of the effect depends on whether the satellite is approaching or moving away from the ground terminal. This shift is first order in the relative velocities $v/c$:
\begin{equation}
    \left| \frac{\delta t}{t} \right| = \left| \frac{\delta f}{f} \right| \approx \frac{v}{c} \qquad \textrm{(Doppler)} ,
\end{equation}
so any series expansion must be kept to second order to match the size of the leading-order $v^2_\textrm{satellite}/c^2$ terms above from special relativity. Note that using the formula for the relativistic Doppler effect double-counts the source's time dilation. To avoid double-counting, it is best to use global coordinates like the Schwarzschild coordinates that asymptote to Minkowski spacetime infinitely far away: 
\begin{itemize}
    \item Fix an infinitesimal time difference between two pulse transmissions.
    \item Calculate the global time difference of reception, which will be longer or shorter by an amount on the order of $v/c$, depending on the orbit and orientation. This is $10^{-5}$ for LEO satellites.
    \item The curved path of light through the Schwarzschild metric can be computed numerically, but its effect is $10^{-15}$, which is small compared to the special and general relativistic time dilation.
    \item Each of these time differences can then be translated from global time to the time experienced by clocks on the satellite and on Earth. This $\approx v^2/c^2$ effect is captured by Equation \ref{eqnGR_clock_slowing} and is of order $10^{-10}$ for LEO satellites.
\end{itemize}

\subsubsection{Polarization Rotation of Photons in General Relativity}

Polarization rotation also occurs when light traverses the warped space around a rotating body. This is the ``frame dragging'' or ``Lense-Thirring'' effect.  An introduction to the effect can be found in Schleich and Scully's Les Houches lecture \cite{Scully1982}. Recent calculations provide numbers relevant for satellite experiments. Brodutch and Terno \cite{Brodutch2011} quotes a rotation of 
55 milliarcseconds ($3 \times 10^{-7}$ rad)
when sending a photon from a satellite out to infinity, but this calculation is ``gauge dependent'' and therefore unmeasurable---the fixed reference frame of stars the satellite is measuring with respect to is also frame dragged.

Brodutch et al. \cite{Brodutch2015} consider a closed loop interferometer, where the polarization is compared at the same point (as differential geometry requires) to make comparisons unambiguous. For an interferometer 100\,km by 10\,km they find a polarization rotation of only 4 pico-arcseconds ($2 \times 10^{-17}$ rad), concluding correctly that the ``minuscule scale of the effect puts it beyond the current experimental reach.''\footnote{For any photon-counting experiment, the statistical uncertainty of photon polarization angle measurement scales inversely as the square root of the number of photons detected: $\Delta \theta = \tfrac{1}{2\sqrt{N}}$ radians for large $N$. Completely disregarding all systematic effects of trying to establish a common reference frame, reaching the required sensitivity requires at least 4 million detection events for the observable ``out to infinity'' case, but $10^{33}$ (a megawatt of 1550 nm photons continuously for 30 years) for the observable ``compare at the same place'' setup.}

\subsubsection{COW Test with Classical Light, Superposition States, Entangled and Hyper-entangled Photons}
 \label{sec:COWtests}
Quantum optics experiments enable a unique window to explore the interplay between quantum mechanics and relativity, both general and special.  The Colella, Overhauser, and Werner (COW) experiment used a neutron interferometer to test the influence of the gravitational field
on a quantum wave function \cite{COW}. Since then many experiments using matter-wave interferometry with atoms and molecules have been performed.  However, all such experiments to date can be interpreted using a Newtonian gravity framework. In contrast, tests with massless particles -- photons -- would require a general relativistic description. There have been several proposals to detect general relativistic effects on single photons using Mach-Zehnder interferometers (MZI) whose arms are located at different altitudes, i.e., at different gravitational potentials. In this section, an overview of the background of photonic COW tests is presented, followed by a detailed discussion of specific experimental concepts.

Both in the original COW experiment and in light-pulse atom interferometers \cite{Kasevich1991,Kleinert2015}, the matter-wave packets are diffracted by periodic gratings (a crystal in neutron interferometers and one or more optical lattices in the atomic case) but freely falling the rest of the time. This implies that the proper-time difference between the interferometer arms is insensitive to gravitational time dilation effects in a uniform field, as can be argued by considering a freely falling frame \cite{Roura2020}. On the other hand, every time a matter-wave packet is diffracted by a grating, it acquires a phase that depends on the central position of the wave packet with respect to the grating, and the total phase shift between the two interfering wave packets depends on the relative acceleration between the wave packets and the diffraction gratings. This can be exploited in high-precision gravimetry measurements \cite{Peters2001} and tests of the universality of free fall (UFF) \cite{Asenbaum2020}.

In contrast, atom interferometers where the wave packets in the two arms are held at two different constant heights through matter-wave guides can be sensitive to gravitational time dilation in uniform fields  \cite{Roura2020}. These kinds of interferometers are, in fact, closer analogs of the optical interferometers considered below, where photons in the two interferometer arms propagate along optical-fiber 
delay lines located at two different heights in a gravitational field.

Indeed, Hilweg et al.\ have proposed a ground experiment that uses an actively switched,  triple-arm MZI, with long fiber loops to extend the interaction time, and thus the size of the effect \cite{Hilweg_2017}. For a vertical separation of 3 m and 100-km fiber delays, their model predicts that an effect might be observed after 2-4 days of integration time. Obviously, maintaining stability over that length of time could be quite difficult, particularly with long pieces of optical fiber that are subject to thermal expansion.  Additionally, to minimize effects of dispersion, narrow-bandwidth photons (<100 MHz) are needed, which can also be challenging.
 
 \begin{figure*}
     \centering
     \includegraphics[width=12cm]{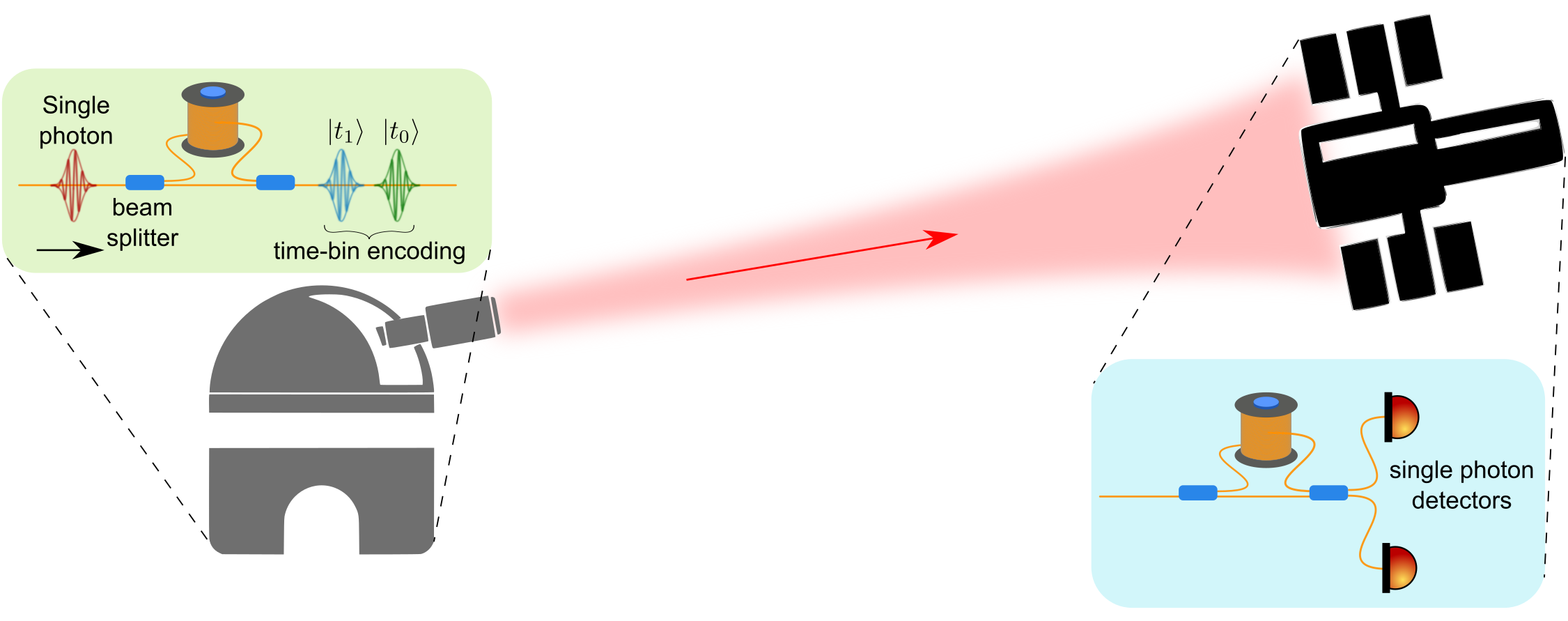}
     \caption{ Simplified scheme of the optical COW experiment in space. A time-bin superposition is generated by injecting a single photon wavepacket into a unbalanced MZIs. The photon is sent towards a satellite, where an identical MZI is located. Interference detected at the satellite reveals the gravitationally induced 
     phase shift.}
     \label{fig:COW_space}
 \end{figure*}
 To increase the magnitude of the gravitational effect, greater variation of the gravitational potential can be employed~\cite{Zych2012}. The idea proposed in~\cite{rideout_2012b} is to use two identical unbalanced MZIs, one located on ground and the other on a satellite (see Fig. \ref{fig:COW_space}). Single-photon wavepackets enter the ground MZI
and at the output a coherent superposition of two wavepackets $|t_0\rangle$ and $|t_1\rangle$, known as time-bin encoding~\cite{Brendel1999}, is generated.
The single photons are then sent toward the spacecraft MZI and at the exit of the second MZI an interference effect can be observed in the ``middle'' time bin (corresponding to photons that took the short path in one unbalanced interferometer and the long path in the other).

The general formalism of the relation of the different optical paths involved in COW tests in space by means of optical interferometry requires a careful treatment of the frequency transformation in each path as well as the relation of time with distance. According to a general analysis by Terno et al. \cite{Terno2019a}, the crucial asset to obtain a closed form is the phase difference due to the difference in the emission times. 
When the two MZIs are properly calibrated, a phase shift due to the gravitational redshift will be observed.
In this case, if the satellite is not geostationary, the main challenge is represented by the necessity to compensate for the first-order Doppler effect.
To solve the above issue, 
an improvement of the above scheme was recently proposed to measure the Doppler shift introduced by the relative motion of the satellite to ground, and to remove it from the final result~\cite{Terno2018, Magnani2019}: in addition to MZIs on the spacecraft and the ground station,
satellite retroreflectors located in space are used to send some portion of the upcoming light back to the ground. 
The latter, detected on ground, can be used to measure the 
first-order Doppler effect, since gravitational effects compensate in the two-way propagation.
 The Doppler shift was assessed in previous experiments \cite{Vallone2016b} and exploited as the modulator of the time-bin qubit phase as a function of the instantaneous velocity of the satellite. This modulation, though passive, may also be seen as a resource when ascertaining the visibility of the interference phenomena. 
 
 The photon temporal superposition state along the space channel may be kept in a single polarization by exploiting suitable corner cube technology for the retroreflectors, as demonstrated for space quantum communications \cite{Vallone2015b}, thus allowing for a large parameter space for the observables under test; the latter use was pivotal in the test in space of the Wheeler ``delayed choice Gedanken-experiment'', addressing the well-known wave-particle duality of quantum physics \cite{Vedovato2018a,Agnesi2019}.
We note that the scheme proposed in~\cite{Terno2018, Magnani2019} exploits classical light in order to test the Einstein Equivalence Principle in the optical domain. However, extending the scheme by using single-photon wavepackets will allow the measurement of a gravitationally induced phase shift on a quantum state.
Finally, the temporal resolution for the discrimination of the interference is a sensible parameter for the experimental design. Present limits in the case of a link to a MEO satellite are of order of a quarter of nanosecond \cite{Agnesi2019}.

The gravitational phase shift measured by the experiment depicted in Fig.~\ref{fig:COW_space} is connected with a small shift, caused by the gravitational field, of the time difference between the modes encoding the time bins. This shift is of the order of $10^{-10}$ times the temporal separation between the two time bins, which implies that the two delay lines (on ground and in space) need to be matched at that level. For example, 100-m delay lines would need to be matched with a precision better than 10\,nm.
One way of achieving this precision is to independently calibrate the length of each delay line with a local frequency reference, such as an atomic clock, and convert to distance using the universal speed of light. The experiment can then be interpreted as a test of UGR where the two frequency references at different heights are compared through quantum states of light.

If time-bin entangled photons~\cite{Brendel1999} are used, the gravitational shift also affects the relative phase of the entangled state: for instance, by generating the entangled pair $|\psi\rangle=(|t_0\rangle_A|t_0\rangle_B+|t_1\rangle_A|t_1\rangle_B)/\sqrt{2}$ on the ground, and sending photon $A$ to the ground MZI and photon $B$ to the satellite MZI, the entangled state will be transformed (before the MZIs) into 
$|\psi'\rangle=(|t_0\rangle_A|t_0\rangle_B+e^{i(\delta\Phi_g + \delta\Phi_D)}|t_1\rangle_A|t_1\rangle_B)/\sqrt{2}$
where $\delta\Phi_g$ is the gravitationally induced phase shift and $\delta\Phi_D = k \ell v/c$, for a wave number $k$ and delay length $\ell$, is the Doppler correction from the fact that the two time bins are emitted and absorbed at slightly different places; here $v$ is the projection of the relative velocities along the optical link, which varies from approximately $-v_{spacecraft}$ to $+v_{spacecraft}$ as the spacecraft flies overhead. For a 10-km delay line and satellite velocity of 10\,km/s, $\ell v_{spacecraft}/c$ = 0.33\,m, corresponding to $\delta\Phi_D = 1.3\cdot 10^6 \,$rad, for a laser wavelength of 1500~nm; as discussed above, any attempt to precisely determine $\delta\Phi_g$ will need to compensate for this much larger value of $\delta\Phi_D$. For example, one could send a classical beacon along (also in a superposition of $t_0$ and $t_1$), and use that to stabilize the remote MZIs by using a fiber stretcher or a piezo-electrically controlled optical ``trombone'', as was done in \cite{Chapman2019space}; comparing the error signal to the expected one would allow one to look for deviations. Alternatively, as discussed above, corner cubes could be used to directly measure the satellite's relative motion, and a local feedback system (with a stabilized laser) could be used to set the unbalanced MZI path length.

More advanced schemes can also be conceived with time-bin entanglement: for instance, if the entangled source is located on a satellite, one photon can be directed towards a ground station and the other photon towards a different satellite,
enhancing the gravitational effect and at the same time allowing Bell tests between frames with large relative velocities (see Section~\ref{BellTest}). 

Finally, one can also conceive experiments that utilize {\it hyper-entanglement} -- photons that are simultaneously entangled in multiple degrees of freedom \cite{Kwiat1997}.
As discussed above, special and general relativity result in predictable time dilation that would affect time-bin entanglement in a measurable way, but would have very little effect on polarization entanglement, with only unmeasurably small frame-dragging effects. Extending the prototypes in \cite{Chapman2019space}, one could repeat the proposed optical COW experiments with a source of photons entangled both in time-bin and in polarization, again with one photon measured on the ground and the other on the space platform. One can thus directly compare the effect on the entanglement in the two degrees of freedom, one of which is sensitive to the relativistic effects and one of which is not. As with the previous suggestions, if precise enough measurements can be made, the Doppler and relativistic effects can be distinguished from each other and from constant offsets by putting the satellite in an elliptical orbit as discussed in Section \ref{sec:equivalence}. 

\subsubsection{Flight Mission Design for Optical COW Tests}
\label{sec:COW_Mission}
For the quantum photonic COW tests that involve classical light, the measurement {\itshape requirements} follow the procedure of \cite{Dequal2016}.  In Appendix \ref{sec:linkAnalysis}, the requirements derived from the parametric model to achieve a statistical confidence level are evaluated using a quantum-channel model. Here we use these to estimate the high-level system performance requirements to achieve the DSQL science objectives.  

\begin{figure}[ht]
   \centering
        \includegraphics[width=7.5cm]{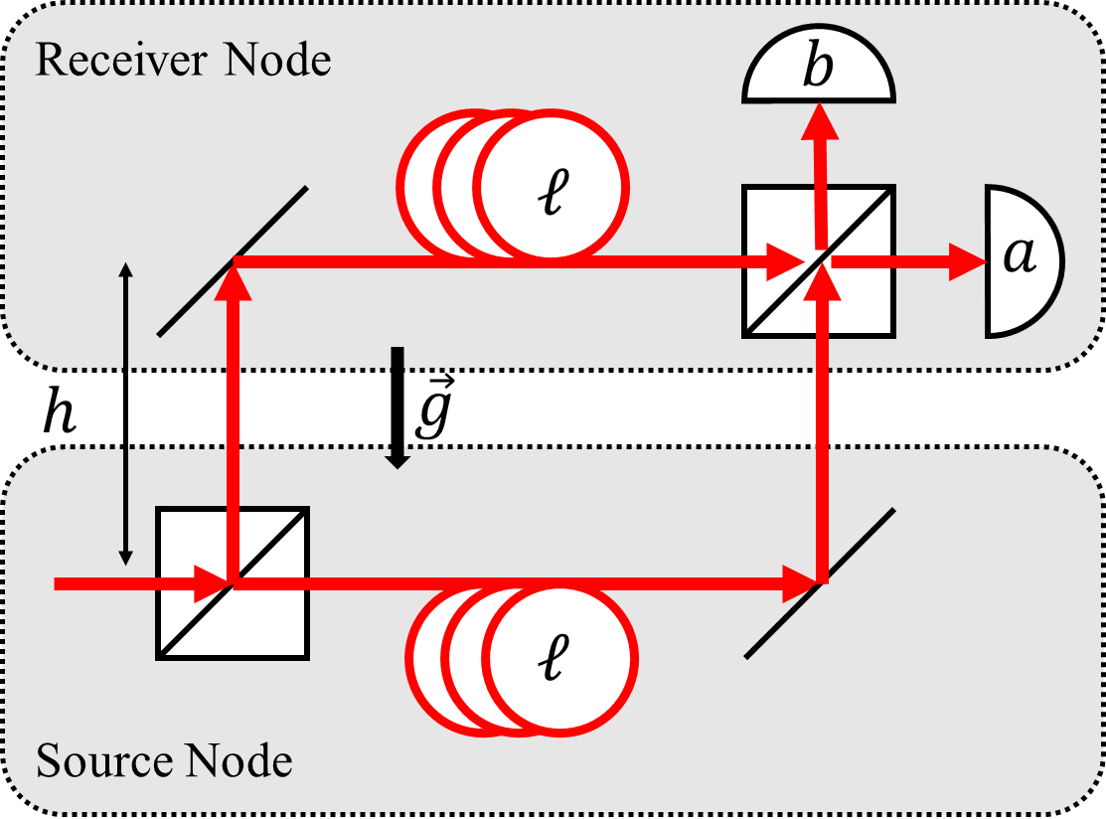}
            \caption{A simplified MZI for quantum optical COW tests.  The source node and receiver node are at different heights $h$ measured along the direction of the gravity vector $\vec{g}$.  Each node has a long optical delay line of length $l$. Note that this is equivalent to the arrangement with two unbalanced MZIs shown in Fig. \ref{fig:COW_space}, which allows photons emitted at different times to propagate along essentially the same spatial mode (modulo whatever lateral shift has occurred due to the motion of the source in the time between the early and late emission times).}
            \label{fig:SimpleCOW}
\end{figure}

The proposed optical COW experiments are carried out by using quantum optical interferometry, for example, by using an MZI with a single photon input into one port (Figure \ref{fig:SimpleCOW}).  In this example, the photon is in a path superposition. The paths are characterized by different values of the gravitational potential; the time dilation between these two paths causes a relative phase shift along the two arms. A number of specific implementations of this concept, describing tests using coherent superposition states, single photons, and different forms of entangled and hyperentangled photon pairs are described in the previous section. 

In a uniform gravitational field, this phase shift is linked to the interferometer dimensions and gravity via the formula:
\begin{equation}
\label{eq:LMalpha}
    \phi_{GR}=\omega_0\tau_{GR}\sim(1+\alpha)\frac{2\pi}{\lambda}\frac{g h \ell}{c^2},
\end{equation}
where $\omega_0$ and $\lambda$ are respectively the central frequency and vacuum wavelength of the photon emitted by the source in its reference frame, $\tau_{GR}$ is the time delay between the two interferometric paths, $g$ is the acceleration due to gravity, $h$ is the orbital altitude difference between source and receiver, and $\ell$ is the horizontal length of optical delay \footnote{More precisely, $n(\lambda) \ell $, if the delay has some non-unity refractive index, as would definitely be the case for an optical fiber as considered here.} in both the source and receiver interferometers.  The parameter $\alpha$, which is equal to zero for general relativity, parametrizes violations of UGR.  

The primary goal of this set of experiments is to validate the predicted gravitational phase shift in quantum light; the expected magnitude of the phase shift for the spacecraft implementation of this experiment is on the order of a few to tens of radians. For example, $ \phi_{GR}\sim 1\,\,$rad for $\lambda= 1550\,$nm, $h=400\,\,$km, and $\ell= 6\,\,$km.  
The other, equally important, objective of this test is to bound the magnitude of parameter $\alpha$, which should be zero according to general relativity, in different regimes of quantum light. Both $\phi$ and $\alpha$ are measured using quantum interferometry.  There is an associated integration time associated with each interferometer measurement, where $N$ signal measurement events are captured.  Qualitatively, the instrument and mission design requirements should enable a sufficiently high rate of signal events, relative to noise events, and a net interferometer stability that yields a phase error smaller than $\phi$. Note that there is a tradeoff, due to  exponential Beers-law decay in the fiber-optic delay $\ell$, and diffractive propagation loss associated with the free-space portion $h$ of the interferometer (assuming the send and receive telescopes are sufficiently small that the received flux falls as $1/R^2$; see Appendix \ref{sec:linkAnalysis} for details); increasing $h$ and $\ell$ both increase the magnitude of $\phi_{GR}$ but decrease the number of photons with which to measure this.

\begin{figure*}[ht]
   \centering
        \includegraphics[width=12cm]{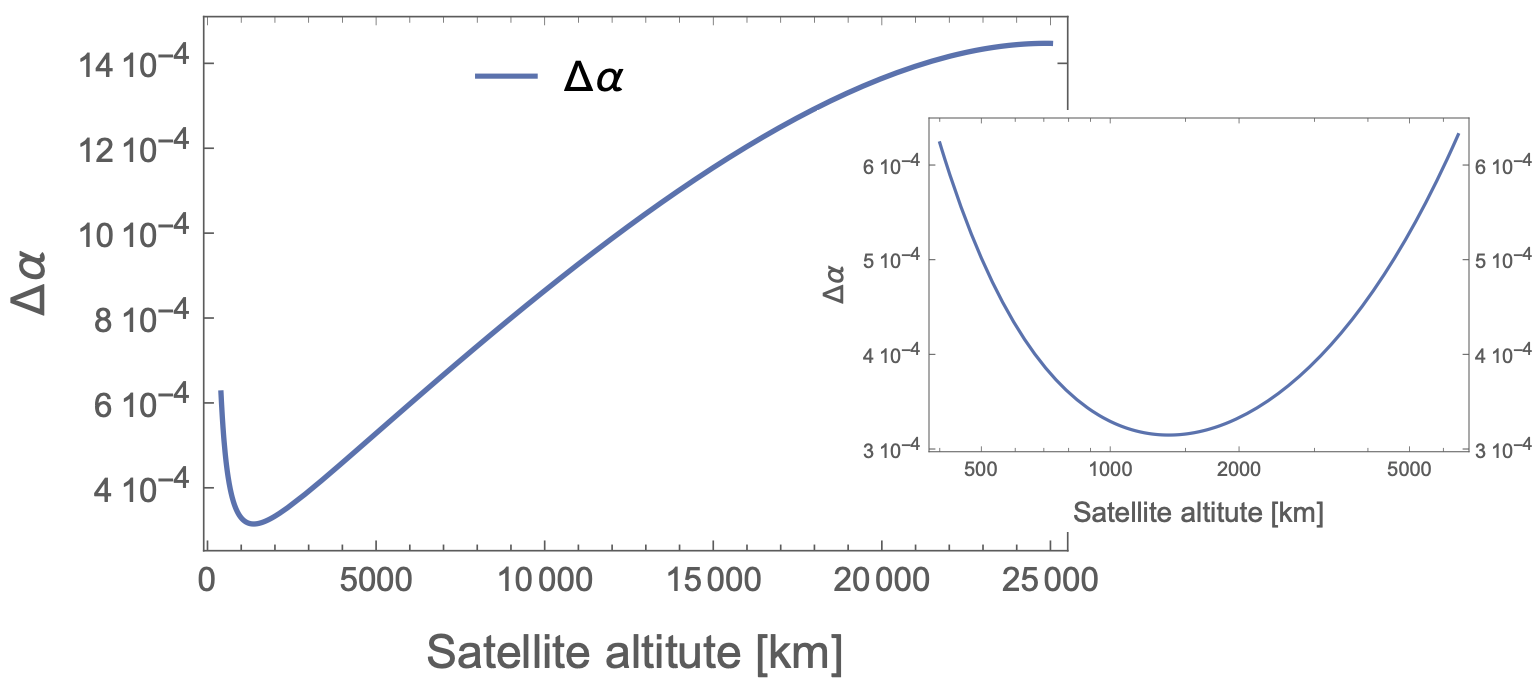}
            \caption{ Plot of the predicted error on $\alpha$ after a satellite passage, as a function of the  satellite altitude in kilometers, assuming circular orbits, and 1.0-m and 0.3-m transmit and receive telescope apertures. The inset shows the trend close to the optimal altitude. }
        \label{fig:COW_Alpha}
\end{figure*}

The development of a quantitative model to express these requirements is described in  Appendix \ref{sec:COW_appendix}.  Our approach treats $N$, the total signal flux collected over some integration period, and the likelihood $p$ that a given measurement is a legitimate signal, as the principle analytical parameters characterizing the link; the key mission instrumentation requirements may then be derived from $N$ and $p$.  As shown in Appendix \ref{sec:linkAnalysis},   $N$ scales directly with the photon production rate and transmission efficiency, and also the integration time.  The integration time itself is a strong function of orbital configuration, requiring line-of-sight between source and receiver nodes. The number of measurements $N$, depends on the number of delivered photons and is a function of the transmitter $D_T$ and receiver aperture $D_R$; in what follows we assume $D_T=1\,$ m and $D_R=0.3\,$ m.  Results of this analysis help constrain the design of a flight mission to measure $\phi$ and bound $\alpha$; representative results are shown in Figure \ref{fig:COW_Alpha}.
We see that there is an optimal altitude, located at about $1,200 \,$km, maximizes the number of received photons, leading to the predicted error on $\Delta\alpha=3\times 10^{-4}$; such a configuration is a reasonable starting point to design a full flight mission to conduct the quantum optical COW test to validate the predicted general relativistic phase shift on various quantum states of light, and bound the parameter $\alpha$ in the quantum optical regime. Similar analyses considering more realistic orbits, spacecraft-to-spacecraft links, and entangled and hyper-entangled photon pairs, are the subject of a future research article under development.

The tests of the Einstein equivalence principle described in this paper all involve long baseline optical interferometry.  Since many of them use a fiber optic delay line, the system wavelength should be within the low fiber-loss optical C, L, or O-bands.  The overall link efficiency follows the description from Appendix \ref{sec:linkAnalysis}, using the single-channel expression (Equation (\ref{eq:SinglePhoton})) for single photons, and the double channel expression (Equation \ref{eq:doubleLink}) for entangled and hyperentangled photons. 

Measuring the gravitationally induced phase shift also requires that other sources of phase shift are accounted for, either through direct measurement or usage of mitigation strategies.  For example, as discussed in \cite{Vallone2014f, Vallone2015b, Vallone2016b}, active measurement of the spacecraft position and velocity can compensate error terms associated with these factors.  
Active stabilization of the fiber coil length, relative to a local stabilized laser system on both the ground system and flight system is also required.  This laser reference should propagate through the fiber delay lines in the opposite direction of the signal photons to facilitate filtering.  In realistic orbit configurations, the time-of-flight between the ground station and receiver is constantly changing.  There is residual error introduced due to the change in time-of-flight along the long arm of the interferometer over the transit time through the optical fiber.  In this sense, there are some fixed noise sources beyond the standard quantum limit that will bound the measurement precision.  A detailed description of these noise factors in the context of a space-mission scenario is the subject of a future publication.

\subsubsection{Gravitational Redshift in a HOM Interferometer with Frequency-Entangled Photons}
\label{sec:HOM_nonDegenerate}
\begin{figure*}[ht]
    \centering
        \begin{minipage}{12cm}
            \centering
            \includegraphics[width=\textwidth]{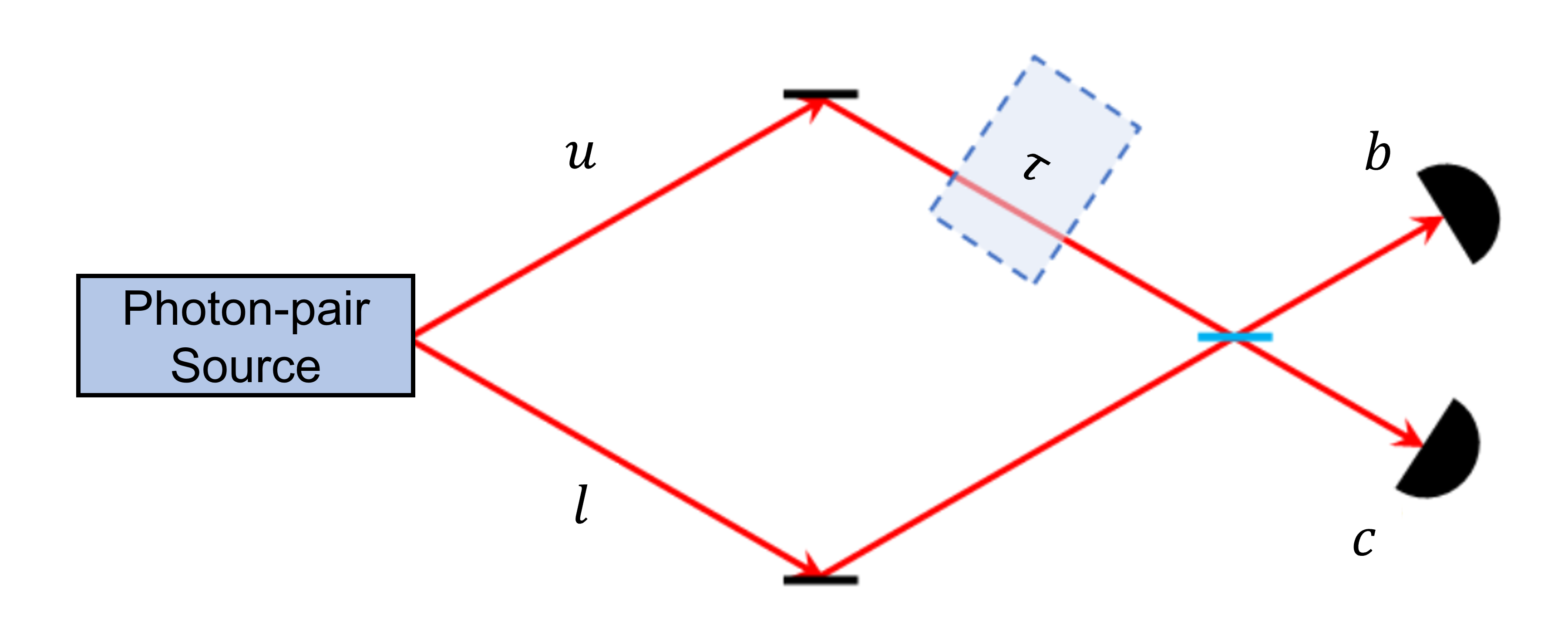}
            \caption{A balanced Hong-Ou-Mandel interferometer.}
            \label{fig:HOM}
        \end{minipage}
\end{figure*}

In this subsection we turn our attention to an experiment using an interference effect that does not have a classical analog, namely the two-photon Hong-Ou-Mandel (HOM) interference  \cite{HOM}.  As shown in Fig. \ref{fig:HOM}, a photon-pair source directs (indistin\-guishable) photons onto a 50-50 beamsplitter. There is then a quantum mechanical interference effect, causing a complete cancellation of the two processes leading to a coincidence between detectors $b$ and $c$: both photons being transmitted, or both being reflected.\footnote{This effect is sometimes loosely described as ``photons interfering on the beamsplitter'', which is not entirely correct. It is possible to construct an HOM interferometer with unbalanced arms such that the photon wavepackets never overlap on the beamsplitter or anywhere else \cite{pittman1996HOM,strekalov1998HOM}, or, as we will see here, where the photons have different colors, and yet the two-photon interference persists; all that matters is the indistinguishability of the underlying {\it processes} that could give rise to coincident detections. 
}

\noindent To describe a balanced Hong-Ou-Mandel interferometer, consider a two-photon state    $|1\rangle_l|1\rangle_u$, with one photon in each mode (see Figure \ref{fig:HOM}). These two photons are now incident on a beamsplitter:

\begin{align}
    & |1\rangle_l|1\rangle_u \rightarrow \bigg( \frac{|1\rangle_b + i |1\rangle_c}{\sqrt{2}}\bigg)\bigg( \frac{|1\rangle_c + i |1\rangle_b}{\sqrt{2}}\bigg)\notag  \\
     &= \frac{1}{2}\bigg( i\sqrt{2}(|2\rangle_b + |2\rangle_c) + |1\rangle_b|1\rangle_c - |1\rangle_c|1\rangle_b\bigg)\notag\\
     &= \frac{i}{\sqrt{2}}(|2\rangle_b + |2\rangle_c),
\end{align}
and photon bunching is seen to occur---both photons go either into Detector b or into Detector c, and the probability of a coincident detection at Detectors $b$ and $c$ is therefore 
\begin{equation}\label{eq:coincidences}
    P_{b,c} = 0.
\end{equation}

The above assumes the two photons are indistinguishable, as otherwise destructive interference between the terms $|1\rangle_b|1\rangle_c$ and $|1\rangle_c|1\rangle_b$ would not occur. However, for a photon with non-zero bandwidth, temporal delays between the photons will introduce distingushability. In this case, the probability of coincidences becomes
\begin{equation}
    P_{b,c}(\tau) = \frac{1}{2}[1 - e^{-2\sigma^2\tau^2}],
\end{equation}
where $\tau$ is the relative temporal delay, and $\sigma$ is the half-bandwidth of the photons. In the limit of zero bandwidth, this reduces to (\ref{eq:coincidences}) for any relative delay. Note that the HOM dip does not depend on the relative phase of the incident photons, only on their relative arrival time. Thus, HOM interferometry has lower resolution when compared with a Mach-Zehnder interferometer, but is robust against certain types of group-velocity dispersion. Specifically, if there is dispersion in one arm of a standard MZI, the visibility will be negatively affected, whereas the HOM interference effect is known to be immune to group velocity dispersion (more precisely, to the odd orders of this) \cite{Steinberg92}. Another key difference between a standard MZI and a HOM interferometer is that the former suffers reduced visibility if there is a relative loss in either of the arms, while the latter does not (in the absence of noise), as such loss equally affects both of the underlying interfering physical processes.

We now consider a modified HOM interferometer, in which the input state is entangled in frequency \cite{Chen2019}:
\begin{equation}
    \frac{1}{\sqrt{2}}(|\omega_1\rangle_l|\omega_2\rangle_u + |\omega_2\rangle_l|\omega_1\rangle_u).
\end{equation}
Adding some temporal delay $\tau$ to the upper path adds an energy-dependent phase $\omega_i \tau$, producing the state
\begin{align}
    & \frac{1}{\sqrt{2}}(e^{i \omega_2 \tau}|\omega_1\rangle_l|\omega_2\rangle_u + e^{i \omega_1 \tau}|\omega_2\rangle_l|\omega_1\rangle_u)\nonumber\\ &=
     \frac{e^{i \omega_2 \tau}}{\sqrt{2}}(|\omega_1\rangle_l|\omega_2\rangle_u + e^{i (\omega_1-\omega_2) \tau}|\omega_2\rangle_l|\omega_1\rangle_u).\label{eq:HOM_state}
\end{align}
Impinging this state upon a beamsplitter as before, the final state at Detectors $b$ and $c$ is
\begin{align*}
    &\frac{1}{\sqrt{2}}(|\omega_1\rangle_l|\omega_2\rangle_u + e^{i (\omega_1-\omega_2) \tau}|\omega_2\rangle_l|\omega_1\rangle_u) \\
   % &\rightarrow \frac{1}{2\sqrt{2}}\bigg[ i(|\omega_1, \omega_2 \rangle_b + |\omega_1, \omega_2 \rangle_c) + (|\omega_1\rangle_b |\omega_2\rangle_c - |\omega_2\rangle_b |\omega_1\rangle_c) + \\
    %&\qquad + e^{i(\omega_1 - \omega_2)\tau}(i(|\omega_1, \omega_2 \rangle_b + |\omega_1, \omega_2 \rangle_c) + (|\omega_2\rangle_b |\omega_1\rangle_c - |\omega_1\rangle_b |\omega_2\rangle_c))\bigg]\\
    &\rightarrow  \frac{1}{2\sqrt{2}}\bigg[i(1 + e^{i(\omega_1 - \omega_2)\tau})(|\omega_1, \omega_2 \rangle_b + |\omega_1, \omega_2 \rangle_c) + \\
    &\qquad + (1 - e^{i(\omega_1 - \omega_2)\tau})(|\omega_1\rangle_b |\omega_2\rangle_c - |\omega_2\rangle_b |\omega_1\rangle_c)\bigg],
\end{align*}
yielding a coincidence probability
\begin{equation}
    P(c, b) = 2 \left| \frac{1}{\sqrt{8}} (1-  e^{i(\omega_1 - \omega_2)\tau}) \right|^2 = \frac{1}{2}\big[1 - \cos((\omega_1 - \omega_2)\tau)\big].
\end{equation}
Accounting for photon bandwidth (and assuming identical bandwidths for simplicity), we have 
\begin{align}
    P_{b,c}(\tau) &= \frac{1}{2}\big[1 - \cos((\omega_1 - \omega_2)\tau)e^{-2\sigma^2\tau^2}\big].
    \label{eq:two-photon_detection}
\end{align}
This interferometer therefore combines the sensitivity of Mach-Zehnder interferometry with the dispersion cancellation and loss-resilience of degenerate HOM interferometry; as such it can be quite useful for precision measurements of phase differences and temporal delays. For example, a frequency-entangled HOM interferometer with wavelengths 800\,nm and 1590\,nm should be able to resolve temporal differences of a few attoseconds with only tens of thousands of detected photons \cite{Lualdi2021}. 

\begin{figure*}[ht]
    \centering
        \begin{minipage}{12cm}
            \centering
            \includegraphics[width=\textwidth]{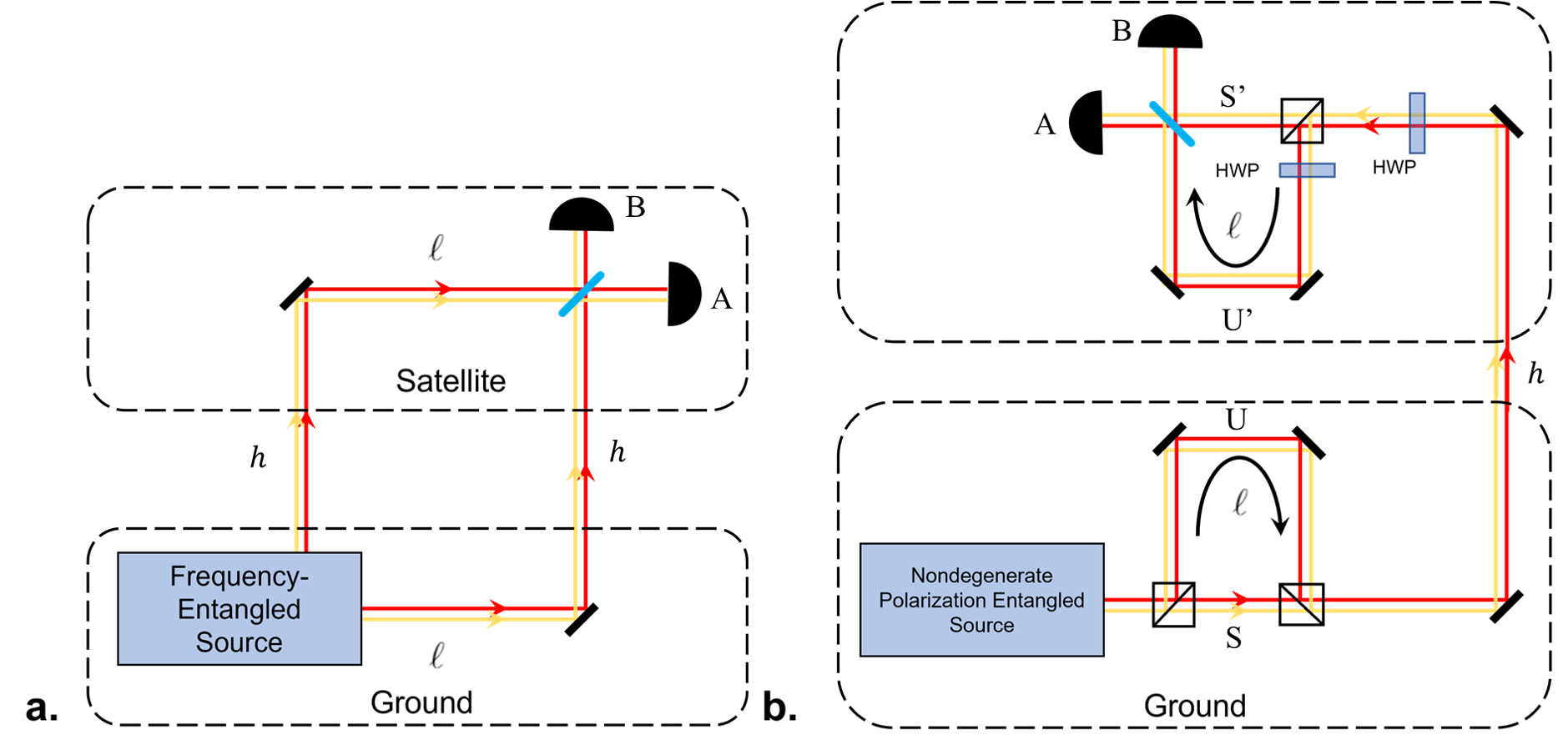}%{HOM_DSQL.png}
                        \caption{Frequency-entangled HOM interferometer sensitive to the gravitational redshift and consisting of a ground station and a spacecraft. a.) Simplified schematic. Photons experience different gravitational potentials, depending on which path they take. b.) Overlapping-path architecture, with a single uplink channel for both photons; polarization entanglement ensures that only interfering processes are present, i.e., due to the polarizing beamsplitters, each photon takes one long and one short path.}
            \label{fig:HOM_DSQL}
        \end{minipage}
\end{figure*}

These techniques may be useful in probing the intersection of quantum mechanics and general relativity, since the HOM effect is truly nonclassical \cite{Brady2020}. Suppose that the two paths of the interferometer are held at different gravitational potentials. This will create %an energy
a frequency- and path-dependent phase difference which can then potentially be resolved by the interferometer, allowing us to study the effects of a curved spacetime on an entangled quantum system. An explicit implementation consisting of a ground station (G), a spacecraft (S) and an optical link is shown in Figure~\ref{fig:HOM_DSQL}. 
A pair of frequency-entangled photons is generated in the ground station and sent through the optical link to the spacecraft, where they are detected. One interferometer arm involves a delay line in the ground station, whereas the other arm involves an analogous delay line in the spacecraft, where the two arms are then recombined on a beamsplitter with single-photon detectors at the two exit ports.
Whereas delay lines with equal physical length $\ell$ would lead to a balanced interferometer for a traditional experiment on Earth, in this proposed setup relativistic effects and the changing distance between the ground station and the spacecraft give rise to the following time shift between the two interferometer arms, obtained up to order $1/c^3$ within a post-Newtonian expansion in powers of $v/c$ and $U/c^2$:
\begin{equation}
\tau = \frac{\ell}{c} \left( \left( \frac{1 + (\hat{\mathbf{n}}\cdot \mathbf{v}_\mathrm{S}) (t_\mathrm{r}) / c}
{1 + (\hat{\mathbf{n}}\cdot \mathbf{v}_\mathrm{G}) (t_\mathrm{e}) / c} \right)
\left( \frac{dt / d\tau_\mathrm{S}}{dt / d\tau_\mathrm{G}} \right) - 1 \right)
\label{eq:time_shift1} ,
\end{equation}
where $(\hat{\mathbf{n}}\cdot \mathbf{v}_\mathrm{G}) (t_\mathrm{e})$ and $(\hat{\mathbf{n}}\cdot \mathbf{v}_\mathrm{S}) (t_\mathrm{r})$ are, respectively, the velocity of the ground station along the direction of the optical link at the time of emission $t_\mathrm{e}$ and similarly for the spacecraft at the time of reception $t_\mathrm{r}$.
The factor containing these velocities corresponds to the Doppler effect associated with the motion of the two stations (ground and spacecraft)%
\footnote{For sufficiently long delay lines or quantum memories, the changes of $(\hat{\mathbf{n}}\cdot \mathbf{v}_\mathrm{G})$ and $(\hat{\mathbf{n}}\cdot \mathbf{v}_\mathrm{S})$ over the time $\ell/c$ would also need to be accounted for.}.
The ratio between $(dt / d\tau_\mathrm{S})$ and $(dt / d\tau_\mathrm{G} )$, which respectively account for the special relativistic time dilation and the gravitational redshift between the two stations, is given by
\begin{equation}
\left( \frac{dt / d\tau_\mathrm{S}}{dt / d\tau_\mathrm{G}} \right)
\approx 1+ \left( \frac{1}{2} \frac{\mathbf{v}_\mathrm{S}^2 - \mathbf{v}_\mathrm{G}^2}{c^2}
- \frac{U(\mathbf{x}_\mathrm{S}) - U(\mathbf{x}_\mathrm{G})}{c^2} \right)
\label{eq:time_shift} ,
\end{equation}
where $U(\mathbf{x})$ is the gravitational potential and we neglect higher-order terms in the post-Newtonian expansion \cite{Misner1973}, as they are suppressed by higher powers of $v/c$. This result, which is also applicable to more general situations such as the Moon-Earth system, reduces to Eq.~(\ref{eqnGR_clock_slowing}) for a Schwarzschild metric.

The probability $P_{b,c}(\tau)$ of two-photon detection in either of the two ports is therefore determined by Eq.~(\ref{eq:two-photon_detection}), with
\begin{equation}
    \tau = \frac{\Delta l}{c} 
+ \frac{l}{c} \left( \left( \frac{1 + (\hat{\mathbf{n}}\cdot \mathbf{v}_\mathrm{S}) (t_\mathrm{r}) / c}%
{1 + (\hat{\mathbf{n}}\cdot \mathbf{v}_\mathrm{G}) (t_\mathrm{e}) / c} \right) - 1 \right)  + \frac{l}{c} \left( \frac{1}{2} \frac{\mathbf{v}_\mathrm{S}^2 - \mathbf{v}_\mathrm{G}^2}{c^2}
- \frac{U(\mathbf{x}_\mathrm{S}) - U(\mathbf{x}_\mathrm{G})}{c^2} \right)
\label{eq:time_shift2},
\end{equation}
where a possible difference $\Delta \ell$ between the proper lengths of the two delay lines (ideally $\Delta \ell = 0$) has been included and we have assumed that $\Delta \ell \ll \ell$.
The third term on the right-hand side of Eq.~(\ref{eq:time_shift2}) corresponds to the relativistic effects that we are interested in and will typically be of order $10^{-10}$. In contrast, the ``classical'' Doppler effect encoded by the second term can be of order $10^{-5}$.
Accurately tracking the trajectory of the spacecraft by means of satellite laser ranging is thus necessary so that the comparatively large contribution of the Doppler effect can be suppressed below the $10^{-10}$ level through postcorrection, as discussed in Section ~\ref{sec:COWtests}.
The length of the two delay lines also needs to be stabilized below the $10^{-10}$ level; moreover, one should guarantee that they are equal (i.e.,\ $\Delta \ell = 0$) at that level, which can be achieved by simultaneously calibrating (and stabilizing) them with identical frequency references on ground and in the spacecraft. Note that using an elliptical orbit would enable one to distinguish the two different relativistic contributions, namely special relativistic time dilation and gravitational redshift, and would also help to separate the small signal of interest from noise sources and systematic effects, as explained below.

A substantial simplification to this complex stabilization and calibration procedure can be achieved by employing classical light at an intermediate frequency between $\omega_1$ and $\omega_2$ for that purpose, which can be combined with an active compensation method involving a tunable delay, e.g., consisting of a movable right-angle prism with a piezo-actuated translation stage, or a piezoelectrically controlled fiber stretcher.
It should be noted that since the classical light acting as a reference is equally affected by the relativistic effects, a HOM interferometer stabilized in this manner will not be able to independently measure such effects. Nevertheless, it could still be regarded as the first experimental confirmation that quantum states of light -- frequency-entangled photons -- experience the same gravitational redshift as classical light, and moreover that purely quantum mechanical interference is similarly affected.

While the implementation displayed in Figure~\ref{fig:HOM_DSQL}a is conceptually simpler, a scheme involving a single uplink channel, depicted in Figure~\ref{fig:HOM_DSQL}b, is preferable in practice. With this goal in mind, it is useful to consider a source of co-propagating frequency-nondegenerate photons in the polarization-entangled state:
\begin{equation}
| \Psi_0 \rangle = \frac{1}{\sqrt{2}} \Big( | \omega_1, H \rangle\, | \omega_2, V \rangle
+ | \omega_1, V \rangle\, | \omega_2, H \rangle  \Big) .
\label{eq:Pol_entangled_state}
\end{equation}
Such a state has been recently demonstrated in the laboratory \cite{Chapman2019space}. By using polarizing beamsplitters (PBS) and a half-wave plate (HWP), as shown in Figure~\ref{fig:HOM_DSQL}b, one can ensure that H-polarized photons will propagate along the short path ($S$) in the ground station and the delay line in the spacecraft ($U'$), while V-polarized photons will propagate along the delay line ($U$) in the ground station and the short path in the spacecraft ($S'$).
 The resulting state (after the satellite PBS) is then
\begin{equation}
    | \Psi \rangle = \frac{1}{\sqrt{2}}
\Big( e^{i \omega_2 \tau} | \omega_1, H \rangle_{BD}\, | \omega_2, V \rangle_{AC}- e^{i \omega_1 \tau} | \omega_2, H \rangle_{BD}\, | \omega_1, V \rangle_{AC} \Big).
\end{equation}
Finally, by inserting an additional HWP in the delay line of the spacecraft, the polarization state in that arm is rotated, $| V \rangle_{SU'} \to | H \rangle_{SU'}$, so that the polarizations are no longer correlated with the interferometer arms and the quantum states from the two different arms can interfere when recombined at the final beamsplitter. Indeed, the polarization state of the two photons, $| H \rangle_{SU'}\, | H \rangle_{US'}$, can then be factored out and one is left with the desired state, cf. Eq.~(\ref{eq:HOM_state}).

We conclude this subsection with an estimate of the actual sensitivity of the proposed HOM interferometry experiment to relativistic effects. As explained in the discussion after Eq.~(\ref{eqnGR_clock_slowing}), the net contribution of these effects is of order $-3 \times 10^{-10}$ for a circular LEO, %orbit,
but just a smaller fraction -- of order $4 \times 10^{-11}$ -- corresponds to the gravitational redshift.
In contrast, for a GEO %orbit
the relativistic effects are dominated by the gravitational redshift, of order $6 \times 10^{-10}$ in that case.
As with the optical COW experiments discussed above, for a highly elliptical orbit, such as that considered in Figure~\ref{fig:GR_epsilon_clock_Molniya_orbit}, these effects are modulated by the orbital period and range from $-5 \times 10^{-10}$ at the perigee, where special relativistic time dilation dominates, to $6 \times 10^{-10}$ at the apogee, where the main contribution comes from the gravitational redshift. Thus, orbital modulation can be very useful to extract the small signal and separate it from noise sources and systematic effects.

Similarly to GEOs, %orbits,
for the Lunar Gateway the effect would be of order $7 \times 10^{-10}$ and dominated by the redshift associated with Earth's gravitational field, because the Moon's mass is 80 times smaller than Earth's, so that contributions from the lunar gravitational field and time dilation due to the orbital velocity %around the Moon
would be much smaller, also implying that there are no significant orbital modulation effects. Hence, if we consider the case in which the platform in Figure~\ref{fig:HOM_DSQL} is the Gateway Spacecraft, the time shift due to relativistic effects (i.e.,\ excluding the classical Doppler contribution) is given by $\tau_\mathrm{rel} = 2.3 \times 10^{-15}\, \mathrm{s}\ (\ell / 1\, \mathrm{km})$ and leads to the following interferometer phase shift:
\begin{equation}
    \delta\varphi = (\omega_1 - \omega_2)\, \tau_\mathrm{rel} \notag= 0.2\, \mathrm{rad}\, \cdot 
 \frac{\Delta \lambda}{100\, \mathrm{nm}}  \frac{1600\, \mathrm{nm}}{\lambda_2} 
 \frac{1500\, \mathrm{nm}}{\lambda_1}   \frac{l}{1\, \mathrm{km}} 
\label{eq:HOM_phase_shift} .
\end{equation}
Quantitatively comparable results hold for GEOs, and also at the apogee of a highly elliptical orbit.
In contrast, for LEOs the result for the total relativistic effect is reduced by about a half and is dominated by the special relativistic contribution, whereas the gravitational redshift is nearly 10 times smaller. Nevertheless, because the transmission rate for {\it pairs} of entangled photons scales inversely with the {\it fourth} power of the optical link baseline (see Appendix \ref{sec:linkAnalysis}), it should be possible to resolve this smaller effect too, potentially with even milder requirements on the telescope size.
Similar conclusions apply at the perigee of the highly elliptical orbit.

\subsubsection{Flight Mission Design for Optical COW Tests using HOM Interference}
\label{sec:HOMCOW_mission}
In this section we present a brief summary of the mission design trade-space; a detailed, rigorous summary of the underlying mathematics is the subject of a future publication.
Tests of gravitational effects on HOM interference are governed by similar processes as described in Section \ref{sec:COW_Mission}; the key difference is that a {\it pair} of photons must be transmitted, which reduces the overall link efficiency, per Appendix \ref{sec:linkAnalysis}.  Furthermore, as described in Section \ref{sec:HOM_nonDegenerate}, the photons comprising the pair may be non-degenerate in frequency. All these factors couple with available spacecraft trajectorys to result in a range of possible mission configurations.  

 The system diagram is shown in Fig. \ref{fig:HOM_DSQL}.    The net timing delay $\tau$ between the upper and lower path of the interferometer is 
\begin{equation}
    \tau = \tau_{GR} + \frac{\Delta\ell}{c} + \tau_{c},
    \label{eq:tau_def}
\end{equation}
where $\Delta\ell$ is the geometric length mismatch between the two paths, due to error terms in the engineering and control of the interferometer, $\tau_c$ is any control signal applied to the interferometer, and $\tau_{GR}$ is the relativistic shift we seek to measure:  
\begin{equation}
\tau_{GR}=\frac{\ell}{c}\left( \frac{\Delta v^2}{2c^2} - \frac{\Delta U}{c^2}\right).
\label{eq:tauGR}
\end{equation}
Here $\Delta v^2$ is the difference in the squares of the velocities of the two interferometer nodes, $\Delta U$ is their difference in gravitational potential energy, $c$ is the speed of light, and $\ell$ is the interferometer path length depicted in Fig.\ref{fig:HOM_DSQL}, equivalent to the arm length of the HOM interferometer\footnote{As before, the effective optical path length should include any non-unity refractive index.}.  

Assume now that with probability $p$ the interferometer contains the input photons and that with probability $(1-p)$ the interferometer is injected with uncorrelated and distinguishable photons, or the detectors have background or noise counts, leading to noise events that can cause an ``accidental'' coincidence count with probability $1/2$. The flux of noise photons $N_{noise}$ can be linked to system parameters such as receiver aperture, spectral filtering, and detector dark counts; see Equation $(\ref{eq:NoiseEq})$. The parameter $p$ can be interpreted as the experiment quality factor, see Subsection \ref{sec:stats_bell_tests}:
\begin{equation}
    p = (1-N_{noise}\Delta t_{R})^2 F ,
\end{equation}
where $(N_{noise}\Delta t_{R})$ is the probability of recording a count due to noise falling within the detector timing window $\Delta t_{R}$ (assumed to be $10^{-9}\, s$), and $F$ is the source fidelity (assumed to be 0.95 in our simulations). Here for simplicity we assume that all the system parameters in $N_{noise}$ are fixed, apart from the spectral filtering bandwidth $\sigma$ that we set to match the signal photon bandwidth; therefore, $p=p(\sigma)$. The  coincidence count probability thus becomes:
\begin{equation}
    P_c=\frac{p(\sigma)}{2}(1-\cos(\Delta\omega \tau)e^{-2 \sigma^2 \tau^2})+\frac{(1-p(\sigma))}{2} .
\end{equation}
 A more complete analysis should also take into account error sources such as path length mismatch and attitude determination error; these and other sources of imperfection will be explored in detail elsewhere.

The error $\Delta\tau$ in a measurement of $\tau$ is computed using the correlated counts $\hat{M} = \hat{M_A}\otimes\hat{M_B}$, for detectors $A$ and $B$ in Figure \ref{fig:HOM_DSQL}:
\begin{align}
    \label{eq:delta_tau}
   & \Delta\tau (\tau,\sigma,\Delta\omega)\equiv \frac{\Delta M(\tau)}{\left|\frac{\partial\left<\hat{M}\right>}{\partial{\tau}}\right|}=\frac{\sqrt{\left<\hat{M^2}\right>-\left<\hat{M}\right>^2}}{\left|\frac{\partial\left<\hat{M}\right>}{\partial{\tau}}\right|}\notag \\
   & = \frac{\sqrt{1-p(\sigma)^2 e^{-4 \sigma ^2 \tau ^2} \cos ^2(\Delta \omega  \tau )}}{p(\sigma) e^{-2 \sigma ^2 \tau ^2} \left| 4 \tau \sigma ^2 \cos (\Delta \omega  \tau ) +\Delta \omega  \sin (\Delta \omega  \tau )\right| ,}
\end{align}
where $\Delta\omega \equiv \omega_1 - \omega_2$ is the frequency difference of the photons, here assumed to have (vacuum) wavelengths $\lambda_1=780 \,$nm and $\lambda_2=1550 \,$nm, corresponding to $\Delta \omega=4\cdot 10^{15} \,$ Hz. In order to avoid excessive broadening of the pulse due to propagation in the atmosphere and to simplify telescope optics, we assume a maximum bandwidth $\sigma \leq 4.7 \cdot 10^{13} \,$Hz ($\delta\lambda_1 \leq 100 \,$nm), i.e., the individual spectral components are still relatively narrow for the non-degenerate HOM implementation.

\begin{figure*}[p]
        \centering
        \begin{subfigure}[b]{0.475\textwidth}
            \centering
            \includegraphics[width=\textwidth]{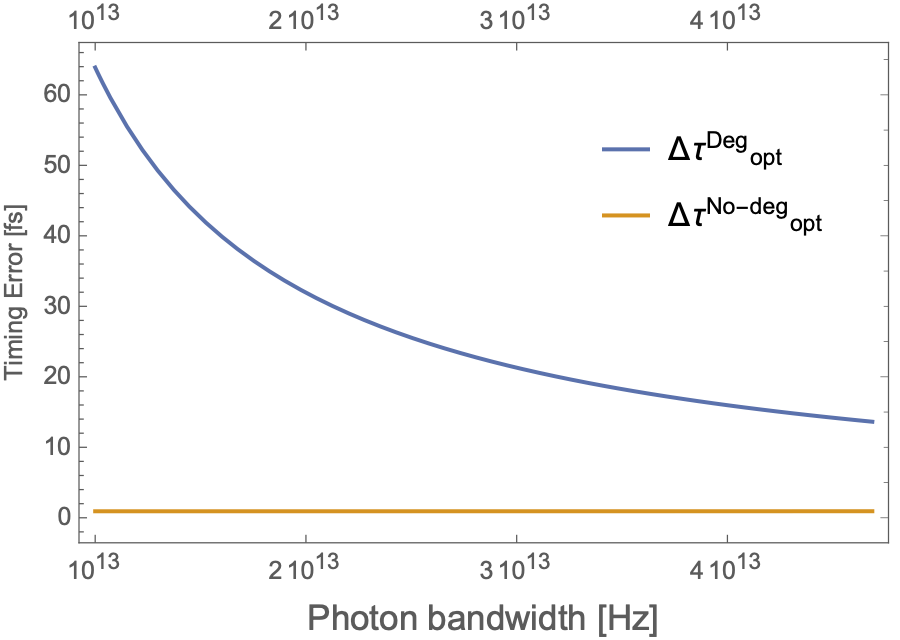}
            \caption{}      \label{fig:errorMinLM}
        \end{subfigure}
        \begin{subfigure}[b]{0.475\textwidth}  
            \centering 
            \includegraphics[width=\textwidth]{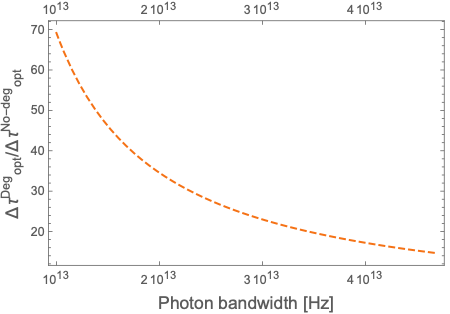}
            \caption{}
              \label{timingratioLM}
        \end{subfigure}
         \caption{\small Contour plot of  $\Delta\alpha$  versus the photon bandwidth $\sigma$ and satellite altitude for a single satellite passage: a) degenerate case (both photons at 780 nm, yielding only the HOM dip), b) non-degenerate case (assuming photons at 780 nm and 1550 nm, yielding a HOM dip with beat-note oscillations).} 
    \end{figure*}

Given an experiment with a certain quality factor, a natural question to ask is which choice of the overall time delay $\tau$ minimizes the timing error. More formally, we want to solve the following optimization problem:
    \begin{equation}
    \Delta \tau_{opt}(\sigma)=\min_{\tau}  \Delta \tau(\tau,\sigma,\Delta \omega).
\end{equation}
The result of the optimization is shown in Figures \ref{fig:errorMinLM} and \ref{timingratioLM}, which shows 
the expected result that the timing error is minimized for the largest bandwidth $\sigma$ (in the degenerate source case $\omega_1 = \omega_2$), but is essentially constant for $\Delta\omega \gg \sigma$.
The minimum timing error for the non-degenerate case is at least an order of magnitude smaller than for the degenerate case in the considered photon bandwidth interval; even when we let $\sigma \approx \Delta\omega$, we still see a $\sim40\%$ improvement for the non-degenerate case.

Assuming that the main source of error is given by  the general relativistic time delay measurement, we can propagate it to obtain:
\begin{equation}
\Delta\alpha=\frac{\partial \alpha}{\partial \tau}\Delta\tau=\frac{\Delta \tau}{- \frac{l}{c}\frac{\Delta U}{c^2}\sqrt{N_c},}
\end{equation}
where $N_c$ is the number of coincidence counts used for the measurement, a function of the optics aperture and of the photon wavelength (which effects diffraction and thereby photon loss; see  Equation \ref{eq:Link}).
     \begin{figure*}[p]
        \centering
        \begin{subfigure}[b]{0.475\textwidth}
            \centering
            \includegraphics[width=\textwidth]{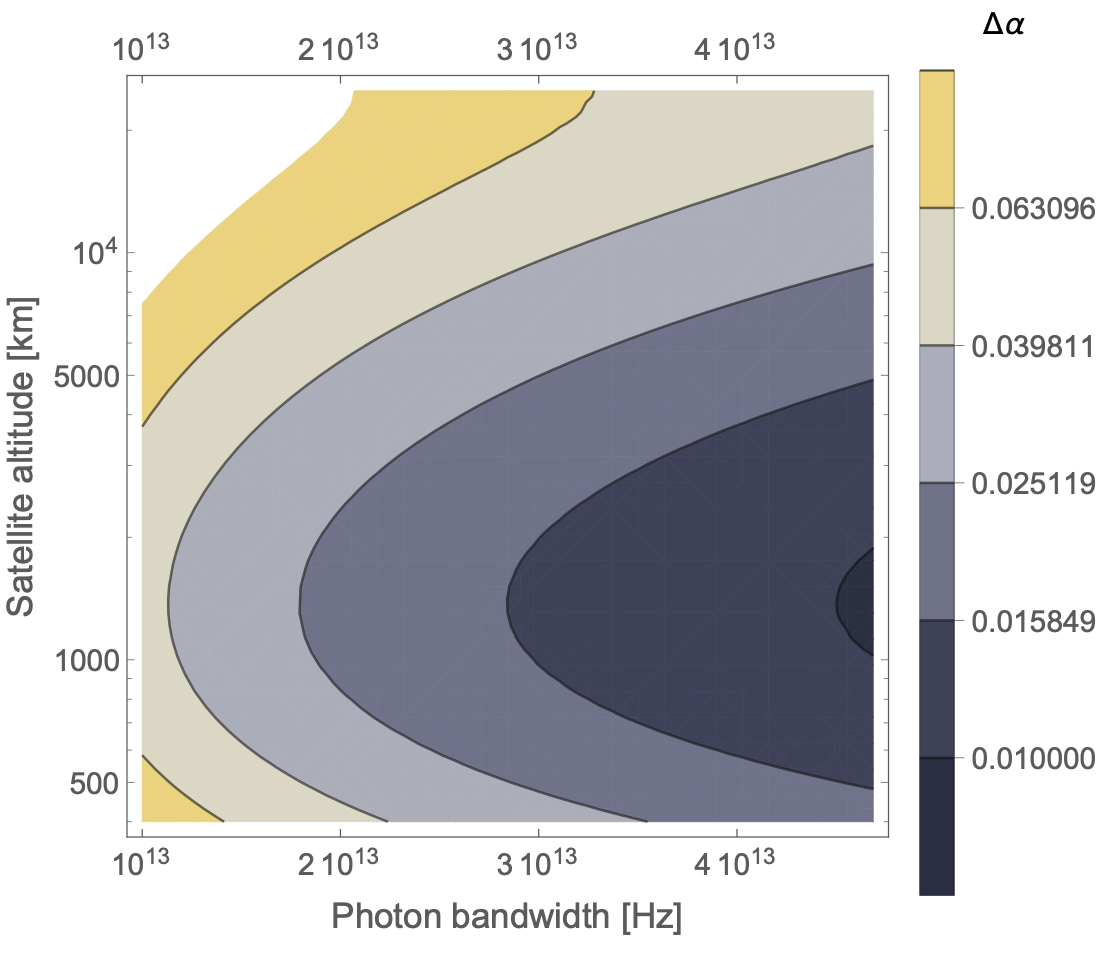}
            \caption{}    
            \label{errAlphaDeg103LM}
        \end{subfigure}
        \begin{subfigure}[b]{0.475\textwidth}
            \centering 
            \includegraphics[width=\textwidth]{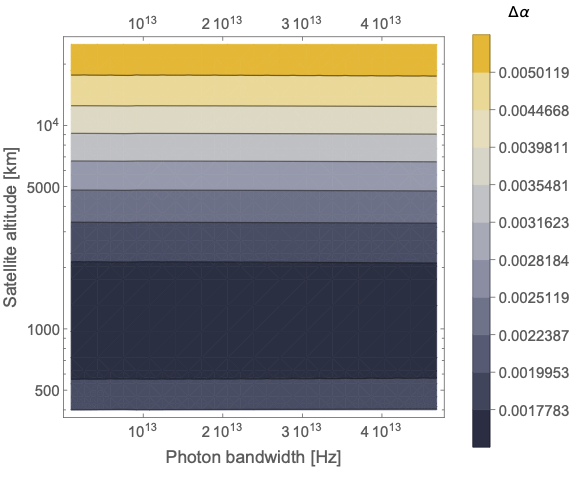}
             \caption{}  
              \label{errAlphaNoDeg103LM}
        \end{subfigure}
        \caption{\small Contour plot of  $\Delta\alpha$  versus the photon bandwidth $\sigma$ and satellite altitude for a single satellite passage: a) degenerate case (both photons at 780 nm, yielding only the HOM dip), b) non-degenerate case (assuming photons at 780 nm and 1550 nm, yielding a HOM dip with beat-note oscillations).} 
    \end{figure*}
    
\begin{figure*}[t]
    \centering
    \includegraphics[width=8.5cm]{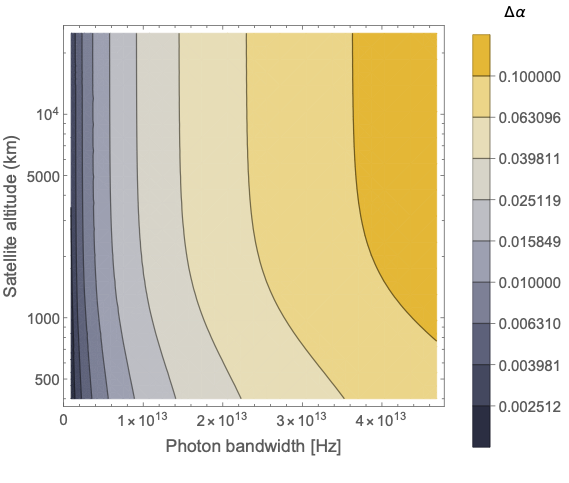}
    \caption{Contour plot of the ratio $\alpha^{non-degen}/\alpha^{degen}$ between the error on $\alpha$ of the non-degenerate and the degenerate case, versus photon bandwidth and  satellite altitude.}
    \label{alpharatioLM}
\end{figure*}

Figure \ref{errAlphaDeg103LM} and  \ref{errAlphaNoDeg103LM}  shows the contour plots for the minimized error on $\alpha$ in a satellite passage, assuming a ground (satellite) aperture of 1 m (0.3 m). For the degenerate case, the error decreases as the photon bandwidth increases, reaching a minimum of $\Delta \alpha = 0.01$ for an altitude around $1,300 \,$km and $\sigma = 4 \cdot 10^{13} \,$Hz.
The non-degenerate case also shows a slight reduction in $\Delta\alpha$ as the photon bandwidth $\sigma$ increases but the effect is much smaller than in the degenerate case, as expected since the timing resolution is dominated by 
$\Delta\omega \gg \sigma$. Here the peak performance of $\Delta\alpha = 0.001$ is reached at an altitude close to 1,000 km \footnote{The optimal altitude depends on the optics apertures; for example, for 0.5-m transmitter and receiver apertures, the optimal altitude would be located lower, at about 500 km, to compensate for the extra diffractive loss of the smaller apertures.}. Figure \ref{alpharatioLM} shows the benefit of using photons with different wavelengths -- the error on $\alpha$ for the non-degenerate case is always at least one order of magnitude smaller than the error for the degenerate case.

The display of an optimal altitude minimizing the error on $\alpha$ was also present for the single-photon interference case (Section \ref{sec:COW_Mission}), in that case at $1,200 \,$km for the same 1-m (0.3-m) telescope apertures. The estimation of $\alpha$ using single-photon interference seems to be advantageous ($\Delta \alpha=3 \cdot 10^{-4}$, i.e., three times smaller than the two-photon case); the lower optimal altitude and higher value of $\Delta\alpha$ in the two-photon case arise from the additional signal loss when both photons have to be transmitted successfully (if one had sufficiently large telescopes that diffractive losses could be ignored, this disadvantage would largely disappear).
Regardless, the HOM-based experiments bring the desirable feature of using purely nonclassical interference.

\subsubsection{Gravitational Dephasing, Decorrelation, and Decoherence}\label{gr:dephasing}

The preceding sections describe a set of experiments to test the gravitationally induced phase shift on quantum photonic modes.  In this Section, the predictions of QFTCST on the {\itshape gravitational decoherence} of quantum states are reviewed. The estimated magnitude of these effects suggest that meaningful tests of this sort are beyond the scope of the proposed DSQL experiments.  However, success of DSQL and related missions will improve the prospect for considering such tests of gravitational decoherence in the future.

Gravitational decoherence first appeared as an attempt to explain the very different behaviors  between the micro and the macro worlds, i.e., we seem not to observe quantum superpositions in the latter. The simplest way is to assume the existence of a length scale which demarcates the micro from the macro --- superpositions can then only persist below this scale; above it, the states naturally decohere in some basis. If the underlying theory attributes the decoherence to new physics (e.g.,  continuous collapse models such as the Ghirardi-Rimini-Weber-Pearle (GRW-P) theories \cite{GRWreview} or Diosi-Penrose (DP) model \cite{Diosi0,Penrose}), then decoherence may happen in the position basis, and it can explain the micro-macro separation in quantum theory (a.k.a. the {\em quantum measurement problem}). In contrast, gravitational decoherence that originates  from known physics---i.e.,  general relativity (GR) and quantum field theory, as in the Anastopoulos-Blencowe-Hu (ABH) theory \cite{AHGravDec,Miles}, occurs in the energy basis.

Here, we  shall focus only on genuine gravitational decoherence 
\footnote{The term ``gravitational decoherence'' is sometimes  used in the literature when the authors actually refer to mere gravitational dephasing.  A key difference is time reversibility: decoherence is an irreversible process, while dephasing  can often be reverted.  These processes follow from the entanglement of the quantum system with additional degrees of freedom that can be taken as an environment. When the environment is traced over, the original quantum system evolves into a mixed state. As a result, phases are randomized. However, if the environment is small, in the sense that it has a small Poincar\'e recurrence time, the phases will be restored after some time. In this case, we have dephasing and not decoherence.  In decoherence, the phase information is lost forever.}, which involves new physics arising from either  A) new phenomena deduced from established GR theoretical foundations (e.g., the ABH master equation) or B) modifications of either a) quantum mechanics (e.g., the DP or GRW-P theories), which we refer to collectively as alternative quantum theories (AQT); or b) the structure of space or time, known as intrinsic, fundamental, or quantum gravity decoherence.

In what follows we shall briefly highlight the salient features of these alternatives, and estimate the magnitude of their effects, as they are directly relevant to DSQL experiments. Models of gravitational decoherence typically involve one or more free parameters. Requirements of theoretical consistency and past experiments on gravitational quantum physics have already excluded regions of the parameter space---see, for example, Refs. \cite{HSMC17,GrSasso}. Deep space experiments can improve such constraints by many orders of magnitude, even if some regions of the parameter space may be beyond   current measurement capabilities.

In the ABH model \cite{AHGravDec, Miles},   decoherence arises from fluctuations of gravitational waves (classical perturbations) or gravitons (quantized linear perturbations); the source of these fluctuations may be cosmological \cite{Miles} (stochastic gravitons produced in the early universe near the big bang or from inflation), astrophysical \cite{ReJa}, or structural when GR is viewed as an emergent theory (e.g., \cite{E/QG}).  The corresponding master equation depends on the noise temperature $\Theta$, which coincides with the graviton temperature if the origin of perturbations is cosmological, but is unconstrained if gravity is emergent; in the latter case $\Theta$ is determined by the deeper layers in the structure of spacetime at the Planck scale.

The ABH master equation is:
\begin{equation}
 \frac{\partial \hat{\rho}}{\partial t} = -i [\hat{H}, \hat{\rho}] - \frac{\tau}{16m^2} (\delta^{ij} \delta^{kl} + \delta^{ik}\delta^{jl}) [\hat{p}_i
\hat{p}_j,[\hat{p}_k\hat{p}_l, \hat{\rho}]] , \label{ABHmaster}
\end{equation}
 where $\tau$ is a constant of dimension time and $\hat{H} = \frac{\hat{p}^2}{2m}$. A similar master  equation can be derived for photons \cite{LagAn21}.
 In the ABH model, $\tau =   \frac{32\pi}{9} \tau_P (\Theta/T_P)$,
 where $T_P = 1.4\times 10^{32}K$ is the Planck temperature and $\tau_P = 5.4 \times 10^{-44}s$ is the Planck time.
If $\Theta$ is regarded as a noise temperature,   it need not be related to  the Planck length, and  $\Theta >> T_P$ is  perfectly acceptable.
 For motion in one dimension, the ABH master equation simplifies to
 \begin{eqnarray}
 \frac{\partial \hat{\rho}}{\partial t} = -i [\hat{H}, \hat{\rho}] - \frac{\tau}{2} [\hat{H},[\hat{H},\hat{\rho}]], \label{1dim}
 \end{eqnarray}
Eq.  (\ref{1dim}) also appears in models by Milburn \cite{Milburn}, Adler \cite{Adler}, Diosi \cite{Diosi} and Breuer et al. \cite{Breuer} from different physical considerations. In these models,  $\tau$ is also a free parameter, but the natural candidate is the Planck-time $\tau_P$.

Diosi \cite{Diosi0} postulates a collapse term with noise correlator proportional to gravitational potential, which leads to a master equation of the form:
 \begin{equation}
 \frac{\partial \hat{\rho}}{\partial t} = -i [\hat{H}, \hat{\rho}] - \frac{G}{2} \int d{\pmb r} d {\pmb r'} \frac{[\hat{f}({\pmb r}),[\hat{f}({\pmb r}'),\hat{\rho}]]}{|{\pmb r} - {\pmb r}'|}, \label{DPME}
 \end{equation}
 where $\hat{f}({\pmb r})$ is the mass density operator. 
 Penrose's \cite{Penrose, Penrose2} idea is not model specific, but leads to similar predictions for decoherence time.   A key point in the DP model is that predictions do not involve any free parameters  (at least in the experimentally relevant regime).  The decoherence rate is typically of the order of $\frac{1}{\Delta E}$ where $\Delta E$ is the gravitational self-energy difference associated to a macroscopic superposition of mass densities.

Other decoherence models that lead to position decoherence have similar properties to DP, but more free parameters, some of which have no intuitive physical interpretation. This includes, for example, 
 continuous collapse models like GRW-P and the Power-Percival \cite{PowPer} decoherence model based on fluctuations of the conformal factor.  In any event, most AQTs have to tolerate a small degree of energy-conservation violation. Experimental tests for such violations lead to significant constraints in the free parameters of some  models \cite{GrSasso}.

Another distinct class of models is based on the  Newton-Schr\"odinger equation (NSE) \cite{NS}. One postulates a non-linear equation for the single-particle wave function,
\begin{eqnarray}
i \frac{\partial \psi}{\partial t} = - \frac{1}{2m} \nabla^2 \psi + V_N[\psi] \psi
\end{eqnarray}
where 
$V_N(\bf r)$ is the (normalized) gravitational (Newtonian) potential given by
\begin{eqnarray}
V_N({\bf r},t) = - G \int d{\bf r'} \frac{|\psi({\bf r'},t)|^2 }{|{\bf r} - {\bf r'}|}.
\end{eqnarray}
Note that the NSE for a single particle is not derivable from GR and Quantum Theory  \cite{AHNSE}.

There are several advantages of carrying out tests of these theories in space experiments. They include the  high quality of microgravity ($\sim 10^{-9}$g),  very long free-fall times ($> 10^4$ s),  and the combination of low pressure ($\sim 10^{-13} $Pa ) and low temperature ($\sim 10$ K) with full optical access.
Here we outline necessary experimental parameters using optomechanical systems, atom interferometry, atomic spatial wavefunction spreading, and photon decoherence.

{\em Optomechanical experiments.} Consider a body brought into a superposition of a zero momentum and a finite momentum state, corresponding to an energy difference $\Delta E$. For the ABH model, the decoherence rate for the center of mass is then
     \begin{eqnarray}
     \Gamma_{ABH} = \frac{ (\Delta E)^2 \tau}{\hbar^2},
     \end{eqnarray}
     where $\tau$ is the free parameter in the master equation (\ref{ABHmaster}). A value for $\Gamma_{ABH}$ of the order of $10^{-3}s$ may be observable in optomechanical systems, as it is competitive with current environment-induced-decoherence timescales. Hence,  to exclude values of $\tau > \tau_P$, we must prepare a quantum state with 
 $\Delta E  \sim 10^{-14}$J.

In the Diosi-Penrose model, the decoherence rate for a sphere of mass $M$ of radius $R$ in a quantum superposition of states with different center of mass position (though the predicted decoherence rate is largely independent of the details of the prepared state) is of the order of
     \begin{eqnarray}
     \Gamma_{DP} = \frac{GM^2}{\hbar \sqrt{R^2+\ell^2}},
     \end{eqnarray}
     where $\ell$ is a cut-off length, originally postulated to be of the order of the size of the nucleus, but recently constrained to $\ell > 0.5 \cdot 10^{-10}m$ \cite{GrSasso}.  Alternative models postulate $\ell$ up to a scale of $10^{-7}$m. For an optomechanical nanosphere with $M \sim 10^{10}$ amu and $R \sim 100$ nm, $\Gamma_{DP} \sim 10^{-3}s^{-1}$, a value that is in principle measurable in optomechanical experiments.

{\em Matter wave interferometry.} The ABH model (but not the 1-d master equation (\ref{1dim}))  leads to loss of phase coherence of the order of
 $(\Delta \Phi)^2 = m^2 v^3 \tau L/\hbar^2$, where $L$ is the propagation distance inside the interferometer\footnote{While the exact derivation of $(\Delta \Phi)^2$ requires a dynamical analysis, its magnitude is of the order of $\Gamma_{ABH} t_{int}$, where $t_{int} = L/v$ is the average time of the particle in the interferometer.}. 
 Setting an upper limit of $L = 100$ km, and $v = 10^4$ m/s, decoherence due to cosmological gravitons requires particles with masses of the order of $10^{16} $amu. If $\Theta$ is a free parameter, experiments with particles at $10^{10}$amu
 will test up to $\Theta \sim 10^{-5}T_P$. For comparison,  the heaviest molecules used to date in quantum mechanical interference experiments are oligoporphyrines with mass of ``only'' $2.6 \cdot 10^4\,$amu \cite{Fein19}. 
 
 The Diosi-Penrose model and other models that lead to decoherence in the position basis can also be tested by near-field \cite{MAQRO} and far-field \cite{MAQRO0} matter-wave interferometry.
A rough estimation for the loss of phase coherence is $ (\Delta \Phi)^2  \simeq \Gamma_{DP}L/v = \frac{Gm^2L}{\hbar Rv}$, where $R$ is the radius of the particles.  In contrast to the ABH model, this loss of coherence is enhanced at low velocities. Assuming $L = 100$ km, $v = 10$ m/s, and $R = 100$ nm, an experiment would require a mass $M \sim 10^9-10^{10}$ amu to observe decoherence according to the DP model.
  
{\em Wave-packet spread.} The intrinsic spreading of a matter wave-packet in free space is a hallmark of Schr\"odinger evolution. ABH-type models predict negligible deviations in the wave-packet spread from that of unitary evolution. The DP model and all other models that involve decoherence in the position basis predict
 a wave packet spread of the form
  \begin{eqnarray}
(\Delta x)^2(t) = (\Delta x)_S^2(t) + \frac{\Lambda}{2m^2} t^3,
\end{eqnarray}
where $(\Delta x)_S^2(t)$ is the usual Schr\"odinger spreading, and $\Lambda$ depends on the model. The changes from free Schr\"odinger evolution become significant at later times.  An exact estimation of this effect depends on properties of the initially prepared state, and  is rather involved. The MAQRO proposal \cite{MAQRO} estimates that for a free-propagation time equal to $100$ s  (accessible in their setup) it is possible to constrain GRW-type models, some models of quantum gravity decoherence, but not decoherence of the D-P type. 

In contrast, the Newton-Schr\"odinger Equation predicts a {\it retraction} of the wave-packet spread   for masses around $10^{10}$ amu \cite{GiGr}. An osmium nanosphere of radius $R \simeq 100$ nm would require a couple of hours of free propagation in order to observe significant deviation from Schr\"odinger spreading \cite{Grossa}.
This effect provides the only
  realistic prospect of directly testing the NSE, and it requires a space environment.
  
{\em Decoherence of photons.} Only the ABH model has been generalized for photons \cite{LagAn21}. For interferometer experiments with arm length $L$, the model predicts loss of visibility of order $(\Delta \Phi)^2 = \frac{8G\Theta E^2L}{\hbar^2 c^6}$. For $L=10^5$km, $\Theta \sim T_P$ and photon energies $E$ of the order of 1eV, this implies a loss of coherence of the order of $\Delta \Phi = 10^{-8}$.  In principle, this would be discernible with EM-field coherent states with mean photon number $\bar{N} > 10^{16}$, though it would be very challenging to suppress all other systematic errors to this degree.

\subsubsection{GR Effects: Summary}
The untested prediction that propagation across a gravitational potential induces a phase shift on a single photon, a photon superposition state, and (hyper)entangled photon pairs was reviewed. A set of experiments involving interferometers distributed between spacecraft and ground nodes designed to test this prediction was outlined.  The order of magnitude of the phase shift on the photon states caused by gravity was determined to be compatible with experimental capabilities.  Preliminary mission systems analysis suggests an optimal regime for a spacecraft mission to achieve these objectives. As such, tests of the equivalence principle using photons are plausible using a future DSQL mission.  In contrast, the magnitude of gravitationally induced {\itshape decoherence}, based on QFTCST models, is likely too small to measure without significant breakthroughs in multiple instrumentation capabilities.

\subsection{Long-baseline Bell tests}
\label{sec:BellTestsection}
\setcounter{footnote}{0} 
The long baseline of DSQL could enable tests of Bell's inequality \cite{Bell1976} up to the lunar orbital radius and between inertial frames with large relative velocity, well beyond what is possible on earth or earth orbit.  As described in the sections below, conducting Bell tests at extremely long baselines between inertial frames opens experimental possibilities and addresses fundamental questions around quantum theory in the regime of general relativity.  Such tests also serve as an important validation benchmark for the implementation of future quantum technologies.

\subsubsection{Verification of Long-baseline Quantum Entanglement}
 A future global-scale quantum network shall be capable of maintaining the fidelity of distributed photonic states in various degrees of freedom, and interact with quantum memory devices, effectively establishing a quantum internet \cite{Boone:2015aa, Khatri2021, Gundogan2021}. This network could be useful for fundamental tests of quantum physics, distributed quantum computing or distributed quantum sensors (e.g., \cite{quantum_telescope}). Based on current technologies, relatively high-rate  entanglement distribution and quantum communication across baselines more than a few hundred kilometers are only possible using spacecraft links \cite{Bedington:17}. Furthermore, all quantum network applications rely on the validity of quantum mechanics, and a complete understanding of long-baseline quantum link behavior, which the DSQL tests could provide.

Entangled quantum systems that are shielded from the environment exhibit correlations that are expected to persist no matter how far apart the systems travel.  An example is polarization-entanglement carried by pairs of photons traveling separately along different optical fibers.
While the local topology of the individual fibers will induce a specific rotation on the polarization of each photon -- changing the specific form of the entangled state but not the {\it amount} of entanglement present -- this transformation can be reversed so that analysis devices at the respective destinations can still measure correlations satisfying the Bell test criterion for demonstrating nonlocality. This reversal is possible in principle, although sometimes it can be quite difficult in practice; for example, if the photon pairs have a large bandwidth, polarization mode dispersion -- wavelength-dependent polarization transformations -- within the fibers can be difficult to correct, resulting in an effective depolarization.

In contrast to propagation through optical fiber, photons propagating across the vacuum of deep space will encounter very few effects known from conventional physics\footnote{Unconventional physical theories predict some polarization rotation and other rotations through proposed coupling mechanisms as photons propagate across changing gravitational potential \cite{ralph14,joshi18, bruschi14}. These predicted effects were not observed by the first battery of on-ground \cite{fink17} and Micius spacecraft experiments \cite{Yin1140,yin17_1,xu19} attempting to measuring them.  In the Micius experiments, entanglement was distributed between a ground station and the spacecraft, and also between two ground-based observatories approximately 1,100 km apart. In these experiments, the entanglement fidelity was 0.907$\pm$ 0.007, only sufficient to determine the Clauser, Horne, Shimony, and Holt (CHSH) parameter to within 2.4 $\sigma$.  The unconventional theories listed above could thus be much better bounded with a test resulting in higher statistical significance on the CHSH parameter, achievable using brighter, higher fidelity sources.
} that could change the entanglement correlations in the various degrees-of-freedom. Kinematic effects may cause small shifts in the polarization state, but these require the detectors (or observers) to be accelerating (e.g., the non-inertial frame experienced when orbiting a massive body).

The evidence and the prevailing theories to date suggest that polarization entanglement correlations should persist in most scenarios involving a deep space communications link. Given the lack of coupling between light and the environment, we also expect that entanglement in other degrees of freedom (such as spatial mode, time-energy,  time-bin, or even simultaneous hyperentanglement {\it across} these degrees of freedom) should also be preserved over long distance propagation.

Consequentially, it is expected that a robust quantum communications link utilizing entanglement correlations should not face fundamental obstacles to realization based on known, conventional physics.  DSQL will help validate the assumption that deep space holds no further surprises in the form of new physics that might invalidate our assumptions of the characteristics of a space-based quantum communications link. In the following sections, we provide a checklist of fundamental experiments that could be performed in order to gain confidence that engineering an entanglement-based quantum communications system is a worthwhile endeavor, while enhancing the distance limits of fundamental tests of quantum physics. At the same time, the \textit{implementation} of these experiments will build up significant know-how and capability that will aid future quantum network engineering efforts.

In order for the DSQL to characterize and validate ultra-long range  Bell tests, the  methodologies to use photon correlation measurements and counting statistics as described in Appendix \ref{sec:stats_bell_tests} would apply.  A violation of Bell's inequality by at least five standard deviations would be considered a viable test. The required number of successfully detected photon pair counts needed depends on the correlation visibility, but typically around  $1,000$ detected pairs should allow the test to be conclusive. Furthermore, a detailed signal-noise analysis that involves quantum optical models of the photon pair source, channel properties, noise sources and photon detectors would be used to further compare the results with the known (conventional) physics, similar to the study in \cite{Holloway:2013aa}.

\subsubsection{Current Status of State-of-the-art Bell Tests}
The most sophisticated ``loophole-free'' Bell tests that have been done up to this point with entangled photons \cite{Giustina15,Shalm2015} close the detector-efficiency loophole, locality loophole, and versions of the freedom-of-choice loophole. These experiments use local sources of quantum randomness, where each random bit comes into being at a point in spacetime that is space-like separated from the measurement on the other side. In \cite{Shalm2015}, three different sources of random bits were XORed together based on the optical phase of a gain-switched laser, sampling the amplitude of an optical pulse at the single-photon level, and a predetermined pseudorandom source comprised of popular movies and digits of $\pi$. Any local-realist explanation for the observed Bell-violating correlations would be required to predict the outcomes of \textit{all} of these processes well in advance of the beginning of each trial.

The assumption that the random bit comes into being when the phase or amplitude is measured is critically important in Bell tests. If the bit on either side is determined (or even influenced) in some way by something in its past that the other side's measurement also has access to, the Bell violation can be explained with a local theory.  To address this, \cite{Gallicchio2014,Handsteiner2017,Rauch2018} used the unpredictable color of incoming astronomical photons from opposite sides of the sky. The latest of these \cite{Rauch2018} used two quasar photons that were emitted when the universe was a half and a tenth as old as today. This forces any local explanation that takes advantage of the freedom-of-choice loophole to have access to the past light cones of these quasar emissions. This access needs to be of sufficient fidelity for one side's measurement to predict the other side's next photon color. These experiments did not simultaneously close the detector efficiency loophole. Also, there is no device-independent way to verify that the clicks registered by the quasar photons were really mostly determined by the distant cosmological past rather than a local conspiratorial random number generator that was, perhaps, matching the average rate of incoming cosmological red and blue photons, but coordinating the specific order with the other side of the experiment.

Recently the BIG Bell Test \cite{BigBellTest2018} created a web-browser game where people around the world were rewarded for acting as unpredictably as possible. For a 24-hour period, their inputs were used  to choose the measurement bases for 13 simultaneous Bell-type tests around the world. This was a heroic effort to close a loophole in previous experiments, where something in each experiment's past could have influenced the settings. However, given the constraints of Earth-bound participants, their choices were electronically recorded and used in such a way that (purely in terms of past light cones) the sources of entanglement and all measurements had access to the choices in advance, i.e., a substantial loophole remains.

As mentioned above, the measurement of entanglement over long distances is expected to follow ``conventional physics'', no matter what distances are traversed, or whether the measurement apparatus (the observers) are at rest relative to each other.  Measurement devices in relative motion can lead to a reference-frame-dependent event sequence, where the expectation of entanglement preservation  becomes less obvious. This is especially striking if the reference frames are physical, if the wavefunction is an element of reality as the Pusey-Barrett-Rudolph theorem (PBR theorem) favors \cite{PBRtheorem}, or if wavefunction collapse is a physical phenomenon caused by interaction with a measurement device in different reference frames. These alternative viewpoints are fundamentally different from general relativity, in which both super-observers and preferred reference frames are impossibilities. Only physical collapse is a non-standard viewpoint. In that sense, conducting experiments along these lines would test the predictions of QFTCST against these alternative ``strawman'' theories.

\subsubsection{Bell Tests Between Frames with Large Relative Velocities}
\label{BellTest}

Consider a Bell test scenario involving three inertial frames.  The entangled photon source is in the center while the two receivers, roughly equidistant on either side, are arranged to travel either towards each other or away from each other. In the rest-frame of the source, the two detection events are simultaneous and therefore space-like separated. In the case where the detectors are traveling away from each other, each detector would ``consider'' itself to be the first to receive the incoming photon and to generate a signal in its own reference frame. When the detectors are moving towards each other, the opposite case occurs and each detector would ``consider'' itself to have received the photon {\it after} its distant counterpart. In special relativity, a reference frame for a local observer is {\it operationally} determined as the radar coordinates, which respect Lorentz symmetry \cite{dInverno92, Lin2020}. With similar operations, the above ``considerations'' by each detector can only emerge after the detector receives an ideal radar signal, emitted earlier by itself, echoed back from the measurement event by the other detector. This is long after both events of receiving photons at the two detectors have occurred in all the reference frames of the detector and the source. Note, however, that such radar signals could have been used prior to the measurements described, so that reference frames are {\it predicted} before the experiment.

Suarez and Scarani \cite{sua97} refer to the above as ``before-before'' and ``after-after'' scenarios, respectively; this nomenclature assumes that, while each observer may not know whether it measures before or after its distant counterpart at the moment of its local measurement, as each observer completes their measurement, the collapse of the wave-function propagates instantly in their respective reference frame. But, since the two observers are in relative motion, it {\it appears} that both reference frames yield contradictory wave-function collapses \footnote{In the theoretical description of a quantum system, a reference frame to coordinatize physical events (spacetime points) is chosen by a specific observer, and wavefunctions to relate the outcomes of physical measurement events are specific to each observer \cite{AharanovAlbert}. In fact, the description of quantum states requires an explicit reference frame choice by an observer, because in the canonical quantization scheme a time-foliation (1+3) scheme needs to be specified  before a Hamiltonian can be written down \cite{AHGravDec, Lin2012}. Making a statement that two wavefunctions supported by different time-slices in two reference frames for two different local observers have contradictory collapses does not make sense, because it presumes that there exists a super-observer who can see the co-existence of two time-slices to support those wavefunctions.}.  Suarez and Scarani identified the polarization analysis-determining beamsplitter in each measurement station as the necessary device that must be moving, and proposed to place them on rotary mounts that spin to achieve relative speeds of 100 m/s.

Experimental studies with such dynamics, as first demonstrated by  Zbinden et al. \cite{zbi01}, utilized rotating absorbers as ``detectors''. In this experiment, part of the entangled photon signal was absorbed on rotators which either moved towards each other, or away from each other. The rest of the signal was routed to conventional, stationary detectors; the results showed no statistically significant deviation from a standard test using only conventional detectors at rest.\footnote{In these experiments the signal was routed by fibers, and the detectors only separated over 10km, so it was very difficult to get the space-like separation needed for the test. Typically, the experiment would begin with slight asymmetrical distances, and then the experiment would run over several hours, so that the diurnal effect on the fiber would sweep through the circumstance of simultaneous measurements at some point. Furthermore, the absorbers used as ``detectors'' gave no active signal (such as a heralding flash upon successful completion of an absorption event) that could directly contribute to the statistics; instead, they served ``null-result'' measurements -- {\it lack} of an absorption event projected the photon wavepacket into the arm with the photon detector.} Another experimental test was reported by Stefanov et al. in 2002 \cite{stefanov02} using time-bin entangled photons separated over 10 km via an optical fiber, and ``moving'' beamsplitters implemented with acousto-optical modulators. Each incoming photon effectively saw a beamsplitter moving at about 2,500 m/s; as in the previous experiment the actual single-photon detectors themselves were not moving \footnote{This is potentially problematic in an experiment whose underlying hypothesis relies explicitly  on  a measurement-induced wavefunction ``collapse''. Specifically, the unitary transformations of the beam splitter could be undone by other (local) optical elements. The measurement basis settings and the results are not truly fixed until the actual detection of the photons, which involves the amplification of a detection signal to the macroscopic scale, involving millions of electrons. Therefore, any convincing test of these simultaneity arguments should have the same collapse-inducing detectors themselves in relative motion.
}. Again, the experiment found that the entanglement correlations were perfectly preserved without need to consider the time-sequencing of events, in agreement with standard quantum mechanics formalism. 
An experimental scenario where the detection events by the two moving observers of entangled photons are in each other's respective future, or past, has yet to be tested -- though it should be noted that Scarani et al. \cite{Scarani2014aa} have shown that the multi-simultaneity model could lead to superluminal communications, in violation of relativity. It is understood that satellites are the best approach to obtain the required relative speeds and distances for such a measurement, and these tests could be considered for the DSQL platform \cite{rideout_2012b}.

 In the previous tests the actual photon detection systems were stationary, leaving an open question if a {\it moving} measurement device is required to represent a moving ``observer", since the beam splitter operation, even while moving, remains coherent (and reversible). Ultimately the entire observer system including the detection process must be in the motion \cite{rideout_2012b, Scarani2014aa}.  Here we consider space-based experiments, where satellites can be distant enough from each other and moving fast enough away from each other such that each detector's measurement can be considered complete in its own reference frame before the other detector even begins its measurement. Similarly, the satellites can move toward each other fast enough such that in each local detector reference frame the distant detector's measurement is completed before the local detector begins its measurement.

\begin{figure}
\centering
\begin{subfigure}{0.40\textwidth}
  \centering
  \includegraphics[height=0.25\textheight]{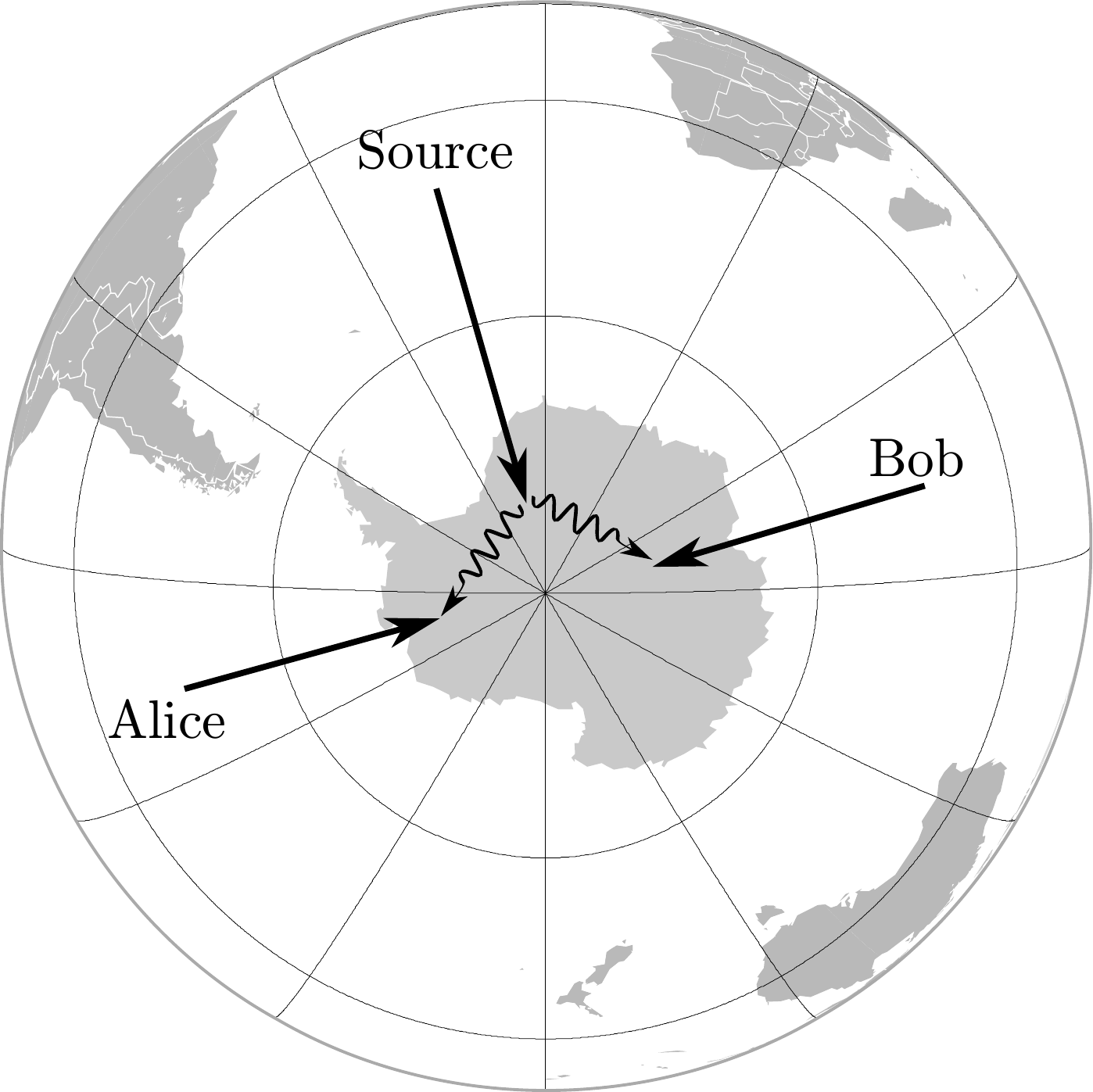}
  \caption{After-After Polar Orbits}
  \label{fig:after_after_satellites_over_Antarctica}
\end{subfigure}
\hspace{1em}
\begin{subfigure}{0.55\textwidth}
  \centering
  \includegraphics[height=0.25\textheight]{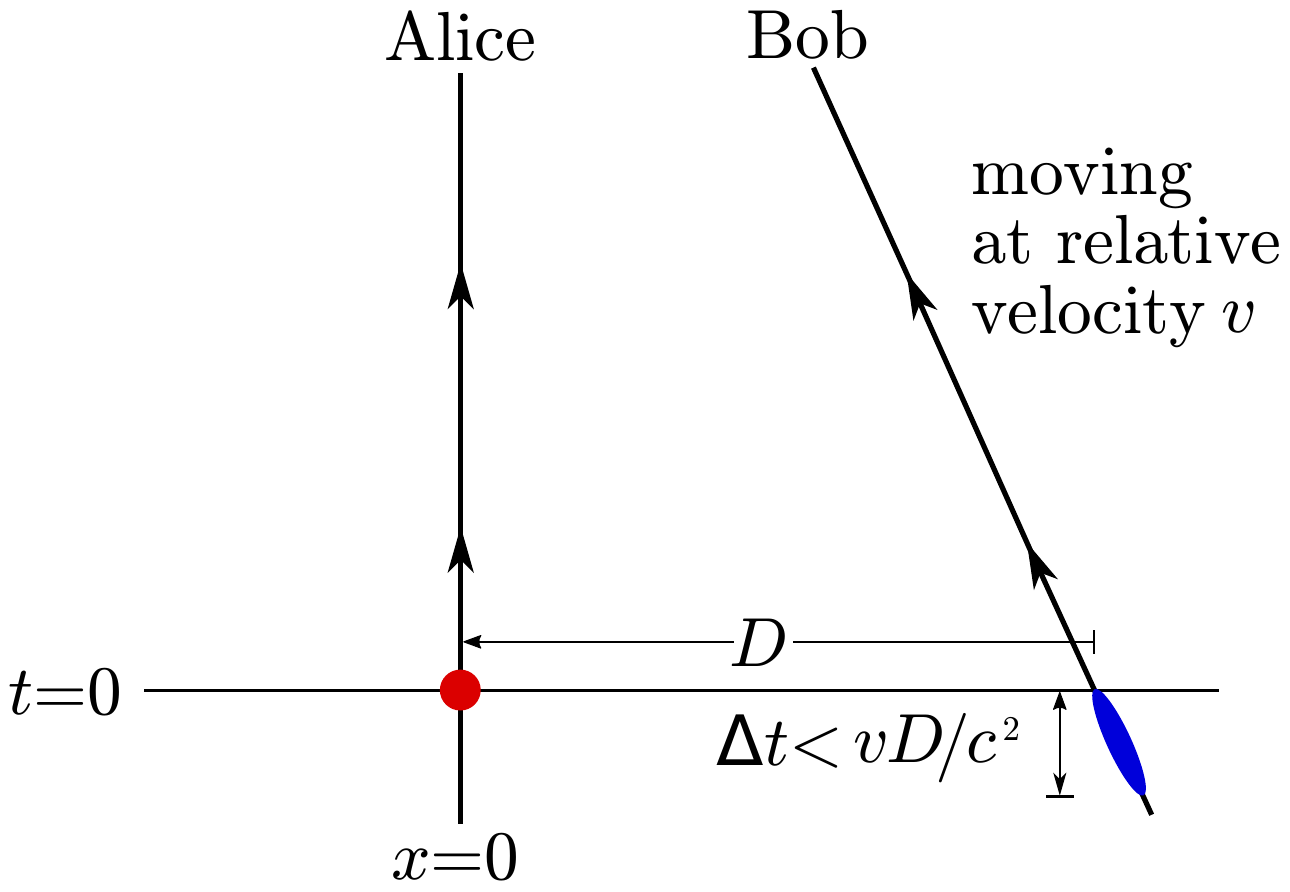}
  \caption{After-After Spacetime Diagram}
  \label{fig:spacetime_after_after}
\end{subfigure}
\caption{(a). One source and two ``After-After'' satellites in polar orbit over Antarctica. An alternative would be to site the source on the ground for double-uplink transmission of entangled photons to the orbiters. (b) Spacetime diagram in Alice's rest frame. Alice's measurement happens at $(t=0,x=0)$. The tight constraint on Bob's allowed measurement window $\Delta t$ is also shown.}
\label{fig:AfterAfter}
\end{figure}

Creating high relative velocity between two moving platforms near enough to each other to maintain high link efficiency over a long enough integration time is the key requirement for such tests of relativistic simultaneity.  The constraints on timing are stricter in these before-before or after-after experiments than in a typical Bell test, where locality dictates only that the two measurements be space-like separated. This space-like separation means that there exists some frame where one measurement is first and some other frame where the other is first. In the before-before experiment, these frames cannot merely exist, but must include the actual rest frame of each detector (see Figure \ref{fig:spacetime_after_after}). For satellite speeds much slower than the speed of light, the measurements must be within $\Delta t < vD/c^2$ of each other, where $v$ is the relative velocity of the detectors and $D$ is the instantaneous separation between the measurements. For reasonable LEO parameters, this is around 10\,m of light travel time, and translates into an accuracy requirement of approximately $10^{-5}$ in the position of the receivers. Note that it is not sufficient to merely \textit{know} the orbits to within 10\,m and the detection times to within 30\,ns---the orbits and timing must be \textit{controlled} to this accuracy for sufficient duration so that a statistically significant number of entangled pairs arrive while this condition holds.  Furthermore, in a LEO constellation, these scenarios only exist for brief periods of time and drive strict orbit determination and station-keeping requirements on the flight platforms. To satisfy the conditions of this scenario, at least two platforms should be on satellites in space. 

 With receivers on two independent spacecraft, the link budgets are likely to improve---in Figure \ref{fig:after_after_satellites_over_Antarctica}, three polar-orbiting satellites would meet regularly over the poles to perform the experiment repeatedly, potentially both while approaching and receding. Some of the challenges in this scenario, however, are the slew rates as the platforms converge, avoiding collisions, and the additional challenge of operating in high-radiation zones. An alternative might be to place the source on the equator and to beam photons to two counter-propagating receiver satellites in an equatorial orbit. To mitigate uplink losses, the source would have to be placed at a relatively high altitude, perhaps even on a balloon.

\begin{figure}[t]
\centering
  \includegraphics[height=0.25\textheight]{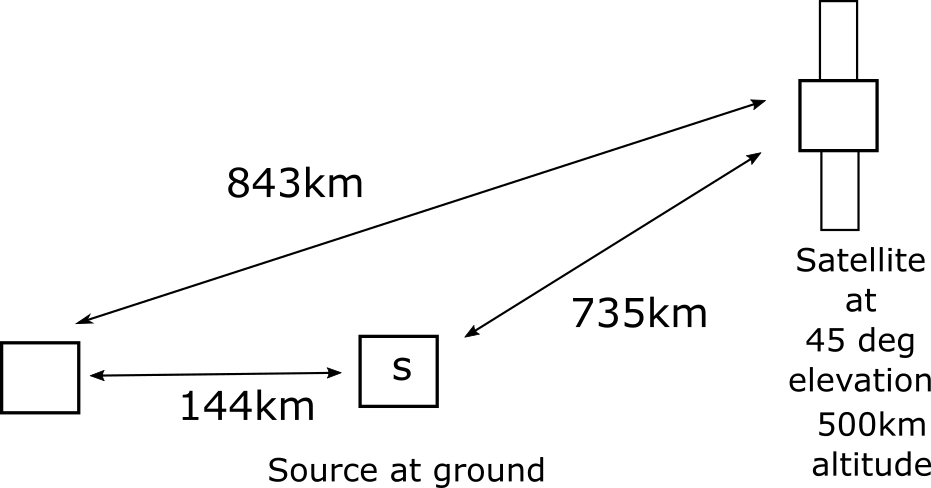}
\hspace{1em}
\caption{The asymmetric Bell test using only one space-based observer, and a ground-based observer with suitable delay (either a fixed path or quantum memory). A ground-based source (e.g., located at the Canary Islands) could transmit one of the entangled photons to another terrestrial receiver, and the other to a receiver located in orbit. The relatively short ranges on Earth require that there be a substantial delay ($ \sim$ 1 to 3 ms) at the terrestrial receiver to achieve nearly similar optical path lengths. The 144-km free-space link would only be sufficient (ca. 0.5-ms delay) for very low altitude satellites. Longer delays could be implemented in fiber-optics, but would entail about -100dB of loss or greater. Ideally, a low-loss quantum memory with finely adjustable readout times would be used on the ground.}
  \label{fig:ground2satellite}
\end{figure}

For comparison, consider the scenario where an entangled pair source transmits one photon to an orbiting platform, and the other to a ground receiver (see Figure \ref{fig:ground2satellite}). The asymmetrical free-space path lengths necessitate a delay line or buffer at the terrestrial receiver that can change quickly by the equivalent of several hundred kilometers, resulting in several technical challenges. Because the detectors are not moving symmetrically, there is only a 1\,ms (10\,m/$v$) time window in each orbit where the alignment can be such that each detector measures first (or second) in its own reference frame.  A huge possible improvement to such an asymmetric scheme would be utilization of a quantum memory to achieve variable read-out times  that correspond to the time-of-flight for the ground-to-space link. If the moving observer were located on a LEO platform, this quantum memory would need to store the ground photon for around 1-3\,ms, and release it at the correct moment, within about 1 ns, to ensure the relativistic separation of the two measurements of the entangled photons.  The low efficiencies of both the quantum uplink and the quantum memory would certainly present challenging low count rates. It is nevertheless encouraging that a first proof-of-concept test of such an asymmetric setup may be possible with a single receiver, such as the Canadian QEYSSAt mission \cite{QEYSSat2014}.

\subsubsection{Human-decision Bell Tests}
\label{sec:HumanBell}
 Quantum experiments that directly  involve human participants are both scientifically interesting and socially relevant, igniting the public interest in fundamental science. In a space-based test of nonlocality, astronaut participants can meaningfully address the free-will loophole of Bell tests in a new regime.  The motivation behind this body of testing is largely philosophical in nature, dealing with the epistemological underpinnings of quantum theory. In fact, John Bell himself suggested letting humans choose the basis in a test of his famous inequality \cite{bell1985exchange}:
``It has been assumed that the settings of instruments are in some sense free variables---say at the whim of experimenters
'' \cite{bell2004speakableTheoryOfLocalBeables} and ``Roughly speaking it is supposed that an experimenter is quite free to choose among the various possibilities offered by his equipment'' \cite{bell2004speakableTheoryOfLocalBeables}. 
Furthermore, Leggett points out \cite{Leggett:2009lr} that any exploitation of a loophole that relies on a complicit role of the process that chooses the random settings (it is either not random, or it is somehow influenced by the entangled photons) can maybe best be settled by having two human observers operate the measurement devices.
Lucien Hardy, in ``Proposal to use Humans to switch settings in a Bell experiment,'' \cite{Hardy2017} sums up this experiment well: ``The radical possibility we wish to investigate is that when humans are used to decide the settings (rather than various types of random number generators) we might then expect to see a violation of Quantum Theory in agreement with the relevant Bell inequality. Such a result, while very unlikely, would be tremendously significant for our understanding of the world.'' In this Section we explore both the feasibility and desirability of doing such an experiment given that all previous Bell tests have vindicated the quantum prediction.

\begin{figure*}
\centering
\begin{subfigure}{0.33\textwidth}
  \centering
  \includegraphics[width=\linewidth]{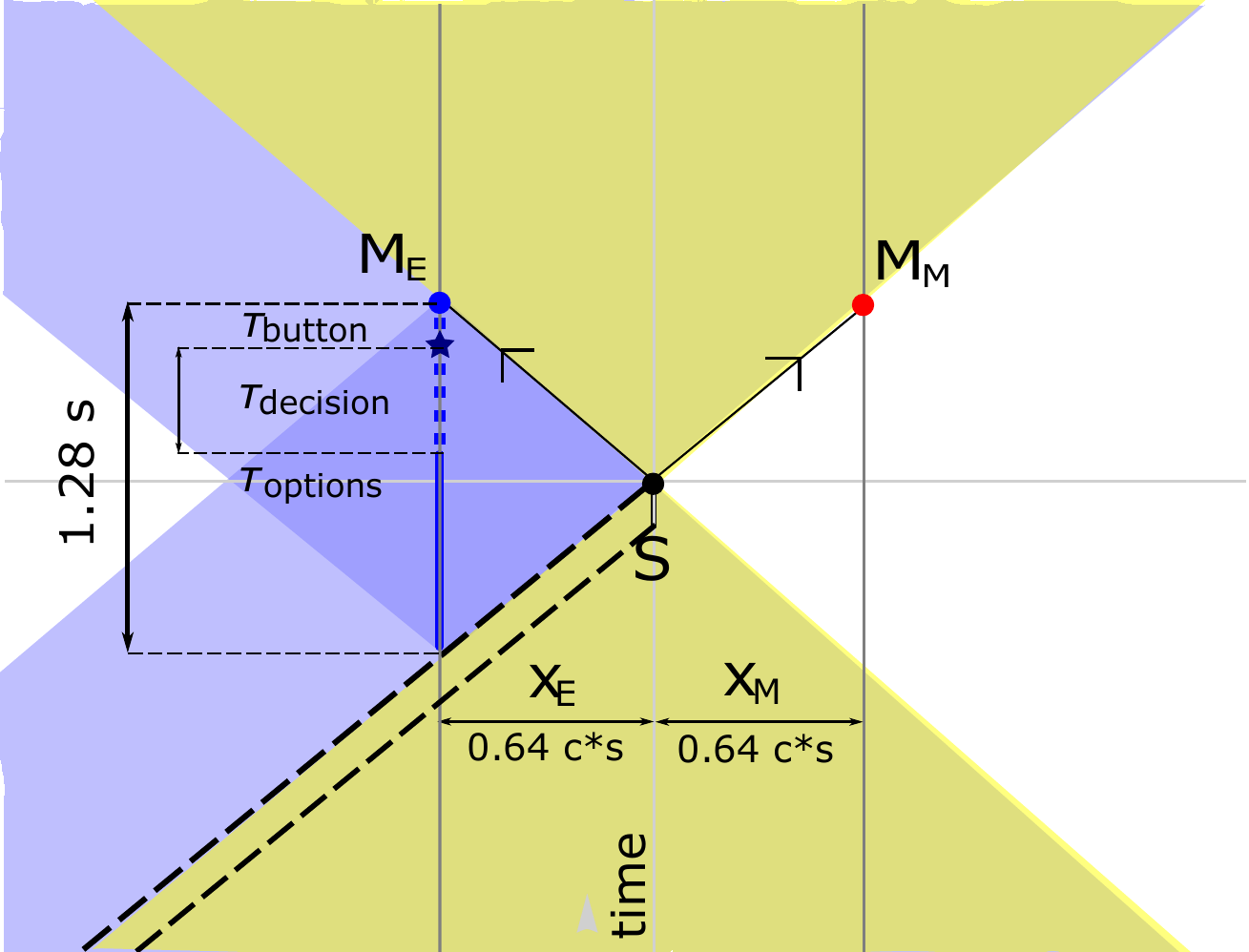}
  \caption{Earth and Source}
  \label{fig:HumanSpacetime_Earth}
\end{subfigure}%
\begin{subfigure}{0.33\textwidth}
  \centering
  \includegraphics[width=\linewidth]{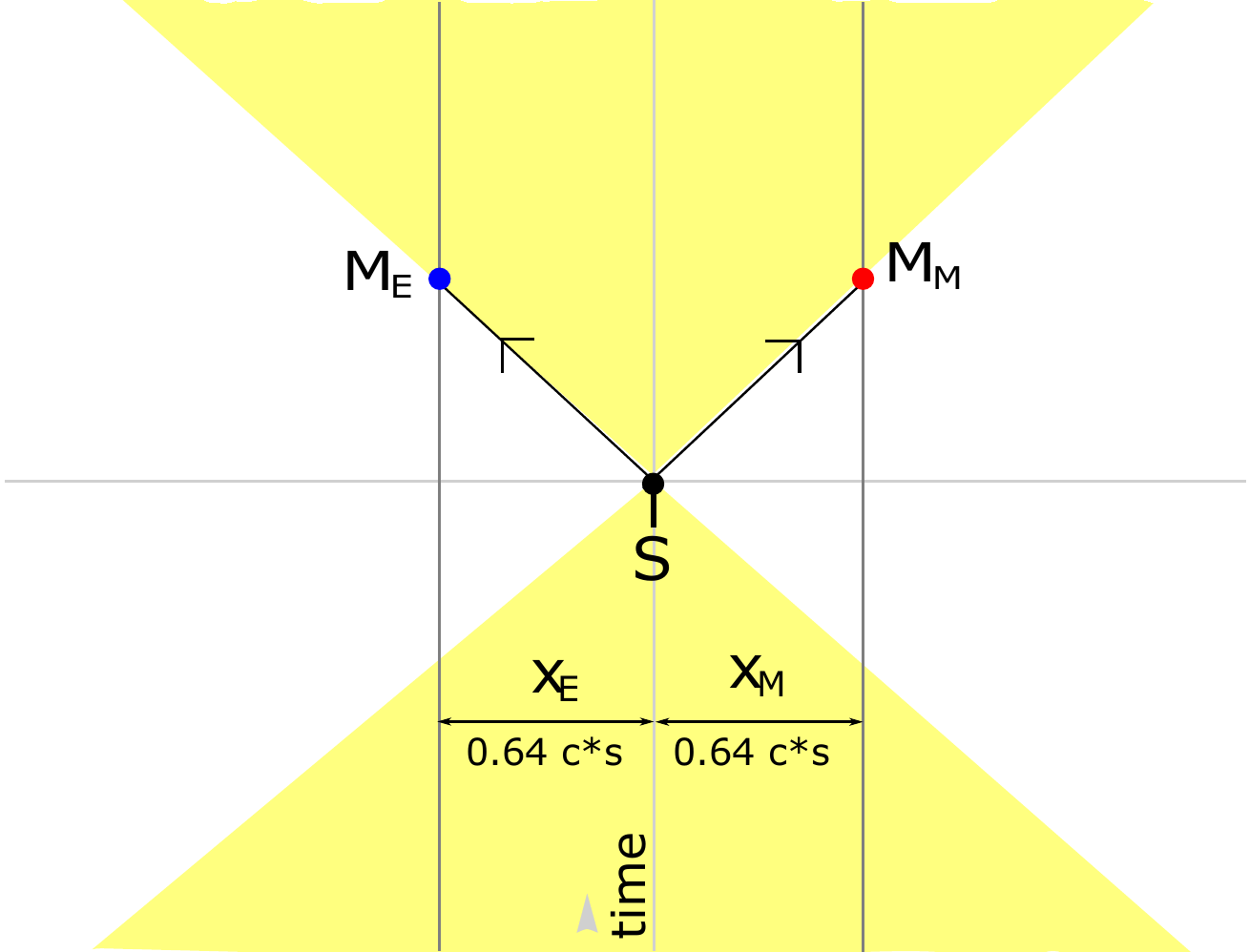}
  \caption{Source Only}
  \label{fig:HumanSpacetime_Source}
\end{subfigure}
\begin{subfigure}{0.33\textwidth}
  \centering
  \includegraphics[width=\linewidth]{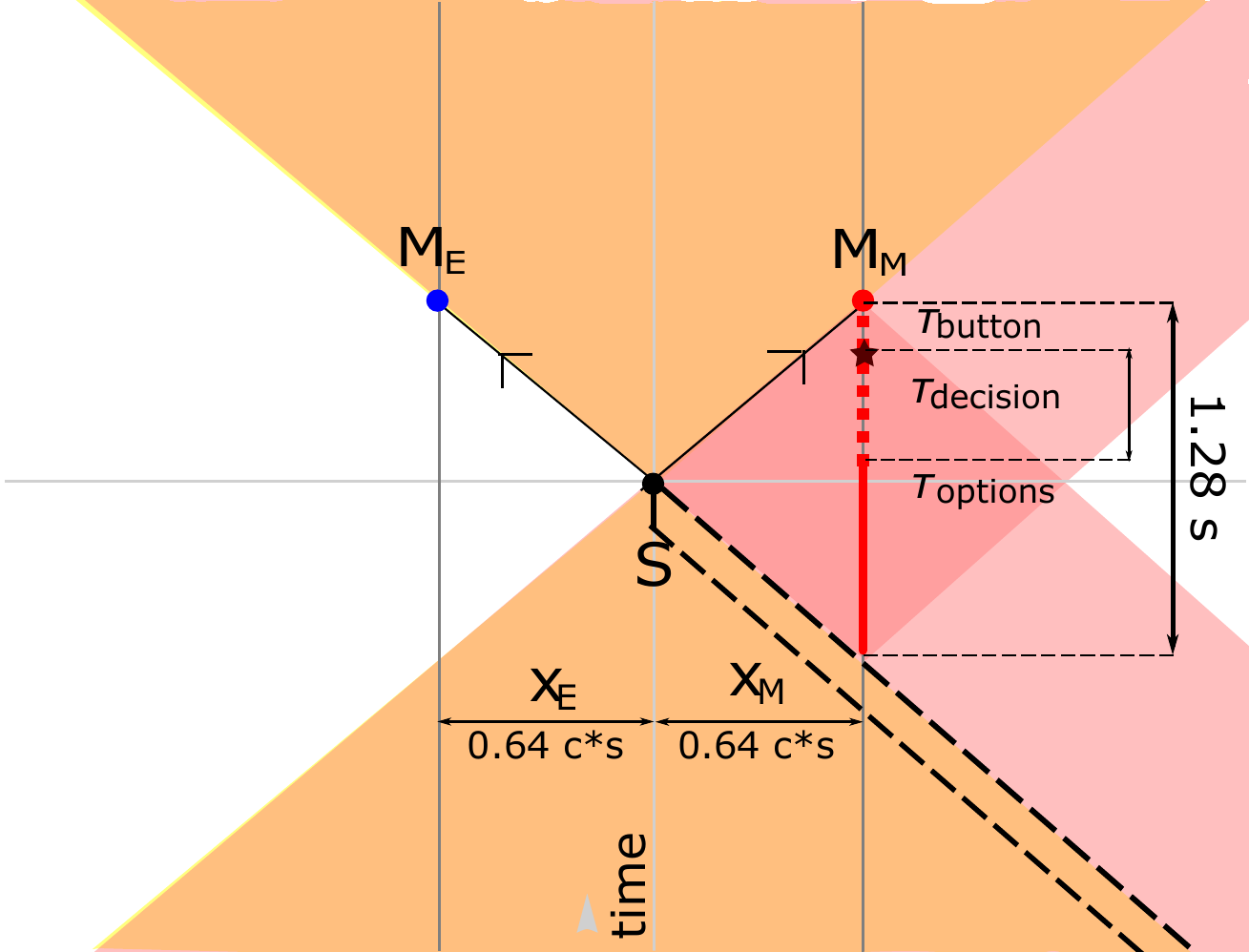}
  \caption{Moon and Source}
  \label{fig:HumanSpacetime_Moon}
\end{subfigure}
\caption{Spacetime Diagrams for a Bell test where the source is halfway between Earth and Moon. In each round, humans on each side have 1.3 seconds to be presented with a choice, make a decision, and register their decision with something like a button, whose activation turns that decision into a polarizer setting.}
\label{fig:HumanSpacetime}
\end{figure*}

Allowing humans to choose the settings while simultaneously closing the locality loophole requires that the experiment be large enough that no information about the measurement choice on one side can be accessible to the measurement on the other side. Libet famously used EEG to measure the Readiness potential 0.3\,s before a person consciously decided to move and 0.5\,s before they pressed a button \cite{libet1993time}. Perhaps specialized training can improve the reaction times, but it takes somewhat less than 1\,s second for a person to be presented with an choice that they did not know about in advance, make a decision using what they perceive to be their free will, and reliably register this choice in a way that can quickly (electronically) change the analysis basis of a polarizer.
This timing makes the 1.2-1.4-s light-travel time between the Earth and the Moon the right scale. The space-time diagrams shown in Figure \ref{fig:HumanSpacetime} lead to the conclusion that with a source half-way in between experimenters on the Earth and Moon, participants would have the full 1.2-1.4\,s to carry out each round of the experiment. Similar timings are relevant for the source at an Earth-Moon stable Lagrange point. Although the link losses over such distances are daunting, laboratory analogs are being used to prepare for quantum communication experiments over such high-loss channels \cite{Cao2018}.
 
\begin{figure}[t]
\centering
  \includegraphics[height=7cm]{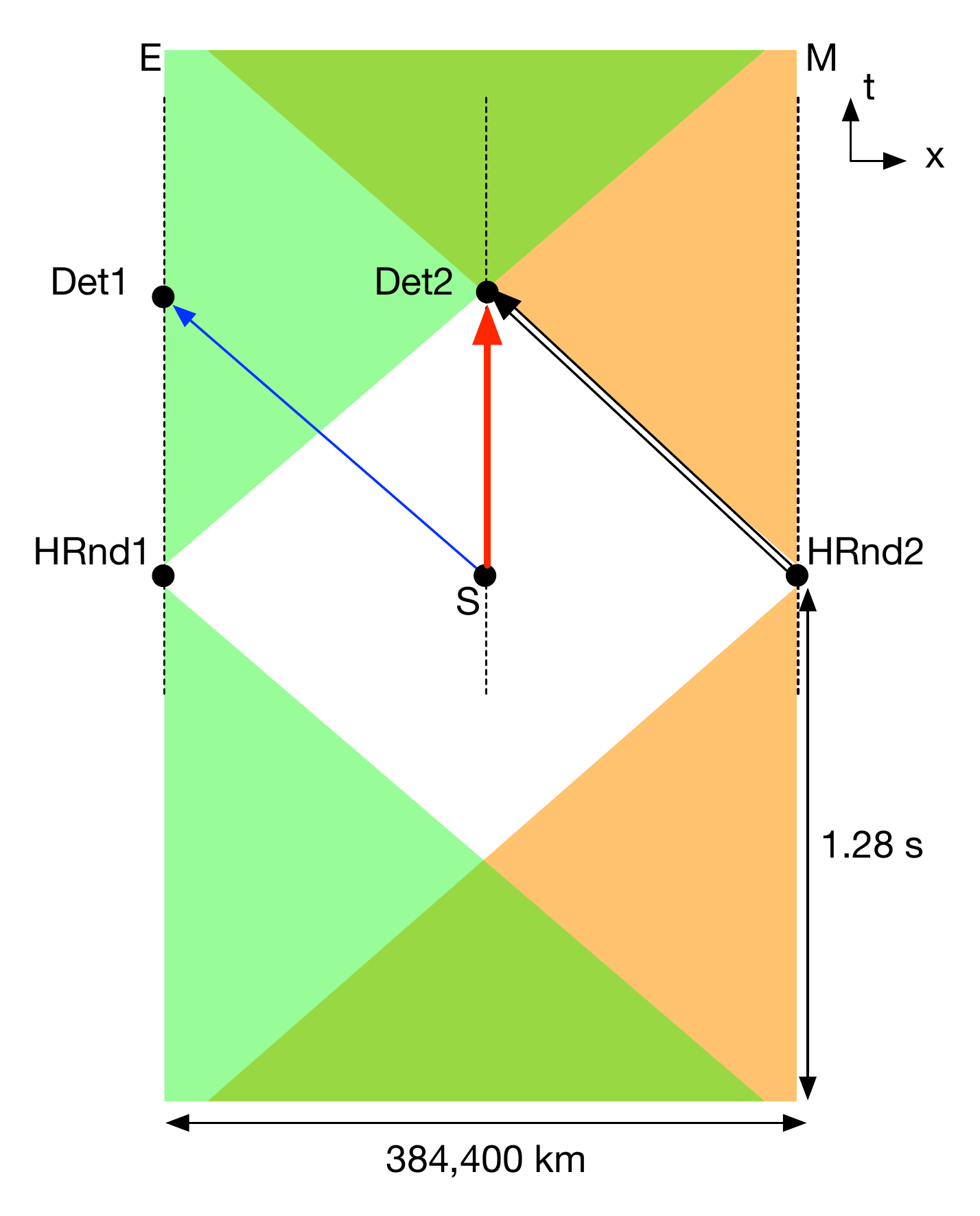}
\caption{Space-time diagram for a Bell test involving human observers. In this asymmetric configuration, while the space-like separation of random choice is made by humans separated by the Earth-Moon distance, only one of the entangled photons is sent over a long distance link to Earth (blue arrow); the other entangled half is stored in a quantum memory at the source (red arrow), awaiting the random choice classical signal transmitted from the Moon (black double-line arrow). When the source is still located on the direct line to the moon, but not necessarily halfway, each participant still has a time to make their choice that is equal to twice their light travel time to the source.}
\label{fig:HumanSpacetime_assym}
\end{figure}

Another scenario is an asymmetric configuration of quantum entanglement, as shown in Fig. \ref{fig:HumanSpacetime_assym}. This involves creating an entangled pair between Earth and Moon, sending one photon to Earth, and storing the other photon until a classical signal arrives from the Moon with the astronaut's basis choice. The random choices implemented by the human random choices are still separated by the Earth-Moon distance, but this configuration has the benefit of requiring only one long-distance link at the cost of half-second quantum storage. The figure shows a source halfway between Earth and Moon, but the optical link can be shortened at the cost of longer storage time. A smaller version of this scheme was implemented with a fiber delay on the Canary islands in 2010 \cite{Scheidl19708}. It was used to study the Freedom-of-Choice Loophole, albeit with measurement settings randomly chosen based on the behavior of photons from an LED on a beamsplitter; no deviations from the predictions of standard quantum mechanics were found. Other similar arrangements are conceivable.

\begin{figure*}[p]
\centering
  \includegraphics[width=11cm]{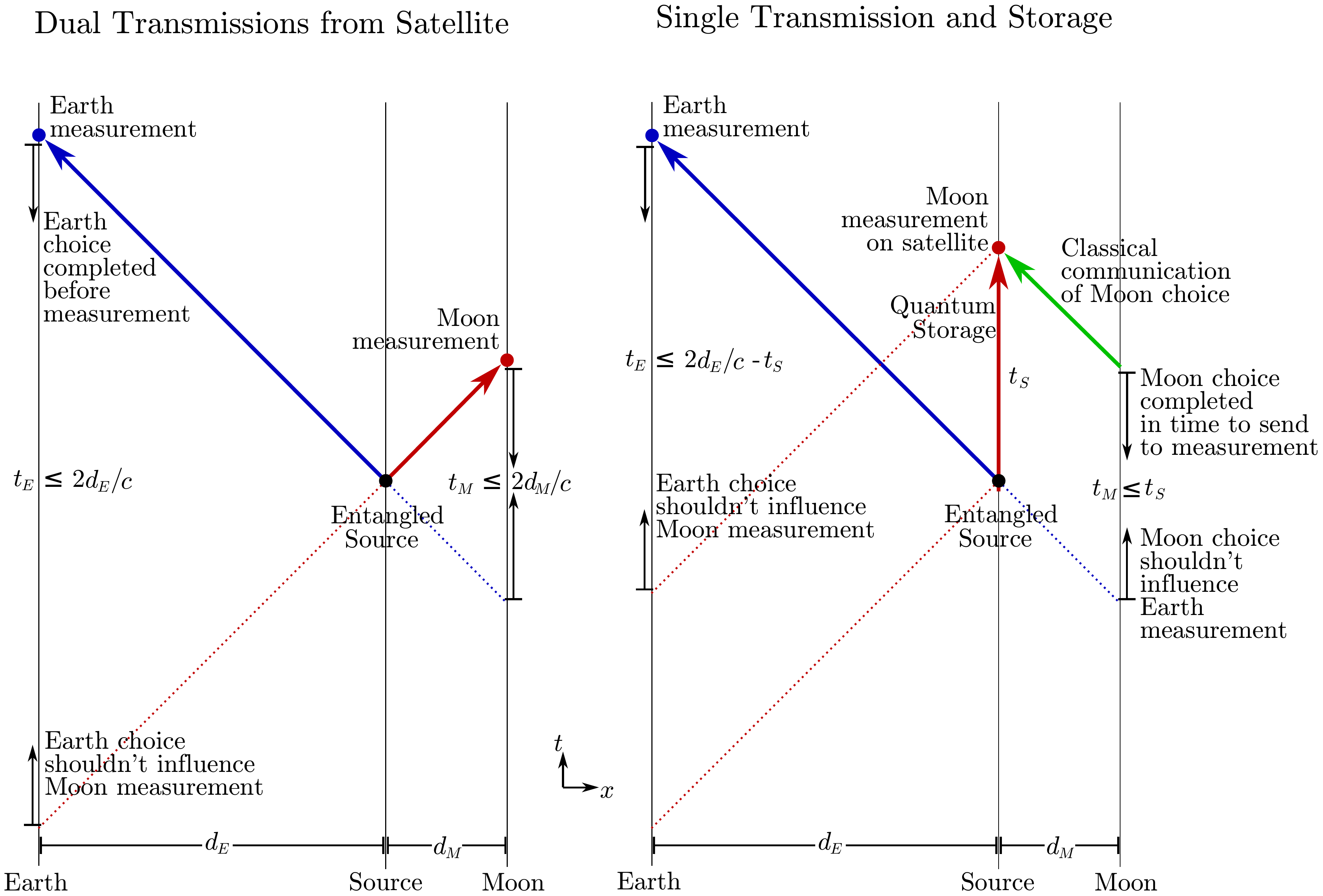}
\caption{Space-time diagrams for two versions of a Bell test involving human choices made by astronauts on Earth and Moon. The source is shown at an arbitrary position between Earth and Moon. There is a limited time window for humans at each location to be presented with options and make a choice, indicated by the black arrows. These windows have maximum duration $t_E$ and $t_M$. Each time window starts after the last event that could influence the other side's measurement and ends when the choice needs to be used or transmitted. These windows are shortened by hundreds of nanoseconds (too small to see at this scale) due to fiber delays, bounces within telescopes, and transmission through atmosphere.  
\textbf{Dual Transmissions (Left):} If the source is midway between Earth and Moon, in each round, humans on each side have $\sim$1.3 seconds to be presented with a choice, make a decision, register their decision with something like a button, and have that decision converted into a polarizer setting. When the entangled source is farther from Earth, people there have more time $t_E$ to make their basis choice at the expense of time on the moon $t_M$. The best situation is when the entangled source is halfway between Earth and Moon, when astronauts at each location have equal time to make a valid choice; that time equals the Earth-Moon light travel time, $\sim$1.3 seconds. 
\textbf{Single Transmission and Storage (Right):} In this asymmetric configuration, while the space-like separation of random choice is made by humans separated by the Earth-Moon distance, only one of the entangled photons is sent over a long distance link to Earth; the other entangled half is stored in a quantum memory at the source, awaiting the random choice classical signal transmitted from the Moon. By the geometry of the causal influences, the maximum time window $t_M$ in which the astronauts on the moon need to make their basis choice is equal to the storage time $t_S$. A longer storage time comes at the expense of Earth decision time $t_E$. The best situation is to divide the decision time equally, ($t_E=t_M=t_S$). In this case, the source near the moon, transmits one entangled photon all the way to Earth, and stores the other entangled photon for the full 1.3 seconds.
}
\label{fig:human_earth_moon_bell_tests}
\end{figure*}

\begin{figure*}[t]
    \centering
    \includegraphics[width=10cm]{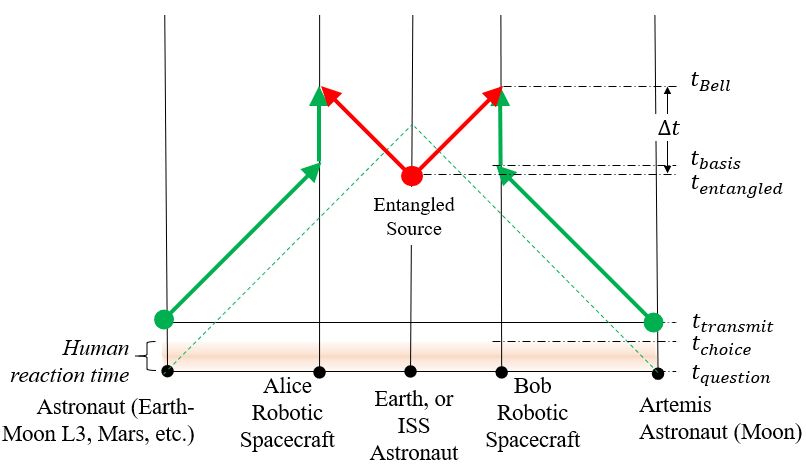}
    \caption{Alternative scheme for human-decision Bell test.  Artemis astronauts on the Moon, and astronauts at the Earth-Moon L3 point or beyond, are queried at time $t_{question}$.  Both astronauts make a decision within the human reaction time $t_{choice}$.  This answer is locally cached until time $t_{transmit}$, when a classical signal is sent to unmanned measurement spacecraft ``Alice'' and ``Bob''.  A light source, located on or near to Earth, emits entangled photons toward Alice and Bob before intersection of the world line of the source and light cones of the query events, i.e., with no chance that the emitted photons could have been influenced by the measurement basis choices of the astronauts.}
    \label{fig:Mod_HBT}
\end{figure*}

And finally, the architecture of the human-decision Bell tests, shown in Figure \ref{fig:Mod_HBT}, does not require deployment of quantum optical hardware in deep space, but ensures that there is no chance the emitted entangled photons nor the distant measurement could have been influenced by the local measurement choice.  The key requirement is that the decision-events of the astronaut participants and emission-events of entangled photons comply with specific timing requirements.

Both astronauts are queried at time $t_{question}$, synchronized using GPS, at a time well before $t_{entangled}$ when the source is activate to (possibly) emit a photon pair.   One set astronaut is here presumed to be located on the Moon, while the other is space-like separated from the Earth, with at least the same distance as the Moon, but in the opposite direction (the third Lagrange point (L3) of the Earth-Moon system provides such a benchmark, but in principle the second set of astronauts can be elsewhere in the solar system).  As described above, each astronaut has some reaction interval $t_{choice}$ during which they can consider and respond to the prompt. There is no reason to expect that both astronauts would submit their answers at precisely the same time, so their decisions are cached locally up to time $t_{transmit} ( > t_{question} + t_{choice})$, at which point their decisions are transmitted classically to unmanned spacecraft with measurement stations ``Alice'' and ``Bob'', and received at time $t_{basis}$.
Completing the Bell test requires entangled photon reception and measurement at the Alice and Bob stations -- at $t_{Bell}$ -- before any possible signal emitted at $t_{question}$ from the other side's basis-choice astronaut station could reach the analysis satellites.  In order to ensure this, the time-of-flight from the entangled photon source to either Alice or Bob (labeled in Figure \ref{fig:Mod_HBT} as $\Delta t$) must exceed $t_{transmit}-t_{question}$. \footnote{For simplicity we have assumed a symmetric arrangement, but clearly the source might be closer to Alice or Bob; in this case $\Delta t$ is taken to correspond to the smaller of the two distances.} Assuming a human reaction time is $t_{choice} = 0.25\,$s, the cache time $t_{transmit}$ should be greater than this; here we assume 0.4\,s.  The corresponding distance between source and Alice (Bob) is about 1/3.25 the Earth-Moon distance.

Because the associated link efficiency of the whole experiment (see Appendix \ref{sec:linkAnalysis}) goes as $1/R^4$ ($R$ being the source-to-measurement station separation), this is $\sim$100 times more efficient than the Earth-Moon human Bell test configuration (Fig. \ref{fig:human_earth_moon_bell_tests}a), and 7 times more efficient than the ``midway'' Earth-Moon Bell test variant (Fig. \ref{fig:human_earth_moon_bell_tests}).  This margin may be traded for making the question and answer interface of the astronauts more relaxed by, e.g., querying them once every two seconds instead of once every 0.4\,s.  Even then, the rate improvement is 22 times higher than the Earth-Moon test, and 1.5 times higher than the midway test. Furthermore, the rate could be further enhanced by using multiple astronauts at each station, e.g., a typical capsule crew of three could supply measurement decisions 3 times as often.

Quantum mechanics itself makes no distinction between basis choices determined by classical randomness, quantum randomness, or human choices---all of these should violate Bell's Inequality equally. If this experiment is performed with all of these sources of randomness and the Bell violation in each case is indistinguishable, we would confirm that quantum mechanics holds, but we would conclude nothing about free will. If we get the unexpected result, which is that the runs using human choices do not match the quantum prediction, what would we conclude? First, we would be forced to concede that quantum mechanics is incorrect. Any local-realist explanation for all of the observations must include a mechanism for predicting or influencing all non-human random number generators used in previous experiments that confirmed quantum predictions in Bell tests. Second, we would be able to say that human choices are at least sufficiently complicated so as not to be predicted or influenced, though exactly how one relates this back to concepts of free will would be up for debate.

In the closing paragraph of \cite{Brans1988}, Brans writes ``It is sometimes said that quantum theory saves free will. In the context of this paper, this might be reversed, so that free will saves quantum theory, at least in the sense of eliminating hidden variable alternatives. In other words, if there are any truly ``free'' events in the experiment, then there can be no classical determinism and hence no classical hidden variables.'' We wish only to caveat this inspirational quote by noting that for an experiment to rule out a local hidden variable theory, a single `free' event would not make a significant difference. Instead, a sufficient majority of the basis choices for each measurement must be `free' in the sense that they are not able to be predicted or influenced by  anything happening on the other side of the experiment.

\subsubsection{Flight Mission Design for Bell Tests}
All of the proposed Bell tests are characterized by the statistical significance of the measured violation of Bell's inequality. This is described mathematically and conceptually in Appendix \ref{sec:stats_bell_tests}. Increasing the number of successful, high-fidelity photon pairs simultaneously detected by Alice and Bob will improve the statistical significance of the test. 
Practically, this experiment could be realized through use of a very broadband entangled photon pair source undergoing dense-wavelength-division-multiplexing, thereby creating many simultaneous channels, each approaching the saturation capacity of the detectors. Leveraging this source architecture requires exceptionally low timing jitter. Reducing the probability of a noise event also improves the Bell test statistics.  Thus, number of succesful photon pairs measured during a measurement campaign, $N$, and purity factor, $p$, are the parameters used to parametrically describe the Bell test mission design. Equation \ref{eq:nsigma_violation_of_Bell}, reproduced here for convenience, relates the statistical significance of violation of Bell's inequality,$\sigma$, to parameters $N$ and $p$: \footnote{The purity factor $p$ used here is not to be confused with the ``$p$-value'' used in the previous loophole-free Bell tests \cite{LoopholeFreeTests,Giustina15} to characterize the significance of the violation; loosely speaking, the $p$-value is the likelihood the observed results are compatible with the null hypothesis, i.e., that a local hidden variable model could explain the results. We acknowledge that a more thorough mission design should use this more sophistocated metric in the analysis, but do not believe the conclusions will be substantially different than our simpler analysis using $\sigma$.}
\begin{equation}
\textrm{\# of $\sigma$ violation} = \left< n \right> = \sqrt{N} \frac{p-\tfrac{1}{\sqrt{2}}}{\sqrt{2-p^2}} .
\end{equation}
The parameters $N$ and $p$ can be expanded into instrument performance parameters, as discussed in Appendix \ref{sec:linkAnalysis}.  Figure \ref{fig:Partitioned_bell} represents the $\sigma$ violation significance achievable with a source of clock rate $f_{clock}$, pair production probability $p(1)$, photon pair fidelity $F$, total link efficiency (the product of efficiencies to Alice and Bob) $\eta_{2e}$, for total integration time $T$, with a receiver with temporal resolution $\Delta t_R$, and total background noise flux of $N_{noise}$.

\begin{figure*}[t]
    \centering
    \includegraphics[height=8.5cm]{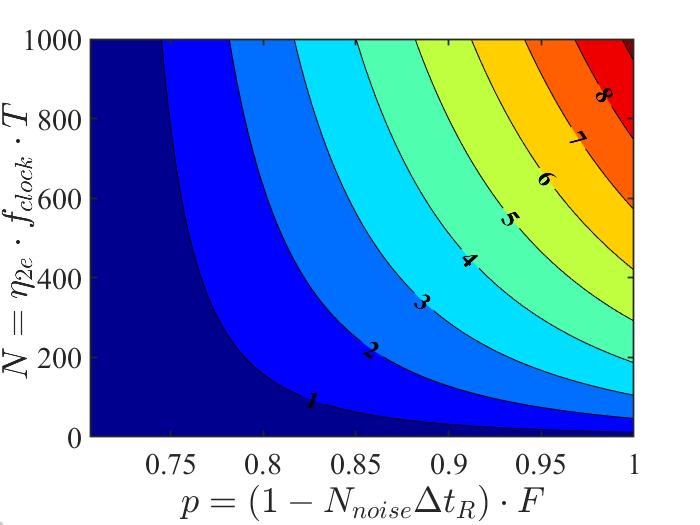}
    \caption{Parametric representation of general Bell test. The indicated statistical certainty on the violation of Bell's inequality, represented on the colormap values, drives the instrument performance requirements.  The colormap corresponds to a statistical confidence between $1\sigma$ and $8\sigma$. No violation at all can be achieved unless $p > \tfrac{1}{\sqrt{2}}$.}
    \label{fig:Partitioned_bell}
\end{figure*}

\begin{figure*}[t]
    \centering
    \includegraphics[height=8.5cm]{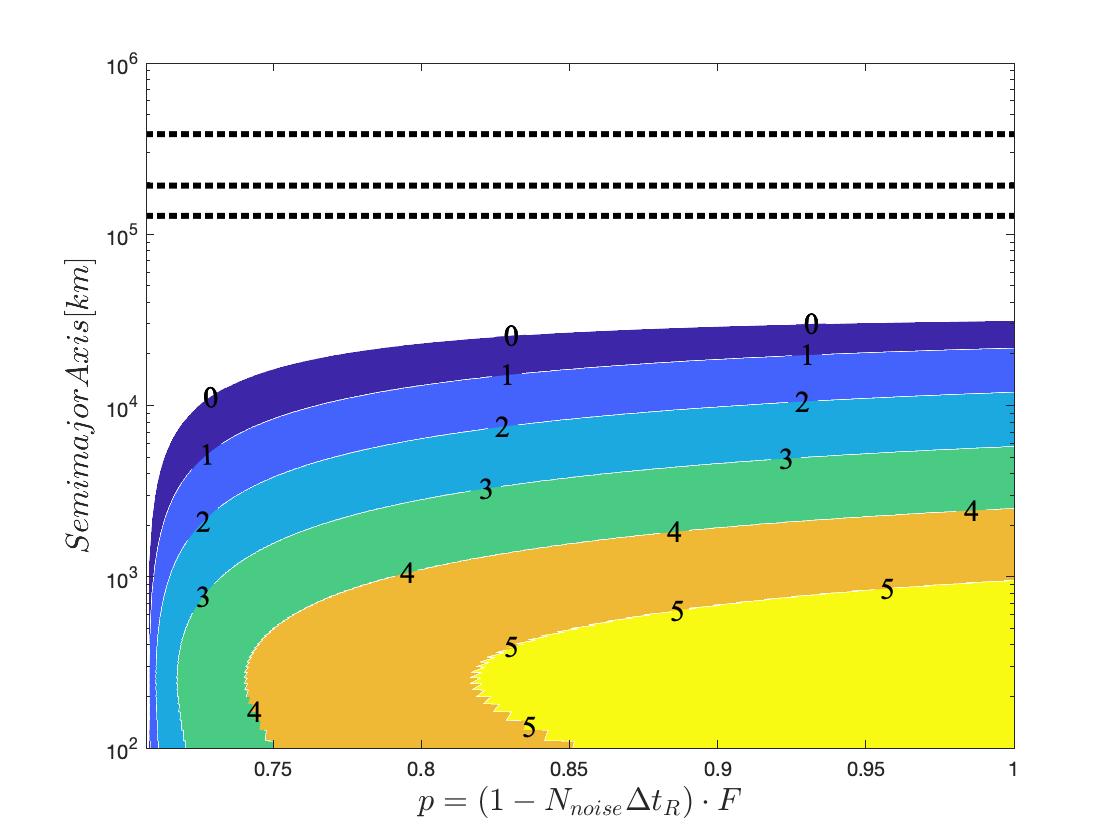}
    \caption{System performance to achieve an $n\sigma$ violation (color coded) of Bell's inequality across orbital distances represented on the y-axis.  An Earth-Spacecraft link is assumed up to Geostationary orbit.  The three dotted lines on the top of the chart represent 1.0, 0.5, and 0.33 times the Earth-Moon mean orbital radius.  For simplicity, the source-to-Alice and source-to-Bob channels are assumed identical.  The color scale represents the achievable statistical significance of the experiment, in $n\sigma$ . Note: for y-axis values greater than Geostationary, the integration time was clamped to 1 hour.}
    \label{fig:Bell_by_orbit}
\end{figure*}

\begin{figure*}[t]
    \centering
    \includegraphics[height=8.5cm]{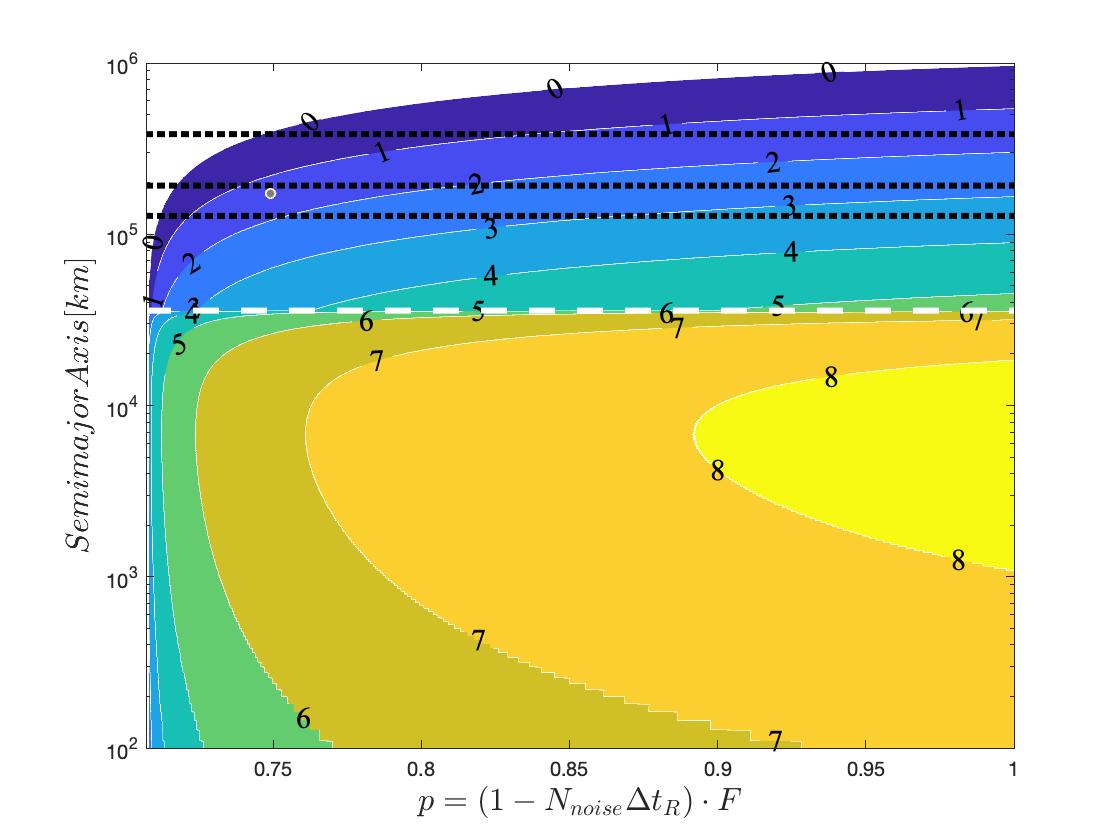}
    \caption{System performance to achieve $3\sigma$ violation of Bell's inequality across orbital distances represented on the y-axis.  Up to Geostationary orbit (the first horizontal white-dashed line), an Earth-Spacecraft link is assumed.  The three dashed lines on the top of chart represent $1.0, 0.5$, and $0.33$ times the Earth-Moon mean orbital radius.  Note there is an optimal contour for high $p$ values near the y-value of $6\cdot10^{3}\,$km.  The shapes of contour boundaries 6, 7, and 8 are due to the integration time associated with the orbital flyby.  This upgraded link scenario will meet requirements for the midway-between-Earth-and-Moon Bell test.}
    \label{fig:Bell_by_orbit2}
\end{figure*}

As a starting point, we consider the design example from Section \ref{sec:COWtests}, which for the COW tests predicted optimum performance around 1,500-km altitude orbits, using $0.3\,$-m  and $1.0\,$-m diffraction-limited telescopes at $1550\,$nm.  This flight system, upgraded with a suitable entangled photon pair source operating at 1\% pair production probability with a $1\,$GHz clock rate, and a second $0.3\,$-m telescope pointed to a second $1.0\,$-m aperture, would perform the Bell tests indicated in Figure \ref{fig:Bell_by_orbit}. 
It is evident that this system as described would be insufficient to support the human-decision Bell tests as described.  An upgraded system, with improved pointing and larger receiver apertures, as well as a higher rate photon source, is required.  For example, Figure \ref{fig:Bell_by_orbit2} shows the predicted performance assuming $2.0\,$-m aperture transmitters, with a $2.0\,$-m Lunar-vicinity receiver and a $10\,$-m aperture on Earth.
In both of the above examples, the x-axis of Figures \ref{fig:Bell_by_orbit} and \ref{fig:Bell_by_orbit2} captures both the fidelity of source and receiver parameters.  Now, we can combine Equation \ref{eq:nsigma_violation_of_Bell} with the contours of Figures \ref{fig:Bell_by_orbit} and \ref{fig:Bell_by_orbit2} to derive system performance parameters.  

The Bell test requires transmission of entangled photons to two receivers, characterized in Eq. \ref{eq:doubleLink}.  For a source-to-receiver separation corresponding to the diameter of the Earth, applying Equation \ref{eq:Link}  with $\lambda = 810$ nm, $D_{Tx} = 0.5$ m, $D_{Rx} = 3.5$ m, $M^2=1.05$, and $\eta_x = 0.1$ leads to a  one-channel link efficiency of $0.003$.  The two-channel efficiency, characterizing the likelihood that both photons from an entangled pair are recorded at their respective (equidistant) receivers, is $0.003^2 \simeq 10^{-5}$ for the assumptions stated.  Assuming a source clock rate of 1 GHz and corresponding pair production probability of 1\%, the rate of success is about 100 transmitted photon pairs per second.   If the source telescope diameter is reduced to 22 cm, and the receiver telescope diameters are reduced to 1.0 m, the success rate drops to 0.03 transmitted photon pairs per second;  the time to measure 500 events would then increase from about 5 seconds to 4.5 hours, the latter of which might require integration over multiple orbital passes.
Per Figure \ref{fig:Partitioned_bell}, with 500 successful measurement events a statistical confidence of $3\sigma$ is achieved for parameter $p \geq$ 0.85.  Assuming a source fidelity of 0.90, Eq. \ref{eq:partition_noise} then constrains the noise probability, $\Delta t_{R} \cdot N_{noise}$, to be less than 0.06 over the measurement interval.   The critical point in evaluating this trade study is that $\Delta t_{R}$ is much less than the integration period.  Using the upper limit of $\Delta t_{R} \approx 1/(0.003\cdot 1$GHz$) = 333 \,$ns, the requirement for $p>0.85$ is satisfied for  $N_{noise} < 170 \,$kcps.  

It is worth noting that, e.g., through the 5 s interval of the experiment, roughly 500 signal counts are resolved against 800 K noise events through the application of temporal filtering -- count events occurring outside the expected time-of-arrival bins of signal events are excluded from further analysis.  This technique requires time synchronization between source and receiver, conveniently represented in units of picoseconds of drift per second of integration.  In the case requiring 5 s of integration, the timing requirement is maintaining temporal synchronization between source and receiver to reduce the net drift to less than $100\, $ps$ / 5\, $s$ = 20 \,$ps/s.  In the link scenario requiring 4.5 hours of total integration, the residual drift is required to be less than $100\,$ps$ / 4.5\, $hr$ = 6\,$fs/s; although this sounds very low indeed, one would incorporate active locking to stay synchronized, e.g., by distributing a classical clock signal.  These figures can be further partitioned into phase noise, long-term stability, and time-transfer requirements for local clock systems on each network node of the Bell test.

We consider 1/2 Earth-Moon baseline for the human-decision Bell tests.  In this scenario, $\lambda = 780 \,$nm, $D_{Tx} = 1.0 \,$m, $D_{Rx} = 1.0\, $m, $M^2=1.05$, and $\eta_x = 0.1$.  The low efficiency (10-100 photons per hour, depending on other link assumptions) drives a further requirement of multiplexing the source to improve the transmitted signal photon count rate to 2-5 photon counts per second (about the maximum rate that an astronaut involved in a human Bell test can reasonably accommodate).   While multiplexed entangled photon pair sources are an active area of research \cite{Migdall2020,Dhara2021}, the high-order multiplexing required to perform the human-decision Bell tests should be considered a new area of research.

The overall efficiency of the human-decision Bell test could be increased by using an even larger aperture receiver on the Earth-side.  If an 8.3-m effective aperture is used (such as NASA's RF-Optical Hybrid telescope \cite{RFO2021, RFO_sun}), the net photon-pair rate improves to 0.2 counts per second.  Without requiring source multiplexing, the detector dark noise requirement to achieve a $3\sigma$ result is 0.095 noise counts per second.  Such detector performance was realized in \cite{AM2021}, with a corresponding detection efficiency of 0.75.

\subsubsection{Bell Tests: Summary}
The proposed Bell test measurement scenarios will test the assumption that there are no local hidden variable processes at work between inertial frames, or across long baselines. Executing highly statistically significant tests across planetary baselines is possible but challenging using existing technologies.  The most ambitious of the human-decision Bell tests require development of multiplexed source technology, and a commitment to deploying large, diffraction-limited telescopes with exceptionally low-noise detection systems in high-earth and Lunar orbits; they would further benefit from ground-based, large-diameter telescope infrastructure.  Performing Bell tests with high statistical confidence in the low-to-mid orbital regimes would be possible using a $10^8$ pair-per-second source of high fidelity entangled photon pairs.

\subsection{Long-Baseline Quantum Teleportation}
\label{sec:QTsection}
\setcounter{footnote}{0} 
The third pillar of DSQL science is to perform quantum teleportation over long distances in space, which acts as a pathfinder for future quantum communication networks as well as a testbed for studying the interplay between quantum entanglement and gravity. Planned and operational space-based quantum optics experiments, most notably {\itshape Micius} \cite{Ren:2017aa}, use long baseline quantum teleportation to test a basic assumption of quantum mechanics: that quantum correlations of entangled photon pairs, shielded from the environment, are maintained across any baseline.  The overarching goal of the proposed DSQL quantum teleportation experiments is to test this assumption across ever-longer baselines, and between inertial frames.  

\textit{Micius} demonstrated the distribution of entangled photon pairs across a 1200-km baseline, and successful uplink of a teleported photon from Earth to space up to a 1400-km baseline \cite{Ren:2017aa}.  How much longer should a new quantum teleportation baseline be to advance the art?  We  consider four phenomenological benchmark distances.  First, in the context of a future network of quantum sensors coupled together using a teleportation swapping system (as described in Ref. \cite{quantum_telescope, quantum_clock_network, quantum_internet}), a global-scale network will require the teleportation protocol to function across a global baseline.  The first baseline benchmark is thus the diameter of the Earth.  
The second benchmark range corresponds to the distance between a geostationary spacecraft and the surface of the Earth.  This range is of practical importance--- if future technology development improves the rate of usable quantum entanglement distribution, geostationary spacecraft could serve a valuable role in tomorrow's quantum networks \cite{Bedington:17}.  

High clock-rate entangled photon pair generation, coupled with high timing resolution photon time-of-arrival detection, are necessary tools to uncover relativistic effects in quantum measurement.  As these timing parameters improve, the departure of the gravitational model of the long baseline link from Newtonian to Schwarzschild, and from Schwarzschild to multi-body, becomes measurable.  Hence, the third benchmark baseline for quantum teleportation occurs where the experimental timing resolution provides sensitivity to multi-body gravitational effects. Roughly speaking, this threshold is reached at the first Lagrange point of the Earth-Moon system.
The final benchmark for teleportation corresponds to the maximum range available given state-of-the-art technology, using the simple link expression described in Appendix \ref{sec:linkAnalysis}.  Currently, this range is on the order of the Earth-Moon mean orbital distance.

Testing teleportation between inertial frames is the other, equally important motivation for this proposed set of DSQL experiments.  Generally, the available timing performance of the system must be high enough to allow sensitivity to these inertial effects.  In analogy to benchmarking quantum teleportation baselines based on phenomenological thresholds, the sensitivity to time dilation between frames is benchmarked against timing resolution.  The first benchmark is when the total predicted time dilation of a given experiment exceeds the available timing resolution of the measurement apparatus.  The second benchmark is when the contribution to time dilation from relative velocity (special relativistic effects) and from gravitational-potential difference (general relativistic effects) are both greater than the system timing precision.  This presents a logical path towards driving future experiments with ever-improving system timing performance, enabling sensitivity to higher-order general relativistic effects, such as frame dragging, Shapiro time delay, and gravitational deflection.

In the Section that follows, the basic concept of quantum teleportation and its variants like entanglement swapping and atomic state teleportation are presented.  Next, we highlight the subtleties of performing such experiments in the Earth-Moon system and review the influence of gravitational effects. We conclude this Section by briefly evaluating the requirements on a flight quantum memory -- which would substantially enhance the viability of the proposed experiments -- and how it could be realized with current technology.

\subsubsection{The Quantum Teleportation Protocol}
\label{subsec:QT-protocol}

\begin{figure*}[t]
    \centering
    \includegraphics[height=7cm]{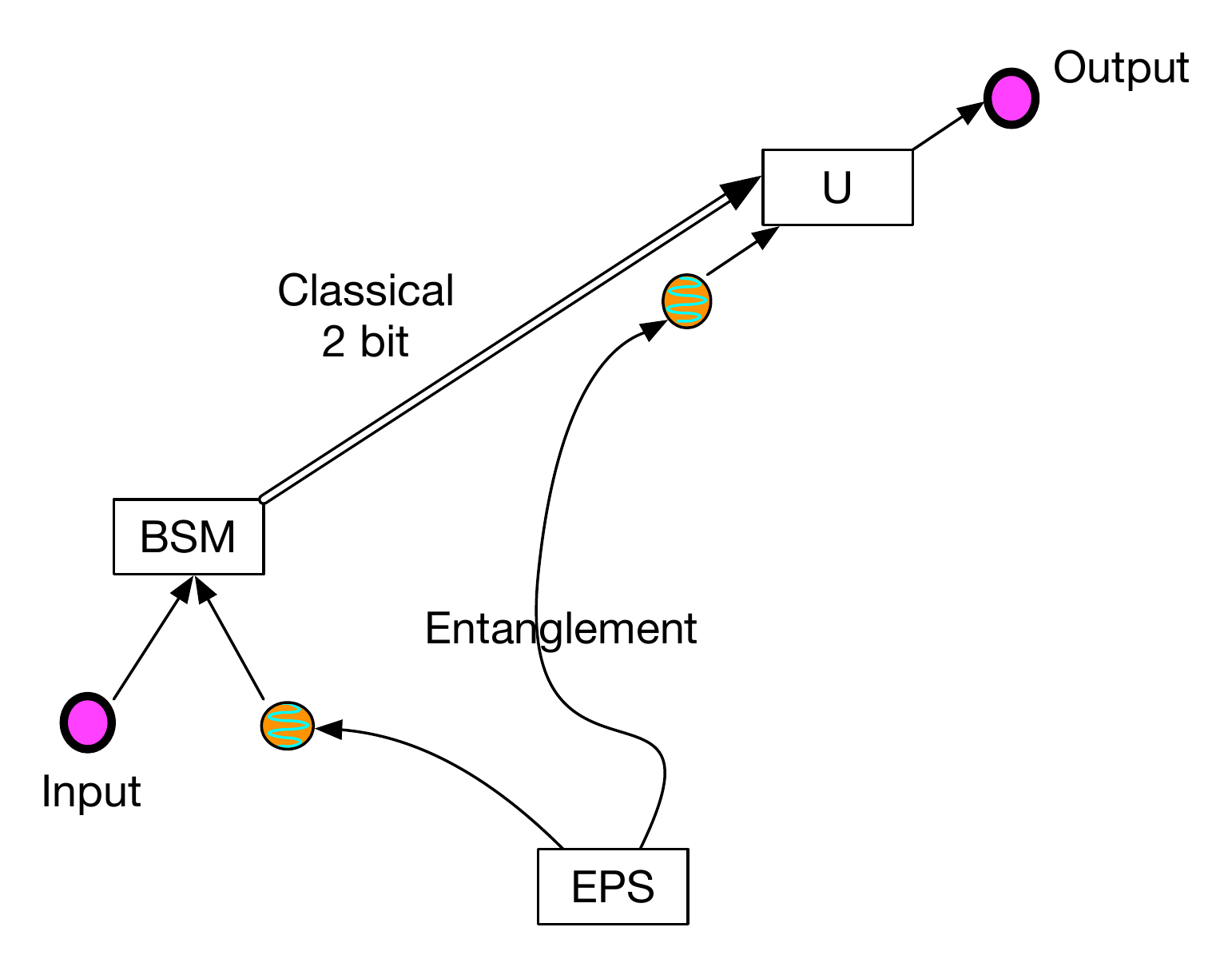}
    \caption{Scheme of quantum teleportation \cite{PhysRevLett.70.1895}. An unknown input state is transferred with perfect fidelity using a combination of distributed entanglement created in an entangled photon source (EPS) and a classical signal conveying the outcome of the Bell-state measurement (BSM). The final step is a  unitary operation (U), which is contingent upon the BSM result, and is applied to the entangled twin particle. }
    \label{fig:q_teleport_scheme}
\end{figure*}

Teleportation \cite{PhysRevLett.70.1895,teleport_exp} is a quantum protocol where an unknown quantum state is transferred from one system to another, possibly far away,  by using two channels: a) a maximally entangled state and b) a classical signal. Because of the use of non-classical resources this protocol does not have an equivalent classical counterpart. In the photonics domain, teleportation can be described as follows \cite{teleport_exp}: first, an entangled pair is generated, one photon is sent to Alice, and the other to Bob (see Figure \ref{fig:q_teleport_scheme}). Alice then performs a Bell State Measurement (BSM) \cite{PhysicsPhysiqueFizika.1.195} on her part of the entangled state and the unknown quantum state, thereby projecting the - innately uncorrelated - two photons into an entangled state. The BSM will project Bob’s part of the state onto one of four different possible states, depending on the BSM result \footnote{Note that with linear optical elements (and no ancilla photons), the projection is not perfect: only 2 of the 4 Bell states give unambiguous experimental signatures, the other 2 giving the same signature as each other \cite{Lutkenhaus2001}. As a consequence, all photonic teleportation experiments to date have been limited to an efficiency of $50\%$. Adding extra single-photon input states to the quantum circuit can boost this efficiency \cite{PhysRevA.98.042323,PhysRevA.84.042331,PhysRevLett.113.140403}, and complete (i.e., $100\%$ efficient) Bell-state analysis can be achieved using various matter-based qubits.}. Alice communicates her BSM result to Bob and based on this information he will suitably rotate the state he has allowing him to recover the original unknown input state.  It is worth noting that neither Alice nor Bob obtain any knowledge on the input state at any time, and that the final unitary transformation does not depend on the input state but only on the result of the BSM. The `original' particle loses all quantum information during this process. This consequently  implies that quantum teleportation does not permit creating a copy or clone of the original state - in accordance with the no-cloning theorem in quantum physics \cite{Park:1970aa,Wootters82a}. 

Quantum teleportation also applies to the transfer of an entangled particle, arguably the ultimate unknown state. This leads to a protocol called entanglement swapping \cite{Zukowski93a,Pan98a}, which generalizes to multi-partite entanglement \cite{Bose98a}. Amazingly, quantum information and quantum entanglement can be seamlessly transferred or relayed through a network of quantum particles using quantum entanglement and classical information as the resources. 

The teleportation protocol is not only highly interesting from a fundamental perspective, but also a crucial concept for multi-user quantum networks which could be used for secure communications, interlinking quantum computers, or distributed quantum sensors. The ultimate application of quantum teleportation or entanglement swapping will make use of quantum memories with ultra-long storage times (hours, days, or even years). For instance, a space craft could carry a register of stored, entangled quantum bits (qubit), and gradually use them up for quantum communication tasks such as quantum networking, secure communications or super-dense-coding \footnote{In superdense coding one is able to transmit up to twice the normal amount of information on a successfully transmitted photon. However, because the protocol must be implemented at the single-photon level (i.e., the maximum attempt rate is limited by the inverse bandwidth of the photons), and long space links imply very high loss, it is nearly always preferable to send bright optical pulses instead, if the goal is simply to transmit classical bits.} proposed by Bennett and Wiesner \cite{Bennett92c}. However, while in principle a quantum memory could store information indefinitely, current quantum memory systems typically operate with storage times of order milliseconds or less; see Section \ref{subsec:quantum_memory}.

Standard quantum theory places no bound on the distance over which entanglement or teleportation may be accomplished. However, the required classical and entanglement channels impose limits on a teleportation protocol. First, the transfer speed for quantum teleportation is limited to luminal signaling due to the classical channel: even if the entangled ``receiving'' particle is already at its destination, the correct input state can only be retrieved once the (classical) information about Alice's Bell State Measurement has arrived. Second, the protocol would suffer from any decorrelation or decoherence in the entangled channel used as a communication resource, and thus any impact of curved space time on entanglement could impact the teleportation fidelity, see Section \ref{gr:dephasing}. The DSQL would allow studies of these unwanted effects in a realistic environment.

\subsubsection{State-of-the-Art Long-Range Teleportation}
\label{teleportation}
Long-range teleportation was achieved outside a single laboratory starting in 2002, when a signal was teleported  between two stations separated by $55 \,$m using optical fiber\cite{Marcikic:2003lr}. In 2003, the first long-range teleportation with an active unitary operation at the receiver was demonstrated over 500\,m \cite{Ursin:2004},  sending the entangled photon through optical fiber at $2/3$ the speed of light, while the BSM result was radioed above ground and ``overtook'' the entangled photon to arrive in time for an electro-optic modulator to rapidly apply the correct unitary operation.   In subsequent years, quantum teleportation was demonstrated over increasingly further distances including demonstrations over 100 km \cite{Yin:2012uq}, over 144 km \cite{Ma:2012fk}, and in 2017 from ground to space \cite{Ren:2017aa} demonstrating the teleportation of independent single-photon qubits from  ground  to a low-Earth-orbit satellite, through an uplink channel, over distances of up to 1,400 kilometres. These demonstrations represent major advances, yet all photons involved in the experiments were generated by the same laser pulse on the same optical bench, and only after their creation was the receiver photon transmitted over a large distance, i.e., the actual entangling BSM operation occurred while all the photons were technically still in or very close to the original lab, which would largely defeat the purpose in a practical quantum networking application. 

A more general implementation of quantum teleportation -- entanglement swapping, which is also used for quantum repeaters \cite{Briegel98a} -- would be implemented with photons generated from {\it independent} sources. This is experimentally much more challenging compared to the previous realizations, because the different photons must be spectrally and temporally indistinguishable in order to achieve a high-quality BSM, which is based on two-photon interference. Typically, entangled photons have a temporal coherence of $\approx 200 - 500$~fs, which is right at the limit of synchronizing lasers. A first demonstration of two-photon interference using two synchronized femtosecond lasers was reported by Kaltenbaek et al.  in 2006 \cite{kaltenbaek-2006-96}, and entanglement swapping shown in 2009 \cite{PhysRevA.79.040302}. 

Another method to realize truly independent optical sources uses two entangled photon sources operated with continuous wave lasers, and very narrow-band filtered photons \cite{gisin_natphys_2007}, with associated coherence times of several hundred picoseconds, longer than the timing resolution of detectors.  A particularly promising approach is to generate entangled photon inside high-finesse optical resonators; such sources are intrinsically narrow-band, and do not require frequency filtering, which otherwise severely reduces the achievable rates. For example, photon pairs were generated in lithium niobate whispering gallery resonators with coherence lengths tunable roughly between 10 -- 20 ns \cite{Fortsch13NC}. Furthermore, these sources can be engineered to match the wavelengths and bandwidths of atomic transitions \cite{Schunk15CsRb}, an important factor for implementation of a quantum repeater \footnote{One important drawback of using such narrow-band sources is that the relative motion of the platforms will lead to spectral Doppler shifts that will need to be compensated for, e.g., a relative velocity of 10,000 km/s leads to a frequency shift of $3.3\times 10^{-5}$, or $\sim$6.4 GHz; if that is comparable to the bandwidth of the light, it will need to be corrected before coupling the photon to a quantum memory, or interfering it with a non-shifted photon.}.

Another interesting aspect of long-range quantum teleportation is the fidelity reduction due to the emission statistics of typical realistic photon sources, including entangled photon sources based on spontaneous parametric down conversion (SPDC) \cite{Kwiat95b, zhang2021} and four-wave mixing \cite{Kumar2005, Camacho2012}. Here the thermal statistics of the source constrain the probability of creating exactly one photon pair in a pulse to be $\leq$ 1/4; the empty pulses lead to inefficiency while pulses with two or more pairs lead to noise; one method to ameliorate this problem uses multiplexing \cite{Migdall2020}. Alternatively, the Jennewein group proposed in 2013 \cite{Bourgoin:12} 
that quantum teleportation implemented with single emitters (e.g., quantum dots) could greatly improve teleportation fidelity for ground and space links. The technical challenges around such emitters make this approach challenging to implement; however, recently high-efficiency coupling of photons from quantum dot sources into optical fibers has been realized \cite{Warburton2021}, as has generation (though not yet efficient extraction) of high quality polarization-entangled photon pairs \cite{Huber2018,Schimpf2020}.

\subsubsection{Teleportation in the Earth-Moon System} %[from ISS to LG]}
\label{subsec:teleportation-Earth-Moon}

Expanding quantum teleportation and entanglement swapping over large distance scales would demonstrate truly quantum communication protocols at unprecedented scales and provide crucial insights into the validity of quantum mechanics, leading the path towards deep-space quantum networking and quantum computing.

As stated above, long range ``passive'' teleportation \cite{Pirandola15} from ground to space was accomplished with the 2017 Micius mission \cite{Ren:2017aa}, transferring one of the entangled photons from an SPDC source from a ground station at very high elevation (around 5000\,m) to a receiver on board Micius, at around 600\,km altitude. With the DSQL we want to extend this range and perform quantum teleportation experiments on the Earth-Moon distance by, for instance, 
connecting the International Space Station (ISS) and the Lunar Gateway (LG) \cite{ISS20, Gateway}. While atmospheric photon scattering can be avoided in outer space, the losses in photons due to the diffraction of optical beams traversing such a distance will be challenging (see sections  Appendix \ref{sec:linkAnalysis}). Furthermore, the travel time of a light signal from Earth's surface to the Moon is about 1.3 seconds. This implies that, to complete the protocol of quantum teleportation from Alice on ISS to Bob on LG, the quantum state carried by Bob's photon entangled with Alice's must be kept longer than a time scale of order 1 second in Bob's quantum memory waiting for the final operation (assuming that Bob already possessed his half of the entangled state before Alice made her Bell state measurement, i.e., assuming that the entanglement was pre-shared as shown in the original picture of quantum teleportation (Figures \ref{fig:q_teleport_scheme}), which has {\it not} been the case for most teleportation experiments to date). Fortunately, this may be achievable by emerging technology \cite{Rancic_2017}. 

\begin{figure*}[t]
 \centering
\includegraphics[width=3cm]{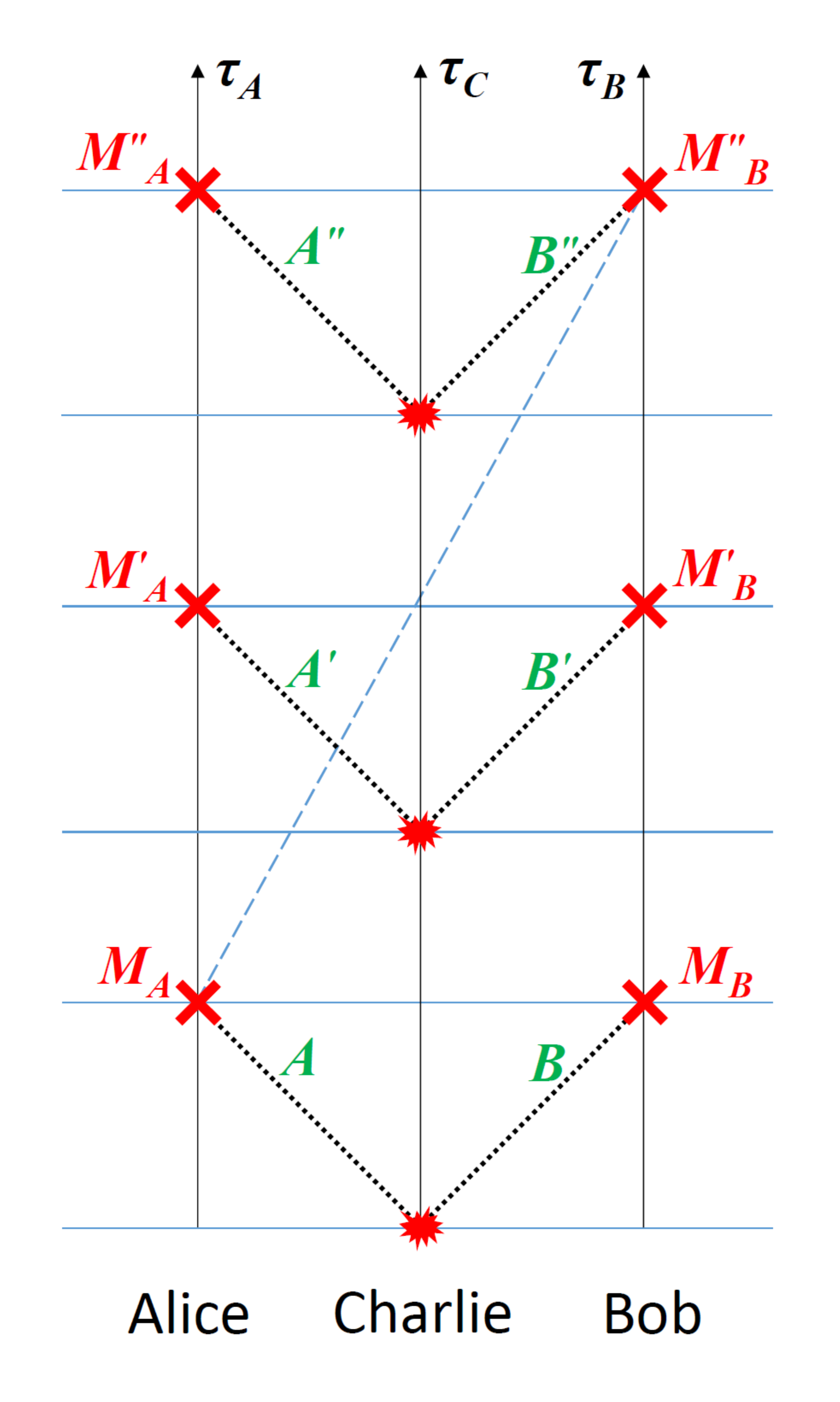}\hspace{.3cm}
\includegraphics[width=5cm]{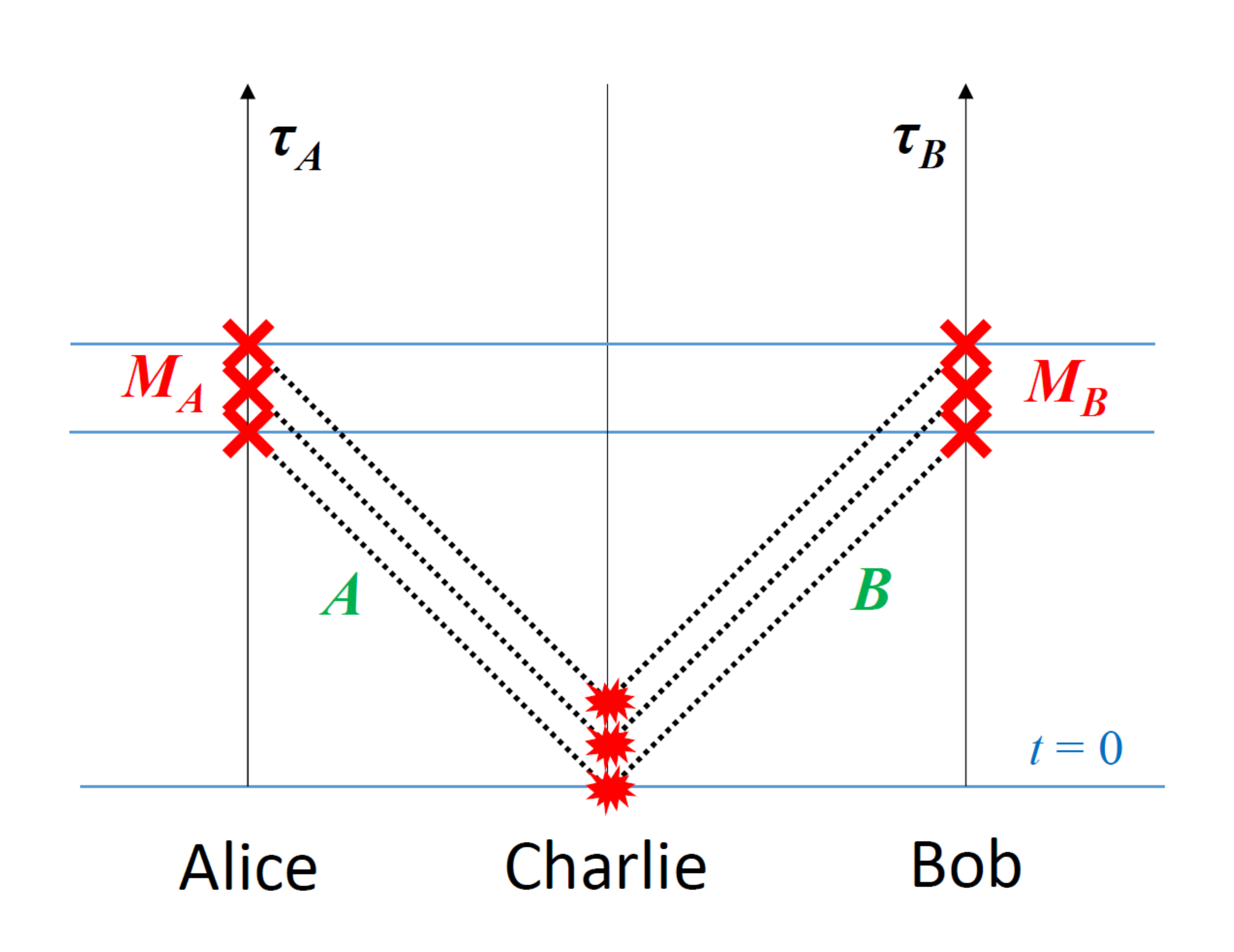}
\includegraphics[width=5.7cm]{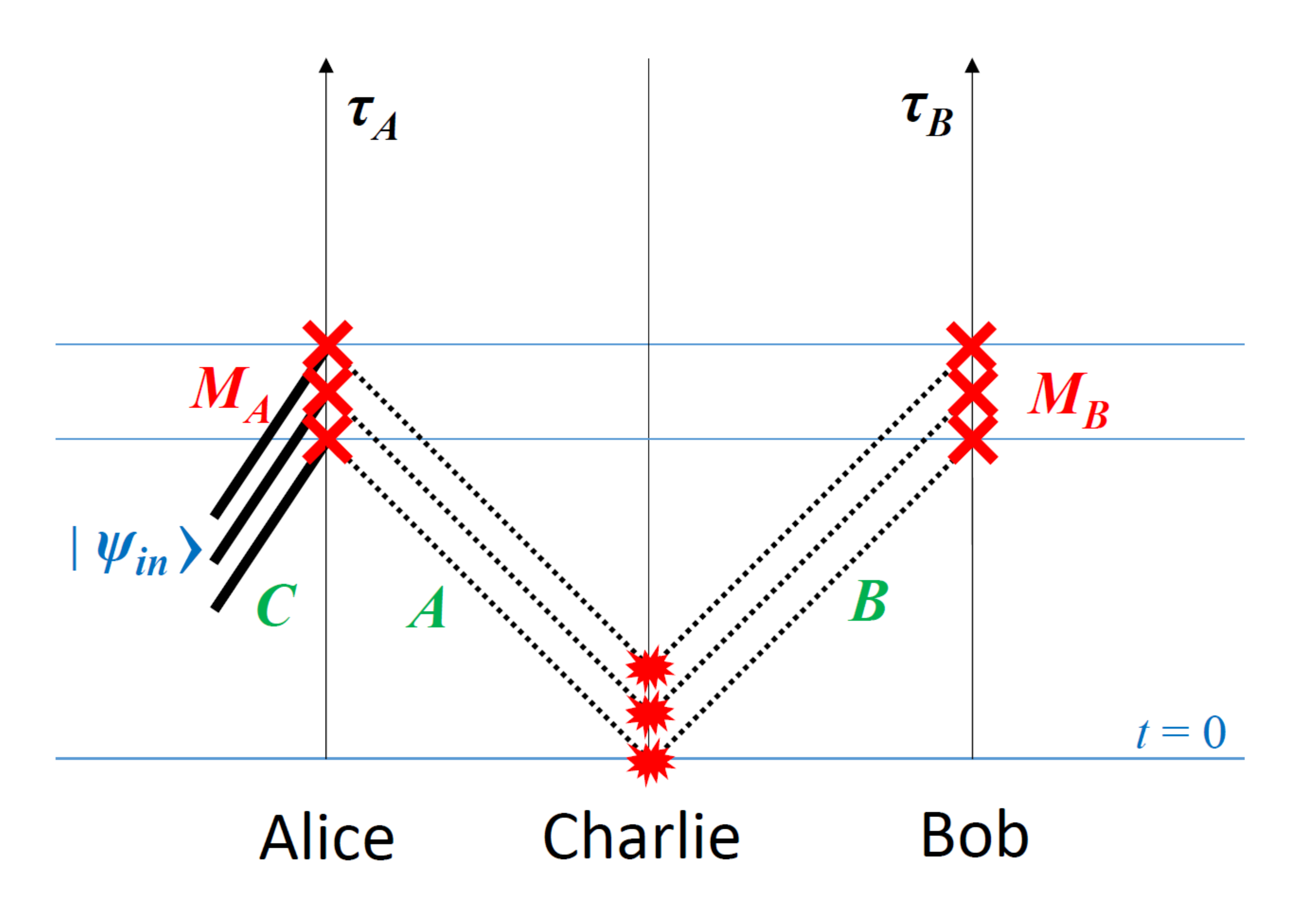}
\caption{In a conventional experiment of the Bell test (left), a series of identical processes are done while a later measurement event can be in the future lightcone of earlier events (e.g., $M''_B$ and $M_A$). Between ISS and LG we may be able to achieve the Bell test (middle) and incomplete quantum teleportation (right) with all Alice's measurements $M_A$ spacelike separated from all Bob's $M_B$ in the period of sampling.}
\label{BellQTel}
\end{figure*}

In the conventional ground-based experiments requiring transmission of multiple entangled photons (e.g., for a Bell test or quantum teleportation verified by full quantum state tomography on a large ensemble of systems), because of the limited spatial separation of the transmitter and receiver, the late events in each agent's worldline would be inside the future lightcones of the early events of the other agent's worldline (e.g., \cite{Giustina15, Shalm2015, Rauch2018}), and thus could causally depend on the outcomes and settings of the early events (e.g., the events $M''_B$ and $M_A$ in Figure \ref{BellQTel} (left)), potentially opening up a ``memory loophole'' \cite{Barrett02}. The O(1)-second travel time of light signals along our long baseline offers the possibility to perform sufficiently many resolvable runs within this travel time, so that a whole set of outcomes by one agent for ensemble averaging can each be spacelike-separated from the measurement events in the same period by the other agent (Figure \ref{BellQTel} (middle) and (right)). To achieve this, however, the photon emission rate of the source of the entangled photon pairs has to be large enough to compensate for the high transmission loss of photons over this large length scale. In the Bell tests this will eliminate the two-sided memory of the early measurements by the other agent, thereby closing the memory loophole \cite{Barrett02} without the need to suppress it by performing sufficiently many runs of an experiment with memory \cite{Gill01, Larsson14}.

The long-term quantum memories that Alice and Bob would ideally use to store their quantum state may experience additional effects which can be treated independently as interactions with their respective environments at finite temperatures.
 For example, if Alice's quantum memory on the (accelerating) ISS is coupled to the vacuum state of quantum fields with respect to the Earth, it will experience the Unruh effect \cite{Unruh76}, seeing an effective temperature due to the acceleration; however, for the ISS acceleration this temperature is only $\sim4 \times 10^{-20} \,$K, which is much lower than the temperature of the ambient environment and thus negligible. With a smaller acceleration, Bob's quantum memory will see an even lower Unruh temperature in the vacuum state of the fields.
 
Since the coupling between photons and gravitational waves is extremely weak, the gravitational effect on photons in the quantum optical experiments at the ISS-LG scale are mainly those for electromagnetic fields in a fixed spacetime background:
1) the gravitational Doppler shift ($\Delta\lambda/\lambda_0 \sim 10^{-9}$) and 
2) the Wigner rotation of polarization, where gravitational field provides a classical background \cite{LinChouHu15, Scully1982, Brodutch2011, Brodutch2015}. These can be negligible compared to similar effects due to the relative motion (the radial Doppler shift $\Delta\lambda/\lambda_0 \sim 10^{-5}$ and the transverse Doppler shift $\Delta\lambda/\lambda_0 \sim 10^{-10}$), which can either be suppressed by executing the experiments during periods when the relative radial motion is minimal, 
or by dynamically correcting according to the reference laser beams from the photon sources \cite{Ren:2017aa}.

\begin{figure}[p]
 \centering
\includegraphics[width=4.9cm]{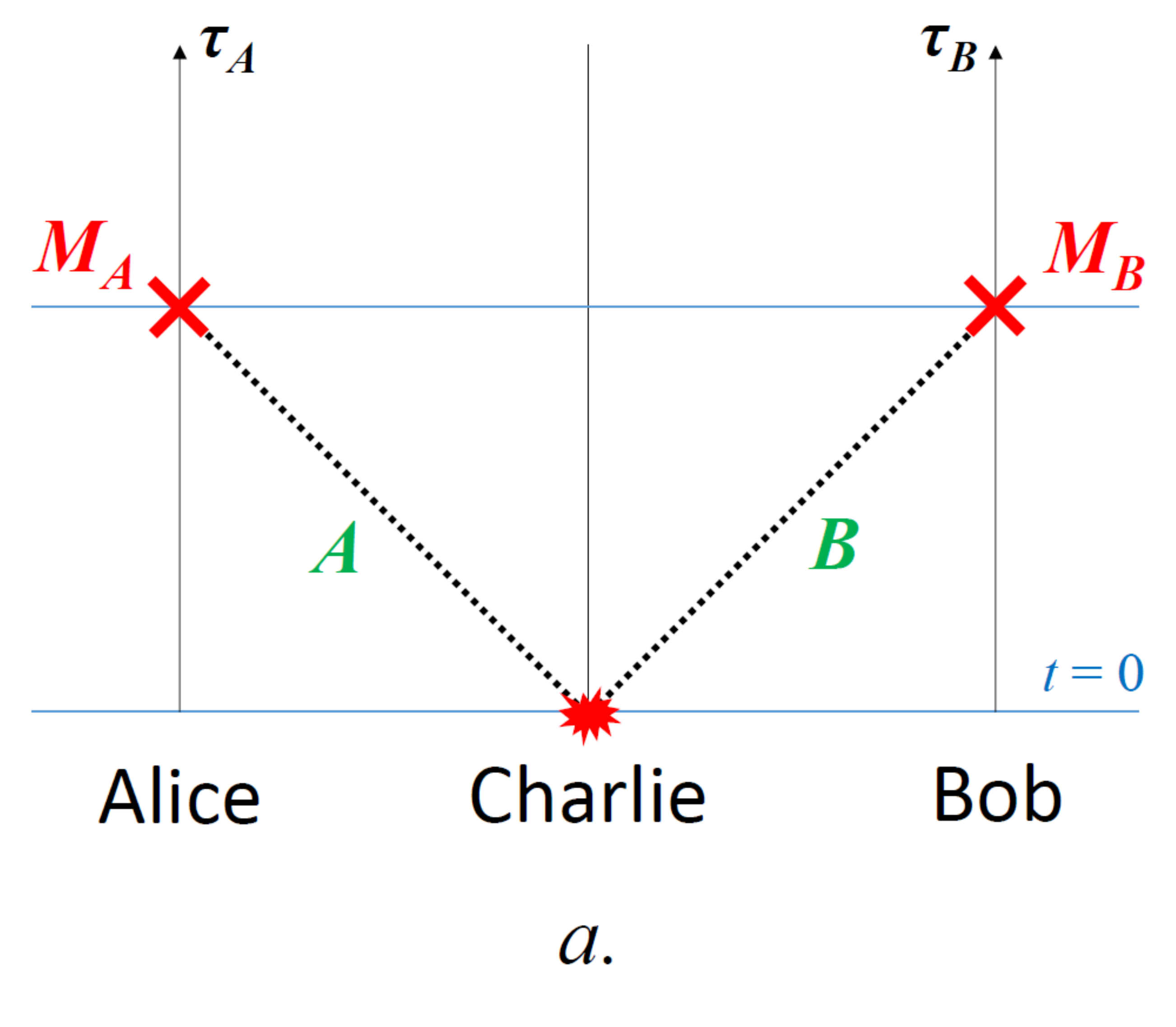}
\includegraphics[width=6.5cm]{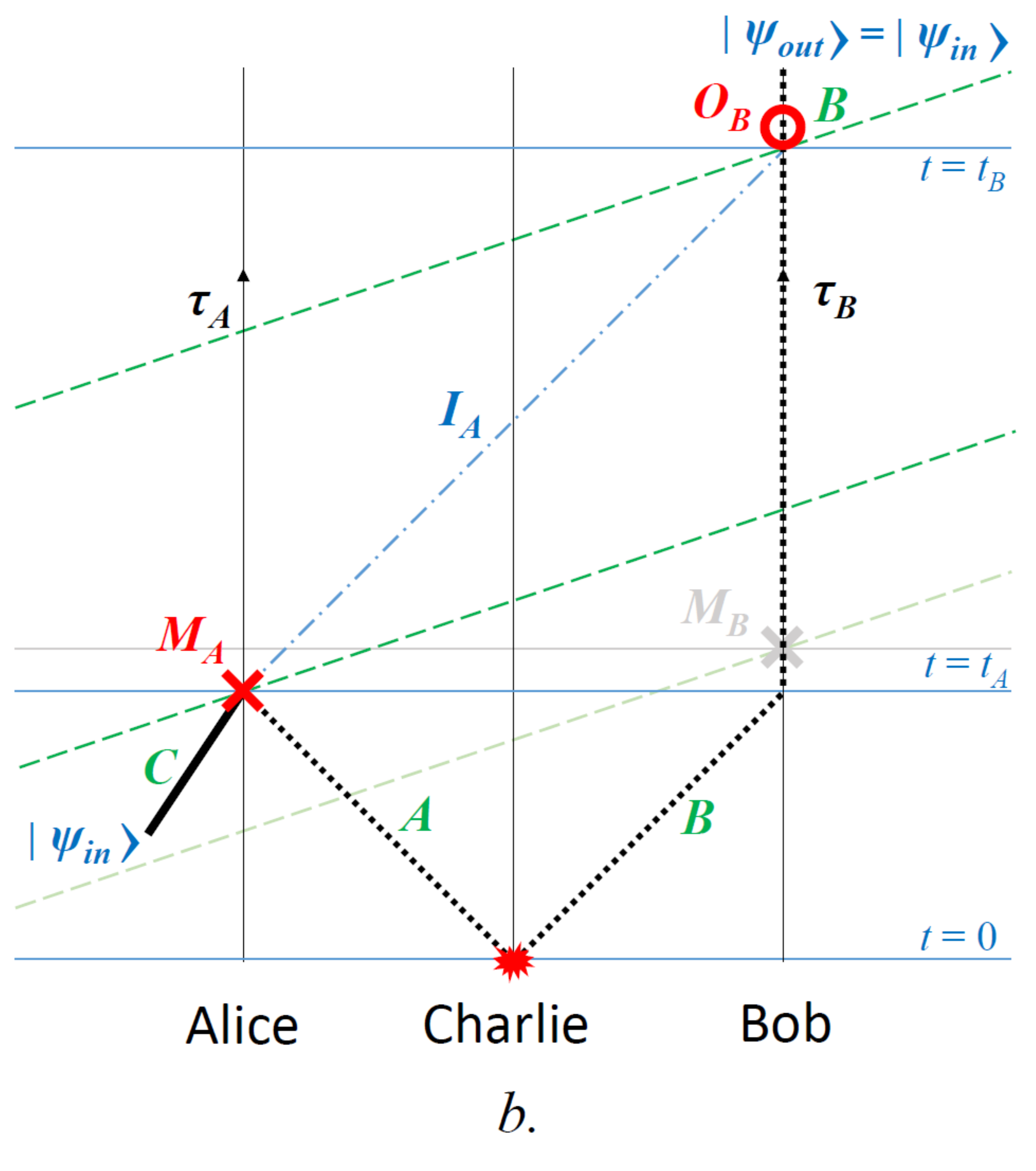}
\includegraphics[width=7cm]{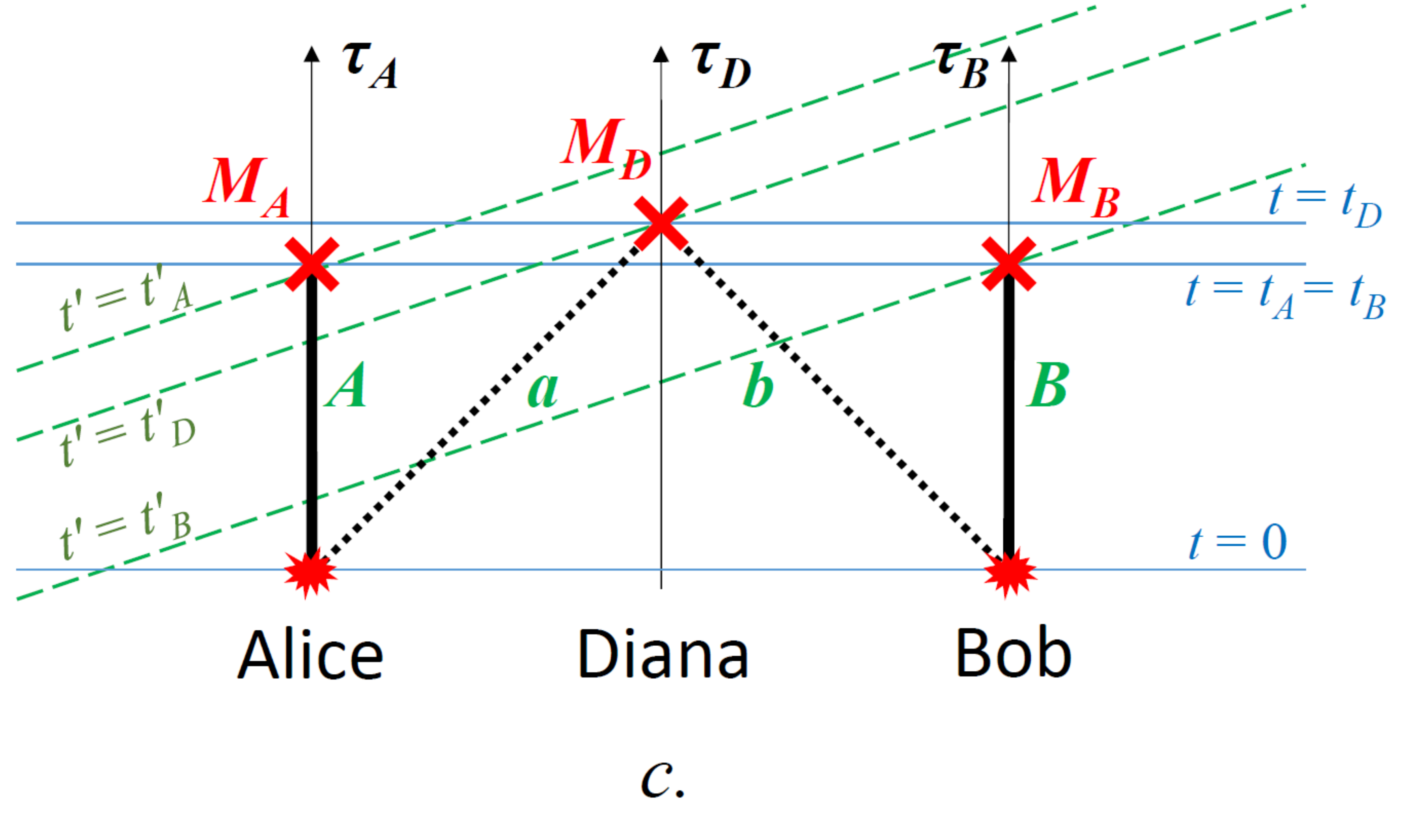}
\includegraphics[width=7cm]{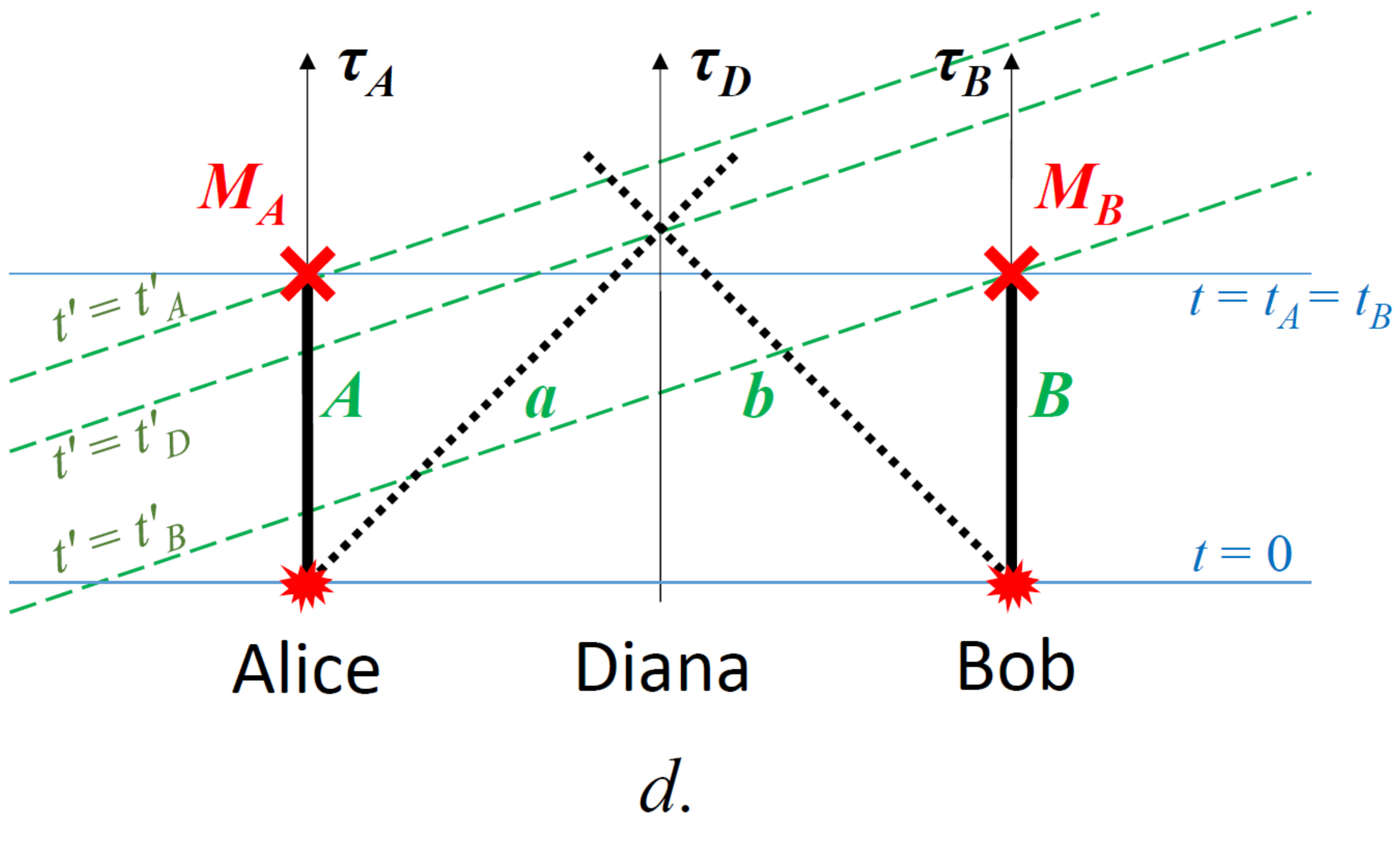}
\caption{Spacetime diagrams of the Bell test ($a$), quantum teleportation ($b$) and delayed-choice entanglement swapping, where Diana may choose to perform a joint measurement ($c$) or not ($d$). Here, ``$\times$'' and ``$\circ$'' represent local measurement and operation events, respectively. The black dotted lines and thick solid lines represent the worldlines of the participating quantum objects.}
\label{NonLocalExpt}
\end{figure}

In the {\it passive teleportation} \cite{Pirandola15} achieved by the {\it Micius} mission, the operation supposed to be done by Bob according to the classical signal from Alice is performed not physically, but virtually via data analysis \cite{Ren:2017aa} to obtain the fidelity of quantum teleportation. In this approach Bob does not need to maintain the quantum coherence of his photon or quantum memory until he receives Alice's classical signals. To obtain the fidelity more efficiently, Bob can perform the measurement immediately [$M_B$ in gray in Figure \ref{NonLocalExpt} (middle)], randomly choosing which measurement to perform on different photons in the ensemble.  Bob can even perform the measurement on $B$ before Alice's joint measurement on $A$ and $C$ in the bookkeeper coordinates. In this case, Alice can also make a ``delayed choice'' on performing the joint measurement on $A$ and $C$ or not \cite{Ma12b}.

{\it Entanglement swapping} is more difficult to achieve than a Bell test or teleportation because more participating photons and observers are involved. In particular, both Alice and Bob may store part of their states [carried by $A$ ($B$) of the entangled $Aa$-pair ($Bb$-pair) in Figure \ref{NonLocalExpt} (right)] in a local memory until they 
perform their local measurements. In each run a joint measurement on two photons [$a$ and $b$ in Figure \ref{NonLocalExpt} (right)], each belonging to an entangled photon pair produced by Alice or Bob, may be performed by Diana. Just like the incomplete passive quantum teleportation, the time order of the measurements by Alice, Bob, and Diana in different bookkeeper's coordinates can be different once they are spacelike separated. In particular, the decision of performing joint measurement or not can be made by Diana later than both local measurements by Alice and Bob in bookkeeper's coordinates, so that quantum entanglement of the photons of Alice and Bob appears to be determined after the fact by Diana's delayed choice, as seen in that reference frame \cite{Peres00}. 

With potentially all parties (Alice, Bob, Charlie/Diana) in high relative motion with respect to each other, it is important to consider the {\it frame dependence}, which is innate with canonical quantization, where one first chooses a coordinate system and specifies the time coordinate. One then writes down the Hamiltonian and the Schr\"odinger equation, and assigns the quantum state evolved accordingly. Note that simultaneity is relative, and quantum states in different reference frames are in general incommensurate when their associated time slices are different. 
Suppose Charlie in our quantum teleportation experiment and Diana in our entanglement swapping experiment are placed on a transfer vehicle between the Earth and the Moon. When the transfer vehicle is going from the Earth to the Moon, two separate events on the ISS and the LG that are considered as simultaneous by Alice on the ISS (or Bob on the LG) will not occur at the same time when perceived on the transfer vehicle: for Charlie and Diana the LG event would occur before the ISS event. If the transfer vehicle is instead returning from the Moon to the Earth, the same events will have an opposite time order, the event on the ISS occurring first according to Charlie and Diana. The time coordinate of each event in the transfer vehicle's frame is determined by the radar signals, namely, after the transfer vehicle receives the echo of the radar signal emitted earlier by itself \cite{dInverno92}. In Figure \ref{NonLocalExpt} the green dashed lines represent the time-slices in the reference frame of a transfer vehicle moving from the Earth (Alice) to the Moon (Bob), while the blue lines represent the time-slices in the reference frame of a bookkeeper who is roughly at rest both for Alice and Bob.

\subsubsection{Flight Quantum Memory} 
\label{subsec:quantum_memory}

Quantum memories are essential ingredients for implementing large-scale quantum networks or long-distance quantum communication channels, where they are critical for quantum repeaters enhancing the transmission range. In addition, quantum memories based on atomic or solid-state systems enable fundamental physics research, like testing atomic teleportation (see Sections \ref{subsec:QT-protocol} and \ref{subsec:teleportation-Earth-Moon}) and performing loophole-free Bell tests (see Section \ref{sec:BellTestsection}) over long distances. In all these applications the role of the quantum memory is to store single or multiple entangled photons in a long-lived state and to retrieve them reliably on demand.

Throughout the last two decades there has been an enormous effort in developing and improving quantum memories for photonic qubits, relying on a variety of physical systems and concepts \cite{Simon_2010, Pirandola15, Heshami_2016}. Most systems studied today can be assigned to one of the following categories: rare-earth-ion doped solids, color centers in diamonds, crystalline solids, hot and cold atomic vapors,  molecules, and switchable optical delay lines. The performance of different quantum memories can be compared by characteristic properties like the efficiency to retrieve a photon when requested, the fidelity that the retrieved photon is in the same state as the previously stored one, and the storage time. Other key parameters include the repetition rate and the ability to store multiple photons simultaneously. Furthermore, when coupling to an optical fiber or transmitting device is required, the wavelength and mode structure of the stored and retrieved photons plays an important role. In general, most systems show good performance in one or more of these aspects but have limitations in others, so that the choice of the optimal quantum memory strongly depends on the planned application. 

In order to support complete atomic state teleportation and Bell tests covering the distance between the Earth and the Moon, the DSQL requires a quantum memory with storage times longer than one second. Since free-space transmission is used, the wavelength and mode structure have to be compatible with the available transceivers (or suitable wavelength/bandwidth converters must be used). The requirements on the efficiency, fidelity and repetition rate depend on the details of the respective protocols and are specified in the corresponding sections.
Moreover, size, weight, and power (SWaP) are limiting resources aboard any spacecraft and have to be accounted for when choosing a quantum memory platform for the DSQL.

Due to the required long storage times, quantum memories based on molecules and crystalline solids (including semiconductor quantum dots) are currently not suited to being used for experiments spanning the Earth-Moon distance. However, systems employing rare-earth-ion doped solids, diamond color centers, and atomic vapors are promising candidates and will be briefly discussed in the following. For a more detailed comparison we refer to the excellent review articles on this topic \cite{Simon_2010, Heshami_2016}. 

Rare earth ion-doped solids combine long coherence times and good optical access to collective electronic and nuclear spins. The energy difference between the ground and excited state of the memory is typically in the low MHz range. Experiments have demonstrated coherence times of the order of one second \cite{Holzaepfel_2020} up to one minute \cite{Heinze_2013}, and even of several hours \cite{Zhong_2015}. With $^{167}\mathrm{Er}^{3+}\mathrm{:Y}_2\mathrm{SiO}_5$ there is also a material available that operates close to the telecom bandwidth \cite{Rancic_2017}. One downside of these systems is the need for cryogenic cooling in the regime of 1-4 K, which could limit its implementation in space missions. However, first steps towards space-compatible cryostats have been made \cite{You_2018}.

Vacancy centers in diamond enable the storage of qubits in single electron and nuclear spins, with the latter providing storage times of up to one second at room temperature \cite{Maurer_2012}, or even one minute within cryostats \cite{Bradley_2019}. In these systems neighboring spins can interact with each other, allowing for multi-qubit storage and the two-qubit operations \cite{Waldherr_2014, Bradley_2019} necessary for advanced quantum repeater applications. In most experiments either neutral or negatively charged nitrogen or silicon are used to create the defect centers, leading to optical wavelengths between 700 and $750\,$nm and frequency differences ranging from 100s of kHz to a few MHz. In order to address single spins the photon-coupling typically needs to be enhanced with resonators and cavities \cite{Sipahigil_2016, Bogdanovic_2017, Haeussler_2019}.

Atomic vapors made of alkali metals provide large optical depths even at room temperature and are therefore well-suited for photonic quantum memories \cite{Yu_2020}. The qubit is stored in the collective excitation of the atoms with energy differences of several GHz between the ground and the excited state of the memory. Since the lifetime is mainly limited by atomic motion -- the photonic state is mapped onto the distributed states of the atoms at the particular locations when the photon was absorbed -- cooling the atomic ensemble and employing dipole traps or optical lattices can improve the storage properties, enabling lifetimes of one second \cite{Katz_2018} and beyond \cite{Dudin_2013}. In addition, mode matching is a crucial step for atomic vapor systems and can be enhanced by employing cavities \cite{Bao_2012} or by placing the atomic cloud inside of nanofibers \cite{Sprague_2014}. 
Fundamental atomic physics experiments generating Bose-Einstein condensates have been realized on a sounding rocket \cite{Becker_2018} and on the ISS \cite{Aveline_2020} demonstrating the general feasibility of such an apparatus in space.

\subsubsection{Teleportation Mission Design}

\begin{figure}[t]
    \centering
    \includegraphics[width=0.70\textwidth]{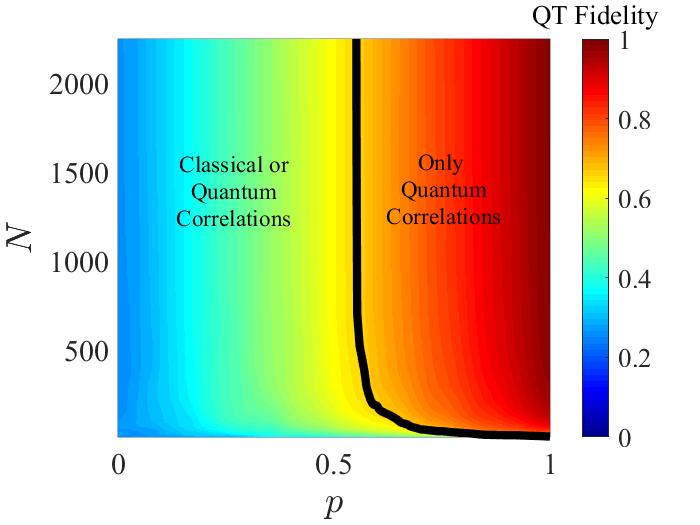}
    \caption{The fidelity of simulated tomographies with an input state being the Werner state in Eq. \ref{eq:wernerBell}
    %\ref{eq:mixed_bell_state}
    , and the target state being a maximally entangled state. The x-axis is parameter $p$ ($p=1$ implies all signal, while $p=0$ implies all noise) and the y-axis is the total number of successful count events (i.e., for all measurements ) integrated over the measurement duration.  The colorscale ranges from a tomographic fidelity of 0 to 1.  Tomographic fidelities greater than 0.66 are only possible through quantum correlations.}
    \label{fig:Statistics_QT}
\end{figure}

The procedure outlined in Appendix \ref{sec:linkAnalysis} is applied to the quantum teleportation process,  characterized by the resultant fidelity of the teleported state compared to the initial qubit state, as determined by state tomography \cite{ALTEPETER2005105}, for an initial entangled resource of the form 
\begin{equation}
\hat \rho = p |\Phi^+\rangle \langle \Phi^+| + (1-p)\frac{1}{4}\hat I  .
\label{eq:wernerBell}
\end{equation}
Using maximum likelihood estimation techniques\footnote{In this specific example, Bayesian estimation did not provide a meaningful benefit to calculation time.}, we calculate the resultant fidelity of the teleportation state in the presence of loss and noise events; see Figure \ref{fig:Statistics_QT}.  The results are an average over 10 tomographies per data point, with the counts  sampled from a Poissonian distribution to take into account normal counting noise. 

The Bell tests described above require up to two simultaneous optical channels.  The single-channel link efficiency expression in Equation (\ref{eq:Link}) characterizes the one-way losses of a teleportation experiment. Figure \ref{fig:Statistics_QT} represents the fidelity of quantum tomography as a function of the noise parameter (x-axis) and the number of successful measurement events (y-axis).  In analogy to the derivation of the instrument requirements for the Bell tests, we start by deciding what tomographic fidelity is required to meet experimental goals, then derive instrument requirements from the corresponding count rate and noise parameter for a successful demonstration of long-baseline quantum teleportation. For example, achieving a quantum tomography fidelity of 0.90 requires, e.g., a noise parameter of 0.95 and signal counts in excess of 1700. 
Consider an $F_{clock}$ = 1-GHz clock rate source generating entangled photon pairs at $810\,$nm  with P(1) = 1\% pair production per pulse efficiency, used to close the link between a $0.5\,$-m transmitter aperture and a $1.0\,$-m receiver aperture across the baseline of the Earth's diameter, with $10\,$dB of additional losses assumed contributing to the net link efficiency (see Appendix \ref{sec:linkAnalysis}).  In this configuration, roughly 250 events per second are expected, requiring $6.8\,s$ of integration to obtain the desired counting statistics. Achieving a purity in excess of $0.95$ means that for every 20 signal events, there is at most 1 noise event.  Accounting for 2-fold and 4-fold noise counts in a detection system with $500\,ps$ resolution, the required noise count rate is about $N_{noise}$ = 11.25 noise events per second, commensurate with the capabilities of state-of-the-art detector systems \cite{AM2021}, with a photon-time-of-arrival resolution $\Delta t_R$ (and residual timing synchronization error) less than or equal to the optical pulse width of signal photons.  Increasing the aperture sizes to e.g., $0.5\,$m and $3.5\,$m, results in a higher flux of signal photons and relaxes the noise requirement commensurately. 

\subsubsection{Quantum Teleportation: Summary}
The experiments propose to perform a completely quantum mechanical process---the teleportation of the quantum state of one photon to another---in a regime where relativistic effects impact the results.  The DSQL will empirically test whether teleportation across long-range links between inertial frames is successful, as predicted by standard theory.  Successful demonstration of the quantum teleportation experiments described in this Section will thus provide critical empirical justification for what are currently untested assumptions of QFTCST in the weak-field regime.  These experimental regimes are not otherwise achievable in laboratory analog experiments, and truly require spacecraft links.  

Using one or more quantum memories may enhance the teleportation system performance. The key figures of merit of a quantum memory are its bandwidth and wavelength, which should be compatible with the signal photons; its read and write efficiency, which need to be high to avoid introducing more loss to an already lossy channel;  its coherence time and storage time, which are linked to the efficiencies and need to be of comparable magnitude to the time of flight between nodes in the network (or the round-trip light time between two nodes); and the number of storable modes, which needs to be high given the high clock rates and low link efficiencies of the long baseline channels.  Furthermore, the quantum memory modes should be individually addressable and exhibit continuous read-out capability.

Ultimately, the performance parameters of a space-qualified quantum memory that would enhance a proposed DSQL experiment are beyond the current state of the art.  Ground station quantum memory systems are marginally more mature. While no specific implementation plan is proposed at this time, the general philosophy of continuing engineering design of the DSQL mission is to ensure system compatibility with future, ground-based quantum memory systems.

\subsection{Potential Applications of Squeezed Light}
\label{sec:squeezelight}
\setcounter{footnote}{0} 
Squeezed states of light are the quantum states that offer a reduced quantum uncertainty in one quadrature of the electromagnetic mode phase space $(x,p)$ while having an increased uncertainty in the conjugate quadrature. If the area in the phase space representing the squeezed state retains the minimal values, the same as for the vacuum or coherent state (such minimum-uncertainty states are sometimes called the {\it intelligent states}), then the squeezed state is said to have a unity purity and it remains a pure quantum state. If, however, the phase space area is increased, i.e., the anti-squeezing exceeds the squeezing, as is often the case in experiment, the squeezed state is (partially) mixed, and is characterized by a purity value below unity. The purity is therefore an important parameter in the context of squeezed states applications in quantum information processing.

One particularly important example of squeezed states is the so-called squeezed vacuum. This is the state centered on the phase-space diagram so that $\langle x\rangle=\langle p\rangle=0$, which means that its Fock representation contains only even photon-number states $|n\rangle$. In spite of having a zero mean field, the squeezed vacuum carries finite optical energy, which is uniquely determined by the degree of squeezing.

It should be noted that squeezed states are fragile quantum states that decohere quickly under loss or other external coupling.
Usually squeezed states are measured with continuous variable measurement techniques, i.e., by measuring the continuous spectrum of the electromagnetic field variables. The field variables are typically measured by optical homodyne or heterodyne detection, while the amplitude squeezing can be measured by directly observing the reduced optical power fluctuations.
The homodyne detection is mode-selective, i.e., it measures the optical mode defined by a reference beam (the local oscillator). Hence, any changes in the classical mode structure of a beam travelling long distances (e.g., red shifts or change of bandwidth or polarization) will be noticed as decoherence leading to a reduced interferometric visibility. By systematic modification of the reference beam, one can deduce a change in the classical mode structure. This technique will enable one to distinguish between various effects that may act on the optical fields along their path in space, and effects that are acting on the field excitation, e.g., the photon statistics.

Squeezed states of light have been thoroughly investigated in the context of sensitive interferometric measurements surpassing the shot-noise limit. The first such application of squeezed vacuum
was done back in 1987 \cite{Xiao87sqz}. A very prominent modern demonstration of this technology geared for gravitational wave detection was performed in the context of the LIGO project \cite{LIGO11sqz}. In this work a 10-dB squeezed vacuum source was coupled to the dark port of the LISA interferometer, leading to a 3.5-dB noise suppression below the shot noise level. This limited noise suppression is due to the loss in the optical system -- a technical problem that needs to be mitigated in all squeezed light applications. More recently both the LIGO and VIRGO projects have demonstrated sensitivity enhancements using squeezed light \cite{LIGO19sqz,VIRGO19sqz}.

Squeezed states and closely related continuous-variable (CV) entangled states of light have been found useful in the area of quantum information processing. For example, these states can serve as a building block for linear quantum computation \cite{Knill01linearqc,menicucci2006qcomp}. The quantum network applications include CV quantum key distribution protocols \cite{Yuen78sqzcomm,Shapiro79sqzcomm,elser_feasibility_2009}, CV quantum teleportation \cite{Yonezawa04teleport,PhysRevA.76.032305,Ma:2012fk} and entanglement swapping \cite{Takei05teleport,Yonezawa07teleport}. The CV quantum teleportation typically has higher efficiency (albeit lower fidelity) then the discrete-value quantum teleportation, and may become the protocol of choice in situations where the efficiency is a stretched resource (e.g., over large distances)\footnote{Note, however, that discrete variable (DV) quantum communication (in which quantum states are encoded on single photons) can often be operated in a {\it post-selected} regime, i.e., only keeping cases in which the photon actually successfully made it through a lossy channnel; in this case the DV protocols can tolerate much higher loss \cite{Boaron18QKD}. }. Another interesting aspect of quantum teleportation, also highly relevant to space applications, is that it can be achieved not only in bipartite but also in tripartite
systems, in which case Bob receives classical communications both from Alice and a third party \cite{Yonezawa07teleport}. There may be an even larger number of communicating parties, which presents an opportunity for building a quantum network and implementing various multipartite quantum communication protocols.

Besides interferometric sensors, quantum communications, and computations applications, squeezed light can be used in spectroscopy  \cite{Polzik92sqz,Ribeiro97spectrscopy}, biological research \cite{Taylor13sqz}, photochemistry, high-resolution imaging \cite{treps_surpassing_2002,Treps03pointer}, and  calibration of light detector and sources \cite{Brida06cal,Vahlbruch16sqz}. 

Spectroscopic and photochemical applications rely on the fact that squeezed light has unusual two-photon absorption properties. It is predicted that one can achieve a linear (rather than quadratic) dependence of the absorption rate on the optical intensity for weak fields, significantly different absorption rates for  phase- and amplitude-squeezed beams of the same power, and the possibility of a decreasing absorption rate with increasing intensity \cite{Gea-Banacloche89two-phot,Javanainen90two-phot}. Conversely, the enhanced intensity fluctuations of an anti-squeezed state can enhance the two-photon absorption compared to coherent or thermal light \cite{Gea-Banacloche89two-phot,Raymer2021}. 

Calibration of photo-detectors is enabled by strong correlation of the intensity fluctuations in two-mode squeezed optical beams. Originally proposed by D.N. Klyshko, and often associated with his name, this method uses two photon states (e.g., from SPDC) to calibrate a pair of photon-counting detectors \cite{Klyshko80calibration}: Treating the detection of one of the photons as a herald guarantees the presence of the other photon directed to the detector being calibrated: after subtracting noise counts, the detection efficiency is simply the coincidence rate divided by the singles rate at the heralding detector (whose efficiency then cancels out) \cite{Rarity87Det,Kwiat97, migdall07Det}.  The method now has been generalized to a pair of analog detectors \cite{Brida06cal,Vahlbruch16sqz}, and even to a CCD array \cite{Brida10cal}, in which cases the two-photon light source is replaced by a two-mode squeezed light source, and the photocounts are replaced by the photocurrents' fluctuations.

Finally, squeezed states can potentially act as ``probe states''. Here the idea is that because of the reduced intrinsic noise, any signature imprinted on such states can be recovered with better fidelity. In the DSQL settings, this may help determine if the fragile quantum states travel long distances and along changing gravitational fields without decoherence or dephasing, and whether it is possible to distinguish different mechanisms of decoherence.
These questions could be investigated by deploying a squeezed light source on a lunar orbit, and a homodyne detection setup on a second satellite around lunar orbit (or on Earth). However, the feasibility of such a measurement depends on whether one can still detect squeezing given the low collection efficiency typical for such distances, which can be viewed as an additional high loss of the quantum link.
Perhaps a positive answer can be obtained by changing the approach to the squeezed states measurement from a quantitative statistical measurement of the squeezed states properties (e.g., variance measurements) to instead merely distinguish the quantum states, i.e., asking whether a measured state is more likely to be a squeezed state (with coherence preserved) or a classical state (e.g., coherent state).

\section{The DSQL Mission}
\subsection{A Sequence of Missions}
The experiments described in this manuscript can be achieved by a phased deployment of spacecraft and ground infrastructure.  
\begin{enumerate}
    \item \textbf{Phase 1:} Elliptical orbit with multiple ground stations
    \item \textbf{Phase 2:} Spacecraft array with multiple ground stations
    \item \textbf{Phase 3:} Lunar node with extremely large aperture ground station
\end{enumerate}

As indicated throughout the text, spacecraft occupying elliptical orbits are well suited for explorations of relativistic effects.  Phase 1 of DSQL could involve a single spacecraft in such an orbit.  The spacecraft would be outfitted with an optical payload consisting of: a pair of independently gimballed telescopes; a high-rate entangled photon pair source \footnote{Note: Many of the COW-type tests described in Section \ref{sec:equivalence} are agnostic as to whether the source is in space with detectors on the ground or vice versa, but there may be technical reasons to prefer one over the other, i.e., not wanting to fly a high repetition-rate pump laser, or detectors that require cryogenic cooling.}; a high performance single-photon detection system capable of performing photonic state tomography; a stabilized fiber optical delay line; and a reconfigurable optical switch array.  The flight terminal requires exceptional pointing accuracy to leverage larger apertures for high efficiency links.  Recent flight missions have demonstrated performance commensurate with the requirements \cite{fesq:19, Israel:16,Yin1140}. A summary of key technology items is provided in Section \ref{sec:technology} below.  

The Phase 1 system would enable COW tests, tests of quantum teleportation, and a subset of the Bell tests between inertial frames.  An array of ground stations, potentially located around the world, could establish quantum communication links with the spacecraft in support of the  experiments described here, as well as supporting new experiments and technology demonstrations by a user community.  

Phase 2 adds additional spacecraft to the network, in complementary elliptical orbits with lower orbital period (greater orbital semimajor axis) than the Phase 1 spacecraft.  This array of spacecraft will perform COW tests at larger baselines, and allow the full range of inertial frames required to achieve the Bell tests and quantum teleportation tests.  One or more of the spacecraft would be located at a point suitable to support a future human-decision Bell test, either in a 9-day period orbit (roughly corresponding to a ``midway between Earth and moon'' configuration), or in orbit about the fourth/fifth Lagrange point of the Earth-Moon system.

Phase 3 of DSQL provides the capability to perform quantum optical tests well into the regime of 2-body gravitational physics, with baselines long enough to finally perform human-decision Bell tests. Astronauts and large-aperture telescopes on or near the Moon are assumed, e.g., the link analysis given in Appendix \ref{sec:linkAnalysis} considers a 1-m aperture, nearly diffraction-limited telescope on/near the moon. A ground system on Earth will need to be established with an extremely large aperture telescope; there are a number of development efforts underway to produce Earth-based 10-m class telescopes, e.g., JPL is engaged in the design and deployment of an 8.3-m class telescope suitable for supporting deep-space classical optical communications \cite{charles:11}.
The ground system further requires a large collection area coupled efficiently to low-jitter single-photon detectors. Superconducting nanowire single-photon detectors are a technology that has demonstrated system detection efficiencies of 98\% \cite{reddy20SNSPD}, photon number resolution \cite{cahall17PNR,schmidt19PNR}, dark count rates below $10^{-4}$ cps \cite{shibata16SNSPD}, and timing jitter below 3 ps \cite{korzh20jitter} (though not yet all in a single device).

\subsection{DSQL Technology Challenges}
\label{sec:technology}

The DSQL experiments considered and discussed here are at the bleeding edge of what current technology can accomplish, as is evident from the photon rates estimated in Appendix \ref{sec:linkAnalysis}.  Therefore, in addition to the scientific research, we would like to summarize some of the quantum technology developments that could immensely improve the feasibility of the proposals discussed in this article.

The challenges involved in technology advances can be divided into three groups:
\begin{enumerate}
    \item Tasks involved with Research and Development
    \begin{itemize}
    \item Deterministic, high-rate sources of  single  and entangled photons. For instance, the currently considered probabilistic sources (e.g. SPDC-based) can only operate at a fraction of the clock rate due to inherent multi-pair emissions (see Appendix \ref{sec:stats_bell_tests}). Solid-state systems such as quantum dots have made huge advances recently \cite{Warburton2021,Huber2018,Schimpf2020}, however their integration into flight-suitable, cryogenic packaging poses challenges.
    
    The long-range teleportation experiments would also benefit from the improved photon statistics from deterministic single-photons and entangled photons; see Section \ref{teleportation}.
    \item High-rate sources for entangled photons suitable for multiplexing many optical channels using spread-spectrum techniques. SPDC entangled photon sources have already been flown in space, and can be tailored for wide-band operations. Combined with multi-channel filtering, a large pair-rate enhancement could be obtained; see Appendix \ref{sec:stats_bell_tests}.
    \item Various of the experiments proposed would highly benefit from advanced quantum memories. While there are many types and configurations of quantum memories that are explored for ground applications, the DSQL requirements may be different than the usual ground-based applications, as a benefit is possible as long as the memory efficiency is higher than a direct link. Although the COW experiments (see Section \ref{sec:COWtests}) may be implementable with shorter delay-line type memories (i.e., in free space or optical fiber delays), most DSQL proposals could benefit from long-term flying memories. For example, long-range teleportation and long-range entanglement Bell-tests would benefit from a fixed, long-term delay memory (about 1 second), see, e.g., Figure~\ref{NonLocalExpt}, while the relativistic observer experiments would require a quantum memory with shorter maximum storage time (few milliseconds) but with adjustable readout times; see Figure \ref{fig:AfterAfter}. 
    \end{itemize}
    \item Tasks dedicated to Technology Development
        \begin{itemize}
        \item Ultra-high-speed and  high-efficiency detectors for single photons (e.g., based on superconducting nanowires (SNSPD)) have achieved tremendous performance \cite{reddy20SNSPD,cahall17PNR,schmidt19PNR,shibata16SNSPD,korzh20jitter}. However, integrating these devices into a space-amenable system requires further advancement in small-scale cryogenics, and advancing  multi-mode optical fiber interfaces.
    \item Large-scale space telescopes for transmitting or receiving the DSQL quantum signals could benefit from larger apertures, and novel approaches including segmented mirrors or ``deployable'' systems could be beneficial. For ground applications, segmented arrays of optical receivers are another option that could be considered, as the large area of a ``photon bucket'' may be more important than precise optical imaging.
            \item 
        The COW tests require sets of fiber-optic delay lines on different communications nodes.  These delay lines are used in a measurement of photon phase shift induced by gravity.  This effect is on the order of several waves, which requires a commensurate length stability of the local delay lines.  Stabilization may be achieved through active feedback to an atomic reference.  We note here that this concept for stabilizing the fiber delay line is very close to an optical clock, which may also be used as a complimentary tool to explore relativistic effects.
  \item 
  Precision measurement of satellite range and velocity are required.  The COW tests, Bell tests between inertial frames, and tests of quantum teleportation all demand precise accounting for the range and the relative velocities between nodes.  As described at length in Section \ref{sec:COWtests}, these kinematic contributions to phase and frequency are orders of magnitude larger than the relativistic effects DSQL aims to investigate; if left uncorrected, they will dominate the measurements;  advanced spacecraft ranging and velocity measurements are thus required to implement these experiments. 
  \item Time synchronization between nodes is required to perform quantum entanglement swapping and teleportation, since the two photons incident on the beam splitter are required to overlap temporally as well as spatially.  The allowable time-of-arrival error is less than the optical pulsewidth.  In the extreme case, an optical pulsewidth of 1\,ps propagating over a 1.3-s light path (between the Moon and Earth) at 1-10\,GHz repetition rate sets the time synchronization requirement.  
  
    \end{itemize}
    
    \item Ground system infrastructure
    \begin{itemize}
        \item Ground-based systems involving multi-meter aperture telescopes need to be adapted and interfaces for quantum subsystems developed and demonstrated. In particular, given the limited access to such facilities, a very efficient and fast DSQL system should be devised.

  \end{itemize}
\end{enumerate}

\subsection{Other Applications of DSQL}
The instrumentation required to achieve the scientific goals described above is useful for other scientific and technical applications.  

For example, a local stabilized laser system is required at all nodes for the Einstein equivalence principle test.  This subsystem could form the basis of an optical clock, opening the doors to classical clock-comparison experiments.  Used in conjunction with the infrastructure required for the quantum teleportation experiments, the fundamental elements of a quantum network of clocks (Re. \cite{quantum_clock_network}) will be hosted by the DSQL. 

As noted above, the extremely long baseline quantum channels require extremely low noise detection systems to achieve meaningful statistical significance.  The low-noise channel could be exploited in a demonstration of purely classical optical communication to achieve performance close to the asymptotic channel capacity limit \cite{Shannon48, Holevo73}.  

The DSQL telescopes could be directed towards astronomical light sources, where the high-rate, single photon-sensitive receiver could be used for narrow-band, high-speed astronomy \cite{highspeedastro}.   Along similar lines, the quantum state tomography system could be used to assess other astronomical sources by testing for correlations in polarized light emission.  The pair of telescopes at each node could also be used for a demonstration of a quantum telescope array \cite{quantum_telescope}, where one telescope from each node is directed towards an astronomical target, and all telescopes are fed by coherent non-local single-photon states. Coincidence counts between different telescopes are then used to determine the coherence of the astronomical light coming to the telescopes (as a function of their baseline separations), and thereby information about the spatial distribution of the source itself.

\section{Conclusion}

The evolution of quantum states can be predicted using a variety of means (e.g., the Wigner function formalism), none of which are fully compliant with Lorentz invariance, as demanded by  General relativity  for all measurements. This fundamental discrepancy is at the heart of modern physics and motivates the body of experiments proposed in this manuscript. 
As stated in the introduction, QFTCST is a successful theoretical framework supported strongly by astrophysical measurement.  The proposed DSQL experiments present a means of testing QFTCST in a complimentary, weak-field setting local to the Earth.  The results of the DSQL tests will have a significant bearing on theories outside of QFTCST that express coupling between gravitation and quantum states \cite{Ralph_2009, Miles, bruschi14}.

We have proposed a set of experiments that conduct quantum optical measurements in a regime where relativistic effects are strong and measurable.  The EEP tests propose to assess a hitherto untested prediction of general relativity -- that quantum states of light accumulate the expected phase when propagating along geodesic paths defined by the local spacetime.  The Bell tests propose to measure violation of Bell's inequality across extremely long baselines, and between relatively moving inertial frames \cite{Lin2020}.  The latency associated with the long baseline is sufficiently high to close the ``free-will'' loophole through the involvement of astronauts; such human-decision Bell tests have important philosophical and psychological implications as well.  Finally, the quantum teleportation tests will validate the prediction that quantum entanglement is maintained over the long baselines associated with proposals to establish global quantum networks \cite{Simon2017,Gundogan2021}.   

One to four spacecraft, and one or more optical ground stations, would be required to execute some or all of the listed experiments.  The technology required to execute the experiments is mostly present.  Key quantum technology development areas are: high-rate, high-purity, multiplexed entangled photon pair sources; simultaneously high efficiency, low dark-noise, high count-rate single-photon detector systems; and addressable, high efficiency, high repeat-fidelity, and long storage time quantum memories.  Key classical technology development areas are large diameter flight telescopes, radiation-hard high-speed read-out electronics, and modified existing optical ground stations, upgraded with the infrastructure required to close quantum optical links. The technology development, instrument development, and mission execution will benefit immensely from international cooperation and long-term strategic planning.

\section{Abbreviations}%% if any
National Aeronautics and Space Administration (NASA), Deep Space Quantum Link (DSQL), Lunar Gateway (LG), International Space Station (ISS), General Relativity (GR), Quantum  Field  Theory  in  Curved  Spacetime  (QFTCST), Colella-Overhauser-Werner (COW), Hong-Ou-Mandel (HOM), Einstein’s Equivalence Principle (EEP), Universality of Free Fall (UFF), Local Lorentz Invariance (LLI), Universality of Gravitational Redshift (UGR), Low Earh Orbit (LEO), Medium Earth Orbit (MEO), Geostationary Earth Orbit (GEO), Global Positioning System (GPS), Mach-Zehnder  interferometers  (MZI),  Ghirardi-Rimini-Weber-Pearle (GRW-P), Diosi-Penrose (DP), nastopoulos-Blencowe-Hu (ABH), Alternative Quantum Theories (AQT), Newton-Schrödinger  equation (NSE), Bell State Measurement (BSM),  Spontaneous  Parametric  Down  Conversion (SPDC), Field Of View (FOW), Clauser, Horne, Shimony, and Holt (CHSH).

\section{Acknowledgments }%% if any
This research was carried out at the Jet Propulsion Laboratory, California Institute of Technology, under a contract with the National Aeronautics and Space Administration (80NM0018D0004).  \copyright 2021. All rights reserved. This work was supported by the Biological and Physical Science Division of NASA headquarters through a contract with the Jet Propulsion Laboratory, California Institute of Technology. B. L. Hu is supported by NASA/JPL grant 301699-00001. A. Roura is supported by the Q-GRAV Project within the Space Research and Technology Program of the German Aerospace Center (DLR).  T. Jennewein acknowledges support by the Canadian Space Agency (FAST).  G. Vallone and P. Villoresi are supported by the Agenzia Spaziale Italiana (2018-14-HH.0, CUP: E16J16001490001, Q-SecGroundSpace, 2020-19- HH.0 CUP: F92F20000000005 Italian Quantum CyberSecurity I-QKD); INFN MoonLIGHT-2; European Union's Horizon 2020 Framework research and development program (857156, OpenQKD), Ministero dell'Istruzione, dell'Universit\`a e della Ricerca (Departments of Excellence, Law 232/2016).

 \section*{Appendix A: DSQL System Design}
 
        \label{sec:linkAnalysis}
  The basic link expressions characterizing optical transmission paths are expressed in this Section.  These are developed following the procedure of \cite{Klein:74,Degnan:74} and partially carried out in analogy to a previously published evaluation of low earth orbiting satellite-enabled quantum key distribution \cite{Bourgoin:12,Polnik20,cryptography4010007}.

The basic one-way link efficiency for coupling a Gaussian mode through perfectly aligned circular apertures is
\begin{equation}
    \label{eq:Link}
    \eta_L(\vec{R}) = \eta_x(\vec{R})\left(1-\exp\left(\frac{-2\pi^2 D^{2}_{Tx}D^{2}_{Rx}}{\pi^2D^{4}_{Tx}+\left(M^2\right)^{2}|\vec{R}|^2\lambda^2}\right)\right). 
\end{equation}
In Equation \ref{eq:Link} \cite{Johnson2021}, light of wavelength $\lambda$ is transmitted through an aperture of size $D_{Tx}$ with diffraction limit factor $M^2 \geq 1$.  The light is directed towards a receiver aperture of diameter $D_{Rx}$ across a displacement vector of $\vec{R}$. Most of the DSQL experiments are reasonably characterized using the assumption   $|\vec{R}|\gg \pi D^{2}_{Tx}/\lambda$, i.e. the diffracted spot at the receiver is much bigger than the receiver aperture. In this limit Equation \ref{eq:Link} then reduces to the ratio of the receiver aperture to the spot size formed by the transmitter telescope at the receiver plane:  
\begin{equation}
\label{eq:SimpleLink}
    \eta_L(\vec{R}) \approx \eta_x(\vec{R})\left(\frac{D^{2}_{Rx}}{\left(M^2\left(\frac{\lambda}{D_{Tx}}\right)|\vec{R}|\right)^2 } \right),
\end{equation}
where the prefactor $\eta_x$ characterizes other loss factors:
\begin{equation}
\label{eq:AddLosses}
\eta_x = \eta_{Rx} \eta_D \eta_{Tx}  \eta_{atm}(\vec{R}) \eta_{margin}. 
\end{equation}
Here, $\eta_{Rx}$ characterizes the receiver efficiency (except for the detection efficiency) \cite{Degnan:74}, $\eta_{D}$ is the total photon detection efficiency of the receiver, $\eta_{Tx}$ characterizes the transmitter efficiency, clipping efficiency and pointing errors \cite{Klein:74}, $\eta_{atm}(\vec{R})$ characterizes absorption through the atmosphere (which is a strong function of horizon angle), and $\eta_{margin}$  accounts for any additional inefficiencies of the link.

\begin{figure}[t]
    \centering
    \includegraphics[width=0.50\textwidth]{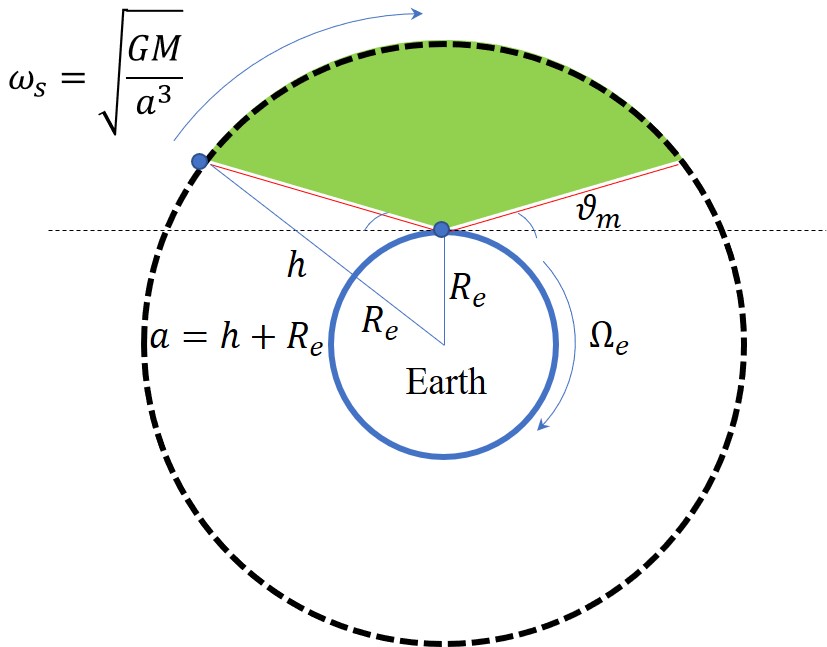}
    \caption{A greatly simplified orbital model to calculate integration time for quantum optical experiments.  A ground station is located on Earth, which rotates at rate $\Omega_e$.  The spacecraft is in an orbit characterized by orbital altitude $h$, or semimajor axis $a=h+R_e$, where $R_e$ is the Earth mean radius. The orbital frequency of the spacecraft is $\omega_s$.  The ground station is limited to view $\theta_m$ above the local horizon angle.  The shaded area in the diagram represents the part of the orbit where line of sight is maintained, in a reference frame rotating with the Earth.}
    \label{fig:Earth_orbit_for_time}
\end{figure}

Equation \ref{eq:Link}  parametrically describes the link efficiency of the quantum channel in terms of the indicated instrumentation performance parameters.  For single photons created at rate $F_{clock}$, the received photon flux $N_{s}$ in units of counts per second is 
\begin{equation}
\label{eq:SinglePhoton}
N_{s} = F_{clock}\cdot \eta_L . 
\end{equation}
Similarly, if only one photon is transmitted from an entangled photon pair source operating at clock rate $F_{clock}$ and a per-pulse photon pair production probability $p(1)$, the number of received photons is
\begin{equation}
N_{e} = F_{clock}\cdot p(1) \cdot \eta_L .  \label{eq:eLinkOneChannel}
\end{equation}
Finally, closing a simultaneous link to a pair of receivers, located at $\vec{Z_1}$ and $\vec{Z_2}$ relative to the source, has an expected total rate of successful events
\begin{equation}
\label{eq:doubleLink}
N_{2e} = F_{clock}\cdot p(1) \cdot \eta_L(\vec{Z_1})\eta_L(\vec{Z_2}). 
\end{equation}
Note that in the limit of perfect spatial and temporal acquisition, the total number of successfully recovered photons is the integral of either Equation \ref{eq:eLinkOneChannel} or \ref{eq:doubleLink} over the time that the spacecraft maintains clear line-of-sight with the other ends of the network \footnote{Due to increased effects of atmospheric turbulence, it is generally assumed that a stable optical link cannot be established for elevation angles less than $20\deg$ above the horizon ($\theta_m$ in Figure \ref{fig:Earth_orbit_for_time}).}.  This can be approximated using the product of the relevant rate equation and the total integration time per orbital passage.  The integration time can be extracted through geometry and numerical analysis of the various orbital configurations described in the sections above \cite{Johnson2021}.  Link expressions such as Equation \ref{eq:doubleLink} require line of sight between the source and two other nodes, which must be satisfied simultaneously.

The integration time is determined by the orbital dynamics.  Our example scenario is an Earth-orbiting spacecraft closing link with a single Earth ground station, though this does not capture the diversity of spacecraft-to-spacecraft links, links between Earth and Moon orbiters, or links to multiple ground stations.  Nevertheless, since many of the experiments described in this report {\it do} rely on links between ground stations on Earth and Earth-orbiting satellites, it is an instructive example.  This situation is depicted in Figure \ref{fig:Earth_orbit_for_time}.  The total integration time $T$ is obtained through geometry and orbital dynamics.  In the regime of perturbation-free, circular orbits about a circular Earth, the integration time can be computed using Equation \ref{eq:integraion_Time}.
\begin{equation}
\label{eq:integraion_Time}
   \centering
T = 2\frac{\sqrt{R_{e}^2+a^2-R_ea\cdot sin(\theta_m)}}{\sqrt{\frac{GM_e}{a^3}}-\Omega_e}\frac{cos(\theta_m)}{a}.
\end{equation}

A separate calculation estimates the total noise count rate incurred during the measurement process. Three sources of noise events are considered: the intrinsic dark count rate of the receiver $D_{r}$; the rate of extra photon events from the source $S_{n}$, and background photons counted by the detection system $B_{sky}$.  The total noise count rate, $N_{noise}$, in units of counts per second, is then:
\begin{equation}
N_{noise} =\eta_{Rx}\cdot\left( W\cdot A\cdot \frac{(FOV)^{2}}{4} \cdot BW + S_{n}\right)+ D_{r},
\label{eq:NoiseEq}
\end{equation}
where $FOV$ is the telescope linear field of view, A is the primary mirror collection area, and $BW$ the filtering bandwidth;  $W$ is the background radiance in units of photon flux per area-solid angle per unit bandwidth.

 \section*{Appendix B: The Human-Decision Bell Tests in the Context of Free Will}
 \label{sec:freeWill}

A human-decision Bell test requires humans on \textit{both} sides of the Bell test- a human ``Bob'' and a human ``Alice''. In the context of a future NASA space mission, this may involve astronauts on the International Space Station (``Bob'') and astronauts on or near the surface of the Moon (``Alice,'' or, perhaps ``Artemis''.)  A local explanation only needs to predict or influence the detector settings on \textit{one} side to violate Bell's Inequality and match the predictions of quantum mechanics, i.e., a local scheme can thus still violate Bell's inequality even if only one side's random number generator is perfectly free will, independent, and unpredictable.  Any amount of unpredictability will suffice to show a Bell violation, but the significance of the violation depends on the predictability. 

While in this experiment, the human choices are not strictly required to pass tests for randomness, these free-will choices must be unpredictable to the measurement on the other side in a way that is different from all prior schemes of generating randomness for Bell tests. At minimum, this requires one of two assumptions, as we now describe.

First, one can take the scientifically accepted materialist view that human choices are results of physical processes in the brain, i.e., some probabilistic combination of deterministic computation and randomness, from the external environment or internal thermal, quantum, or chaotic processes. For human (or, perhaps animal) choices to make a difference, one must assume that the complexity of the process that links physical inputs in the brain's past light cone to a decision must exceed some threshold such that the other side cannot compute, predict, or influence the decision.

Alternatively, one can drop the assumption that human choices are purely results of physical processes in the brain, and instead adopt a stance like Cartesian dualism\footnote{\url{https://plato.stanford.edu/entries/dualism/}}, where one invokes some external non-physical mind that is somehow able to inject events into our 3+1 dimensional spacetime while not being part of it. This is similar to the assumption that is required to close the freedom-of-choice loophole when quantum random number generators are used in the ``loophole-free'' experiments cited above---a truly novel bit of information enters the world such that a 0 and 1 are both perfectly compatible with identical past light cones. The only difference here would be that somehow ``will'' is involved, not just ``freedom''.

The experiment must be implemented in a way that ensures the participants feel their choices come from their own free will, while simultaneously being unpredictable;  otherwise, the physical state of their brains might already be zeroing in on a choice. For example, the participants could be asked a series of questions that they did not know or even consider in advance.  For an astronaut on the moon, it could be questions about their next series of meals---something that matters to them enough to feel they are exercising free will, but something where their answers are not predictable, e.g., ``instant coffee or instant tea,'' to use an astronaut version of Sam Harris' opening question in his book about free will \cite{harris2012free}.  Their answers would need to be registered and turned into polarizer settings in a way that is space-like separated from the source and the measurement on the other side.  The composition and phrasing of the questions asked of the astronauts would need to be carefully considered by experts in behavioral psychology and philosophy of the mind.

Other than choosing basis settings, there is a second way in which the human Bell test addresses foundational principles of quantum theory, going back to Wigner's suggestion \cite{wigner1961remarks} that quantum collapse could somehow be caused by conscious minds. There is barely enough time in an Earth-Moon experiment for measurement results to be shown to participants such that each becomes conscious of the results in a way that is space-like separated from the other. If collapse only happens in conscious minds, no experiment to date has actually closed the locality loophole. One may even consider moving macroscopic masses based on the measurement results, to address Penrose's suggestion \cite{penrose1990emperor} that collapse takes place between macroscopically distinct gravitational fields.

 David Hume states that the question of free will is ``the most contentious question of metaphysics'' \cite{Hume}. An age-old discussion, beginning in the ancient Western philosophical texts of Plato and Aristotle, and continuing to the present day, the question of whether human decisions are based in genuine free will or are deterministic, is still open.
Determinism is the idea that everything in the universe is determined by causal laws. This means that anything in the universe that happens at any given moment is the result of some antecedent cause. Thus, determinism maintains that there is no such thing as an uncaused event. The idea that every event is caused, is one of the fundamental presuppositions of science \cite{Pojman}. According to determinism, since human actions are events, no human action is uncaused, and therefore are not \textit{free}, and instead are simply the result of some causal process. Additionally, determinism precludes randomness -- since everything is the effect of some previous cause, nothing is truly random.

If human beings can exercise free will, then humans are able to perform uncaused acts. For instance, at time $t_{1}$, an agent $S$ can perform either act $A_{1}$ or act $A_{2}$. So, $S$'s action at $t_{1}$ will determine what the world looks like after $t_{1}$, regardless of the pre-existing conditions at $t_{1}$ \cite{Pojman}.  An uncaused act should be unpredictable, non-hysteretical, and otherwise stochastic.  There is a subtle difference between making a ``knee-jerk'' or instinctual reaction to some stimuli versus taking time to internally deliberate decision before action \cite{lamont}.  This distinction underscores the timing latency requirement in human-decision Bell tests.  It also suggests alternative testing schemes with either shorter decision time intervals (forcing an instinctual decision) or longer time intervals (allowing some level of thought before decision making), or even using non-human animals as the decision-making agents.  

Free will may be non-random and predictable by the local observer themself.  Free will is independent of the observed system, assuming that our universe consists of 1) free will of observers and 2) the observed world; both assumptions follow from causality. 

The question of free will versus determinism is based in the question of causality: whether every event must have a cause, or if there are events that are causally undetermined, a question that  impacts the value of scientific inquiry. In this broader context, a Bell test involving human decision-making creates an empirical framework with which to assess the idea of free will, and to explore the relationship between human decision making and the causal trajectory in nature leading to the moment of decision.

\section*{Appendix C: Obtaining the Parametric Model of the COW Tests}

\label{sec:COW_appendix}
The state of the interferometer output (see Figure \ref{fig:SimpleCOW}) can be written as:
\begin{equation}
     |\psi\rangle=\frac{1}{2}\left((e^{i(\phi_{GR}+\theta)}-1)|0,1\rangle+i(e^{i(\phi_{GR}+\theta)}+1)|1,0\rangle\right),
\end{equation}
where the expression for $\phi_{GR}$ is given in Equation \ref{eq:LMalpha}, and $\theta$ is an additional controllable phase that can be tuned to improve the measurement precision, by biasing to the linear part of the fringe. We can model the experimental imperfection by assuming the preparation contains our target state $|\psi\rangle$  with probability $p$, and a noise photon  with  probability $(1-p)$.  The flux of noise photons $N_{noise}$ has been defined in Equation $(\ref{eq:NoiseEq})$. The parameter $p$ can be interpreted as the experiment quality factor and can be linked to the system parameters such as timing, receiver aperture, and spectral filtering bandwidth as follows:
\begin{equation}
    p = (1-N_{noise}\Delta t_{R}) F .
\end{equation}
Here $(N_{noise}\Delta t_{R})$ is the probability of recording a count due to a noise photon within the expected coincidence time window $\Delta t_{R}$, and $F$ is the fidelity of the source, assumed here to be 0.95. We further assume all the system parameters in $N_{noise}$ to be fixed apart from the spectral filtering bandwidth that needs to  change according to the signal photon bandwidth, so that $p=p(\sigma)$. Keeping into account the finite bandwidth $\sigma$ of the photon having frequency $\omega_0$, the probability of having a detection at detector A is given by:
\begin{equation}
    P_A=\frac{p(\sigma)}{2}(1- e^{-2\sigma^2\tau^2}\cos(\phi))+\frac{(1-p(\sigma))}{2}  ,
\end{equation}
where $\phi=\phi_{GR}+\theta$ and $p$ should be regarded as an overall quality factor for the experiment. A more complete analysis should also take into account error sources such as path length mismatch and attitude determination error; these as well other source of imperfection will be explored in detail elsewhere.  

\begin{figure}[t]
    \centering
    \includegraphics[width=10cm]{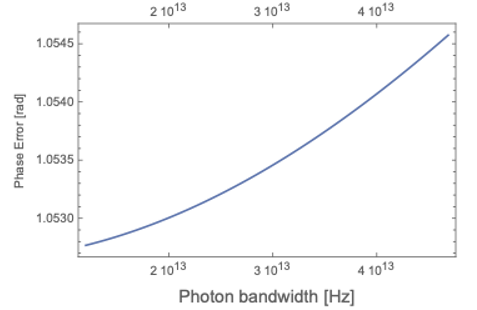}
    \caption{Plot of the phase error, $\Delta \phi_{opt}$, versus the photon bandwidth.}
    \label{timingMZI}
\end{figure}

Let $M$ be the operator denoting a count at detector A; the error on the gravitationally induced phase shift between the interferometric arms is then given by:
\begin{align}
    \label{eq:delta_phi}
 &   \Delta\phi (\phi,\sigma)\equiv \frac{\Delta M(\phi)}{\left|\frac{\partial\left<\hat{M}\right>}{\partial{\phi}}\right|}=\frac{\sqrt{\left<\hat{M^2}\right>-\left<\hat{M}\right>^2}}{\left|\frac{\partial\left<\hat{M}\right>}{\partial{\phi}}\right|}\notag\\
 &= \frac{\sqrt{1-p^2 e^{-4 \frac{\sigma^2\phi^2}{\omega_0^2}} \cos ^2(\phi )}}{p e^{-2 \frac{\sigma^2\phi^2}{\omega_0^2}} \left| 4 \frac{\phi^2}{\omega_0^2} \sigma ^2 \cos (\phi ) + \omega_0  \sin (\phi )\right| }.
\end{align}
Where we have used $\tau=\phi/\omega_0$ .The quantity $\Delta M(\phi) \equiv \sqrt{\left<\hat{M^2}\right>-\left<\hat{M}\right>^2}$ is often referred to as an ``estimator'' for the unknown quantity $\phi$.
Given an experiment with a certain quality factor $p(\sigma)$, we can choose the overall relative phase $\phi$ in order to minimize the previous expression. More formally, we can solve the optimization problem:
\begin{equation}
    \Delta \phi_{opt}(\sigma)=\min_{\phi}  \Delta \phi(\phi,\sigma).
\end{equation}
The optimized phase error can then be propagated in order to obtain the error on the parameter $\alpha$, characterizing violations of UGR. The result of the optimization is shown in Figure \ref{timingMZI}, showing that $\Delta \phi_{opt}$ is essentially constant for the range of the photon bandwidth considered as expected -- as long as the interferometers are well matched, so that the path imbalance is less than the coherence length, the fringe visibility will be very high. 

\section*{Appendix D: Obtaining the Parametric Model for Bell Tests}
\label{sec:stats_bell_tests}
A simple but somewhat old-fashioned way to estimate the statistical significance of a Bell test is to assume uncorrelated trials along with unbiased and uncorrelated random basis selection, though this approach does not address the detector efficiency or memory loopholes. Under these assumptions, there are two types of effects: systematic imperfections which lower the intrinsic and/or measured entanglement fidelity of the produced entangled Bell state, and purely statistical fluctuations on the measurements due to Poisson photon counting statistics. A lower measured entanglement fidelity means a CHSH parameter $S$ that is less than the quantum mechanical maximum of $2\sqrt{2}$, but still hopefully greater than 2, the maximum value that local realism allows. Statistical fluctuations would lead to a measured value of $S$ drawn from a distribution centered around an expectation value (proportional to the entanglement fidelity of the measured quantum state) with a standard deviation of $\sigma$. A 5-sigma result would mean that the measured value of $S$ was 5$\sigma$ above the local-realist limit of 2.

Many effects can degrade the measured entanglement fidelity: intrinsic entanglement fidelity of the source, detector dark counts, sky background, and noise from multi-photon events, i.e., having photons from different entangled pairs each arrive within the coincidence window. In all of these cases, %especially if polarization is used, 
the measurement results on each side of the experiment are completely random and uncorrelated with each other. Experimentally, they produce results that are indistinguishable from those produced by a completely incoherent ``mixed'' state.
As in the previous section, to model the mixed state as measured by the pair of detectors, we take a fraction $p$ of a particular Bell sate, say $|\Phi^+\rangle$, and a fraction $(1-p)$ of the completely incoherent state of dimension 4. The mixed state that is actually measured, where the $(1-p)$ contribution includes both the source's intrinsic incoherence and the external noise, is
\begin{equation}
    \hat \rho = p |\Phi^+\rangle \langle \Phi^+| + (1-p)\frac{1}{4}\hat I .
    \label{eq:mixed_bell_state}
\end{equation}

Given this mixed state, we will first calculate the expected value of the CHSH parameter $S$ and its statistical uncertainty $\sigma$, assuming $N$ total coincidence measurements. % $\sigma$ will be proportional to $1/\sqrt{N}$.
To this end, we first define an experimentally measured correlation coefficient $E(a,b)$ as a function of measurement basis settings $a$ and $b$; its value ranges from $-1$ to $+1$:
\begin{equation}
      E(a,b)\equiv \frac{
    N(a,b) + N(a_\perp,b_\perp) - N(a,b_\perp) - N(a_\perp,b)
    }{
    N(a,b) + N(a_\perp,b_\perp) + N(a,b_\perp) + N(a_\perp,b)
    } ,
\end{equation}
where $N(a,b)$ is the number of coincidences where Alice's photon passes through an analyzer with setting $a$ (e.g., for polarization entanglement, a polarizing beam splitter oriented at angle $a$), and Bob's photon passes through an analyzer with setting $b$. Similarly, $N(a,b_\perp)$ is the number of coincidences where Bob's photon is instead detected in the $b_\perp$ output of his measurement apparatus (e.g., his polarizing beam splitter). There will be 16 such coincidence measurements; the sum of all 16 counts is $N:=\sum_i N_i$.

The CHSH parameter is then defined as the sum of four correlation coefficients, with the sign flipped on the coefficient with the widest separation between the measurement basis angles:
\begin{equation}
   S= E(0,\tfrac{\pi}{8}) + E(0,-\tfrac{\pi}{8}) + E(\tfrac{\pi}{4},\tfrac{\pi}{8}) - E(\tfrac{\pi}{4},-\tfrac{\pi}{8}) . 
\end{equation}
For the mixed state in Eq. (\ref{eq:mixed_bell_state}), the expected CHSH parameter $\langle S \rangle=2 \sqrt{2}p$ \footnote{One could also consider state imperfections of the form $p|\Phi^+\rangle\langle \Phi^+|+(1-p)(|HH\rangle\langle HH|+|VV\rangle\langle VV|)/2$, i.e., a partially entangled state, with perfect correlations only in one basis. In this case the CHSH parameter has an expected value of $\langle S \rangle=2 \sqrt{2}p+2(1-p)$.}. This reaches the quantum mechanical maximum of $2 \sqrt{2}$ for the pure Bell state and 0 for the completely incoherent state, where there are no correlations.

Assuming each of the 16 counts $N_i$ is drawn from a Poisson distribution whose standard deviation $\sigma_i$ is $\sqrt{N_i}$, the variance of $S$ can be calculated through propagation of error as
\begin{equation}
\sigma^2 
= \sum_{i=1}^{16} \sigma_i^2 \left(\frac{\partial S}{\partial N_i}\right)^2 
= \sum_{i=1}^{16} N_i \left(\frac{\partial S}{\partial N_i}\right)^2 ,
\end{equation}
whose expectation value for the state in (\ref{eq:mixed_bell_state}) is
\begin{equation}
\left< \sigma^2 \right> = \frac{8}{N}(2-p^2).
\end{equation}
To claim an $n\cdot \sigma$ violation of Bell's inequality, $S_\textrm{measured}-2>n\sigma$. The expectation value of $n$, the number of $\sigma$'s of Bell violation, is then
\begin{equation}
\textrm{\# of $\sigma$ violation} = \left< n \right> = \sqrt{N} \frac{p-\tfrac{1}{\sqrt{2}}}{\sqrt{2-p^2}} .
\label{eq:nsigma_violation_of_Bell}
\end{equation}

A contour plot of this expected number of $\sigma$ violation as a function of Bell-state fraction $p$ and total coincidence counts $N$ is shown in Figure \ref{fig:Partitioned_bell}. If $p \le \tfrac{1}{\sqrt{2}}$, no Bell violation would occur, and the result would be compatible with local realism no matter how large $N$ is. Near this threshold, quality (high $p$) wins over quantity (high $N$). Above this threshold, the significance of the result scales as $\sqrt{N}$ as might be expected. In practice, once the entanglement fidelity threshold is crossed, accumulating enough data does not take very long. This is qualitatively different than many other physics experiments, where one can ``average down'' the noise.

Now we proceed to estimate $p$ for various space scenarios, determine how many coincidences $N$ are required to achieve a given $\sigma$-level of Bell violation, and estimate the time required to achieve this. An important parameter is $t$, the time window inside of which two photons will be counted as a coincidence. This should be as short as possible to avoid accidental coincidences from dark counts, sky background, or incorrectly paired entangled photons. However, it does not help for $t$ to be shorter than the combined effect of the  time resolution of the detectors, the jitter in the amount of atmospheric delay, and the timing jitter of the optical and electronic systems; otherwise, the signal of true coincidences is also reduced. The intrinsic jitter of the superconducting nanowire single-photon detectors (SNSPD) is 0.1\,ps, determined in part by material parameters of the nanowire. The best reported performance is 3\,ps, but more typically 30\,ps for optimized nanowire and electronics. The atmospheric jitter is typically 10\,ps \cite{Kral:05}. Thus, accounting for 10-100\,ps of excess atmospheric jitter captures the expected dynamics. Next we will see that this window $t$ sets the scale for maximum source brightness and maximum allowable background.

If photons (entangled or not) arrive at a rate $r$ and are Poisson distributed,
\begin{equation}
P(k \textrm{ photons in time window } t) = \frac{(r t)^k e^{-r t}}{k!} .
\end{equation}
For small rates such that $rt \ll 1$,
\begin{equation}
    P(\textrm{1 or more photons in time window } t)  =1-e^{-r t} \approx rt .
\end{equation}

If Alice is receiving photons at a rate $r_a$ and Bob at a rate $r_b$, the probability of recording (though not necessarily detecting) an accidental coincidence within the small coincidence window $t$ is $P_\mathrm{accidental}=(r_a t)(r_b t)$. The rate that accidental coincidences occur is $r_\mathrm{accidental}=r_a r_b t$. 

Next we turn to an imperfect source of entangled photons. Without knowing more about the source itself and any potential dependence on measurement basis choices, we model the source itself as producing pure entangled photons at a rate $r_e$ along with completely incoherent photon pairs at a rate $r_i$. For good sources, $r_i$ will be <5\% of $r_e$. If noiseless detectors were to measure the source directly with no other background, $p$ would be $r_e/(r_e+r_i)$. Both of these rates are reduced by $\eta_a$ for Alice's link and by $\eta_b$ for Bob's link.

Finally, we turn to backgrounds counts and dark counts, whose rates can be added together. For nanowire detectors at the wavelengths of interest, dark counts can be less than 1 per second, assuming (great) care is taken to exclude thermally excited blackbody radiation. Depending on the wavelength, observatory location, phase of the moon, and size of the telescope, the total rates of noise counts $n_a$ and $n_b$ are somewhere between 0.1 and 1000 counts per second.

The full accidental-pairing rate should take into account mispairings between any combination of photons from different entangled pairs, incoherent photons emitted in pairs by the source, background photons, and dark counts. 
Each observer's photon detection rate is the sum of background noise counts $n_a$ or $n_b$ and photon pairs, which are created at the source at a rate $(r_e + r_i)$, and which reach the observers with efficiencies $\eta_a$ and $\eta_b$. This makes
\begin{eqnarray}
    r_a &=& (r_e + r_i) \eta_a + n_a, \quad \textrm{and} \nonumber \\
    r_b &=& (r_e + r_i) \eta_b + n_b . \nonumber 
\end{eqnarray}
The accidental coincidence rate is then
\begin{equation}
    r_\mathrm{accidental} = 
        (r_e \eta_a + r_i \eta_a + n_a)
        (r_e \eta_b + r_i \eta_b + n_b)t,
\end{equation}
where different terms on each side might dominate for asymmetric links.

This is to be compared to the rate of proper entangled pairs arriving. Assuming the coincidence window is set so that the vast majority of entangled pairs that do make it arrive within the window, that rate is just
\begin{equation}
    r_\mathrm{entangled} = r_e \eta_a \eta_b .
\end{equation}
With these definitions, the purity of the state $p$ as measured at the noisy detectors exposed to background light is
\begin{equation}
    p
    = \frac{r_\mathrm{entangled}}{r_\mathrm{entangled}+r_\mathrm{accidental}}  = \frac{1}{1+r_\mathrm{accidental}/r_\mathrm{entangled}} .
\end{equation}
Unless $p > \tfrac{1}{\sqrt{2}}$, no amount of data taking will violate Bell's Inequalities. This corresponds to $r_\mathrm{accidental}/r_\mathrm{entangled}<0.41$. If this holds and $p$ is high enough, the number of photon pairs required to achieve an $n$-$\sigma$ result can be computed using Equation (\ref{eq:nsigma_violation_of_Bell}) or looked up in Figure \ref{fig:Partitioned_bell}. For example, per Figure \ref{fig:Partitioned_bell}, conducting a Bell test to demonstrate a CHSH inequality violation of $3\sigma$ requires 500 counts with entanglement fidelity of $p=0.85$.  To conduct the same test to a violation of $6\sigma$ requires 1000 counts with  $p=0.90$.  Achieving this level of performance requires increasingly stringent instrumentation performance requirements as the baseline is increased, since the signal falls off while the detector noise does not.

The two parameters represented in Figure \ref{fig:Partitioned_bell}, $N$ and $p$, are partitioned into high-level subsystem performance requirements sufficient to achieve the desired level of statistical confidence.  The y-axis value, $N$, is computed by multiplying Eq. \ref{eq:eLinkOneChannel} or Eq. \ref{eq:doubleLink} by the duration of the experiment.

Next we consider, the effects of noise counts.  Equation \ref{eq:mixed_bell_state} characterizes measurements in terms of a noise parameter $p$.  This parameter is composed of a contribution of external noise events (Eq. \ref{eq:NoiseEq}) and the quantum fidelity $F$ of the source.  First, the a source with fidelity $F$ is: 
\begin{equation}
    \hat \rho_{in} = F |\Phi^+\rangle \langle \Phi^+| + (1-F)\frac{1}{4}\hat I .
    \label{eq:mixed_source_F}
\end{equation}

The receiver sums photon counts over a time interval $\Delta t_{R}$, corresponding to a frame-period, word-length, or user-defined integration window.  The smallest value $\Delta t_{R}$ could take would be the total timing jitter of the receiver.  The largest value would be the effective frame-rate of the receiver system, which would be no smaller than the inverse of the product of transmitter clock rate and single-side channel loss.  The probability of counting a noise event  within this period is $P_{N} = \Delta t_{R}\cdot N_{noise}$.  A limiting case describing the worse-case effect of noise events on the measurement process is expressed in Equation \ref{eq:noise_on_rho}:
\begin{equation}
    \hat \rho_{out} = P_{N} \hat \rho_{in} + (1-P_{N})\frac{1}{4}\hat I .
    \label{eq:noise_on_rho}
\end{equation}

Combining Eq. \ref{eq:mixed_source_F} and Eq. \ref{eq:noise_on_rho} with Eq. \ref{eq:mixed_bell_state}, the noise parameter $p$ can be reduced to source fidelity, noise count rate, and coincidence window time using
\begin{equation}
    p = (1-N_{noise}\Delta t_{R}) F.
    \label{eq:partition_noise}
\end{equation}

 \bibliographystyle{apsrev4-1} 
 \bibliography{bibsdt,bibhumans,bibCOW,bibQOW_I,bibsimul,bib_teleport,bib_gravdec,bib_clock,bib_quantum_memories,bib_tech,biblio}

%\end{backmatter}
\end{document}